\def\arxivVersion{}

\ifdefined\arxivVersion
\documentclass[11pt]{article}
\usepackage[margin = 2.6cm]{geometry}
\else
\documentclass[preprint,12pt]{elsarticle}
\fi

\def\arxiv{}

\pdfoutput=1

\usepackage{graphicx}
\DeclareGraphicsExtensions{.pdf,.jpeg,.png}
\usepackage{amsmath}
\usepackage{amsfonts}
\usepackage{amssymb}
\usepackage{amsthm}
\usepackage{stmaryrd}
\usepackage{algpseudocode}
\usepackage[tight,footnotesize]{subfigure}
\usepackage{url}

\theoremstyle{plain}
\newtheorem{theorem}{Theorem}[section]
\newtheorem*{apptheorem}{Theorem}
\newtheorem{lemma}{Lemma}[section]
\newtheorem*{applemma}{Lemma}
\newtheorem{proposition}{Proposition}[section]
\newtheorem*{appproposition}{Proposition}
\newtheorem{corollary}{Corollary}[section]
\theoremstyle{definition}
\newtheorem{definition}{Definition}[section]

\theoremstyle{remark}
\newtheorem{assumption}{Assumption}
\newtheorem{remark}{Remark}[section]
\newtheorem*{exu}{Example}

\newcommand{\bbN}{\mathbb{N}}

\newcommand{\bbR}{\mathbb{R}}
\newcommand{\bbP}{\mathbb{P}}
\newcommand{\bbE}{\mathbb{E}}

\newcommand{\eps}{\varepsilon}

\newcommand{\vr}[1]{\mathbf{#1}}

\def\qed{$\hfill\blacksquare$\newline}

\newcommand{\calC}{\mathcal{C}}

\newcommand{\calT}{\mathcal{T}}

\newcommand{\calX}{\mathcal{X}}

\newcommand{\calD}{\mathcal{D}}

\newcommand{\calP}{\mathcal{P}}

\newcommand{\calS}{\mathcal{S}}

\addtocounter{page}{0}

\newcommand{\N}{^{(N)}}

\newcommand{\calXN}{\mathcal{X}^{(N)}}

\newcommand{\ncalT}{\hat{\mathcal{T}}}
\newcommand{\ncalX}{\hat{\mathcal{X}}}
\newcommand{\ncalXN}{\hat{\mathcal{X}}^{(N)}}

\newcommand{\X}{\mathbf{X}}
\newcommand{\XN}{\mathbf{X}^{(N)}}
\newcommand{\x}{\mathbf{x}}
\renewcommand{\v}{\mathbf{v}}
\newcommand{\y}{\mathbf{y}}

\newcommand{\z}{\mathbf{z}}

\newcommand{\Y}{\mathbf{Y}}

\newcommand{\nX}{\hat{\mathbf{X}}}
\newcommand{\nXN}{\hat{\mathbf{X}}^{(N)}}
\newcommand{\nx}{\hat{\mathbf{x}}}

\newcommand{\FN}{F^{(N)}}

\newcommand{\flow}[2]{\Phi(#1,#2)}
\newcommand{\flowi}[2]{\Phi^{-1}(#1,#2)}
\renewcommand{\phi}{\varphi}

\newcommand{\prjN}[1]{\nu\N(#1)}
\newcommand{\lm}{\llbracket}
\renewcommand{\rm}{\rrbracket}

\newcommand{\lts}[1]{\ensuremath{\stackrel{#1}{\longrightarrow}}}

\newcommand{\until}[2]{\mathbf{U}^{[#1,#2]}}
\newcommand{\next}[2]{\mathbf{X}^{[#1,#2]}}

\newcommand{\bP}{\bar{P}}

\begin{document}

\ifdefined\arxivVersion

%\title{Model Checking Single Agent Behaviours\\ by Fluid Approximation}
\title{Fluid Model Checking}

\author{Luca Bortolussi \\
Department of Mathematics and Geosciences\\
University of Trieste, Italy.\\
CNR/ISTI, Pisa, Italy.\\
\texttt{luca@dmi.units.it}\\
\and Jane Hillston\\
Laboratory for the Foundations of Computer Science,\\ 
School of Informatics,  University of Edinburgh, UK.\\
\texttt{jane.hillston@ed.ac.uk}}

\date{}

\maketitle

\pagenumbering{arabic}

\begin{abstract}
In this paper we investigate a potential use of fluid approximation techniques in the context of stochastic model checking of CSL formulae. We focus on properties describing the behaviour of a single agent in a (large) population of agents, exploiting a limit result known also as fast simulation. 
In particular, we will approximate the behaviour of a single agent with a time-inhomogeneous CTMC, which depends on the environment and on the other agents only through the solution of the fluid differential equation, and model check this process. We will prove the asymptotic correctness of our approach in terms of satisfiability of CSL formulae. We will also present a procedure to model check time-inhomogeneous CTMC against CSL formulae. 

\

\noindent\textbf{Keywords:} Stochastic model checking; fluid approximation; mean field approximation; reachability probability; time-inhomogeneous Continuous Time Markov Chains

\end{abstract}

\else

\journal{Information and Computation}

\begin{frontmatter}

\title{Model Checking Single Agent Behaviours\\ by Fluid Approximation}

\author[TS,PI]{Luca Bortolussi}
\address[TS]{Department of Mathematics and Geosciences\\
University of Trieste, Italy.}
\address[PI]{CNR/ISTI, Pisa, Italy.}
\ead{luca@dmi.units.it}
\author[ED]{Jane Hillston}
\address[ED]{Laboratory for the Foundations of Computer Science,\\ School of Informatics,  University of Edinburgh, UK.}
\ead{jane.hillston@ed.ac.uk}

%\def\titlerunning{Fluid Model Checking}
%\def\authorrunning{L. Bortolussi \& J. Hillston}

%\maketitle

%*****************************************************************************

\begin{abstract}
In this paper we investigate a potential use of fluid approximation techniques in the context of stochastic model checking of CSL formulae. We focus on properties describing the behaviour of a single agent in a (large) population of agents, exploiting a limit result known also as fast simulation. 
In particular, we will approximate the behaviour of a single agent with a time-inhomogeneous CTMC, which depends on the environment and on the other agents only through the solution of the fluid differential equation, and model check this process. We will prove the asymptotic correctness of our approach in terms of satisfiability of CSL formulae. We will also present a procedure to model check time-inhomogeneous CTMC against CSL formulae. 
\end{abstract}

\begin{keyword}
Stochastic model checking \sep fluid approximation \sep mean field approximation \sep reachability probability \sep time-inhomogeneous Continuous Time Markov Chains 

\end{keyword}

\end{frontmatter}
%*****************************************************************************
\fi

%\linenumbers

\section{Introduction}
\label{secIntro}

In recent years, there has been a growing interest in fluid approximation techniques in 
the formal methods community 
\cite{PA:Bakhshi:2009:meanFieldGossip, PA:Bakhshi:2011:meanFieldPushPullGossip, PA:Remke:2011:MeanFieldP2P,  PA:Hillston:2005:ODEandPEPA, My2009TCSBjournalSCCPandODE, PA:LeBoudec:2008:MeanFieldContinuousTime}. 
These techniques, also known as mean 
field approximation, are useful for analysing quantitative models of systems based on 
continuous time Markov Chains (CTMC), possibly described in process algebraic terms. 
They work by approximating the discrete state space of the CTMC by a continuous one, and 
by approximating the stochastic dynamics of the process with a deterministic one, expressed 
by means of a set of differential equations. The asymptotic correctness of this approach is 
guaranteed by limit theorems 
\cite{STOC:Kurtz:1970:ODEandCTMC, 
STOC:Darling:2002:PracticalFluid, STOC:DarlingNorris:2008:DifferentialEquationsCTMC}, 
showing the convergence of the CTMC to the fluid ODE for systems of increasing sizes.

The notion of size can be different from domain to domain, yet in models of interacting agents, 
usually considered in computer science, the size has the standard meaning of population number. 
All these fluid approaches, in particular, require a shift from an agent-based description to a 
population-based one, in which the system is described by variables counting the number of 
agents in each possible state and so individual behaviours are abstracted. 
In fact, in large systems, the individual choices of single agents have a small impact, hence the 
whole system tends to evolve according to the average behaviour of agents. 
Therefore, the deterministic description of the fluid approximation is mainly related to the 
average behaviour of the model, and information about statistical properties is generally lost, 
although it can be partially recovered by introducing fluid equations of higher order moments 
of the stochastic process (moment closure techniques \cite{PA:BradleyHaiden:2010:FluidFrameworkPEPA, 
SB:Hespanha:2006:momentClosure, My:QAPL:masterEquationSCCP:2008}).

Differently to fluid approximation, the analysis of quantitative systems like those described by 
process algebras can be carried out using quantitative model checking.  These techniques have a 
long tradition in computer science and are powerful ways of querying a model and extracting 
information about its behaviour.  As far as stochastic model checking is considered, there are some 
consolidated approaches based mainly on checking Continuous Stochastic Logic (CSL) formulae 
\cite{MC:Hermanns:2000:MCofCTMCtransient, MC:Aziz:1996:VerifyingCTMC, 
MC:Kwiatkowska:2004:ProbabilisticModelChecking}, which led to widespread software tools 
\cite{MC:Kwiatkowska:2004:PRISM}.  All these methods, however, suffer (in a more or less relevant 
way) from the curse of state space explosion, which severely hampers their practical applicability. In 
order to mitigate these combinatorial barriers, many techniques have been developed, many of them 
based on some notion of abstraction or approximation of the original process 
\cite{MC:Norman:2008:GameBasedAbstractionPRISM, MC:Norman:2009:AbstractionRefinement}. 

In this paper, we will precisely target this problem, trying to see to what extent fluid approximation 
techniques can be used to speed up the model checking of CTMC.  
We will not tackle this problem in general, but rather we will focus on a restricted subset of system 
properties: We will consider population models, in which many agents interact, and then focus on 
the behaviour of single agents.  In fact, even if large systems behave almost deterministically, the 
evolution of  a single agent in a large population is always stochastic.  Single agent properties are 
interesting in many application domains. For instance, in performance models of computer networks, 
like client-server interaction, one is often interested in the behaviour and in quality-of-service metrics 
of a single client (or a single server), such as  the waiting time of the client or the probability of a time-out. 

Single agent properties may also be interesting in other contexts. For instance, in ecological models,  
one may be interested in the chances of survival or reproduction of an animal, or in its foraging patterns 
\cite{PA:Sumpter:2000:WSCCSinsects}. In biochemistry, there is some interest in the stochastic properties 
of single molecules in a mixture (single molecule enzyme kinetics 
\cite{SB:QianElson:2002:singleMoleculeEnzymology, SB:GillespiePetzold:2009:legitimacyMichaelisMenten}). 
Other examples may include the time to reach a certain location in a traffic model of a city, or the chances 
to escape successfully from a building in case of emergency egress \cite{PA:Massink:2012:EmergencyEgress}. 

The use of fluid approximation in this restricted context is made possible by a corollary of the fluid 
convergence theorems, known by the name of \emph{fast simulation} \cite{PA:Gast:2010:workStealing, 
STOC:DarlingNorris:2008:DifferentialEquationsCTMC}, which provides a characterization of the 
behaviour of a single agent in terms of the solution of the fluid equation: the agent senses the rest of the 
population only through its ``average'' evolution, as given by the fluid equation. 
This characterization can be proved to be asymptotically correct. 

Our idea is simply to use the CTMC for a single agent obtained from the fluid approximation instead of 
the full model with $N$ interacting agents. In fact, extracting metrics from the description of the global 
system can be extremely expensive from a computational point of view. Fast simulation, instead, allows 
us to abstract the system and study the evolution of a single agent (or of a subset of agents) by decoupling 
its evolution from the evolution of its environment. This has the effect of drastically reducing the dimensionality 
of the state space by several orders of magnitude.

Of course, in applying the mean field limit, we are introducing an error which is difficult to control (there are 
error bounds but they depend on the final time and they are very loose 
\cite{STOC:DarlingNorris:2008:DifferentialEquationsCTMC}). However, this error in practice will not be too 
large, especially for systems with a large pool of agents. We stress that these are the precise cases in which 
current tools suffer severely from state space explosion, and that can mostly benefit from a fluid approximation. 
However, we will see in the following that in many cases the quality of the approximation is good also for small 
populations. 

In the rest of the paper, we  will basically focus on how to analyse single agent properties of three kinds:
\begin{itemize}
\item Next-state probabilities, i.e.\ the probability of jumping into a specific set of states, at a specific time.
\item Reachability properties, i.e.\ the probability of reaching a set of states $G$, while avoiding unsafe states $U$.  
\item Branching temporal logic properties, i.e.\ verifying CSL formulae. 
\end{itemize}

A central feature of the abstraction based on fluid approximation is that the limit of the model of a single 
agent has rates depending on time, via the solution of the fluid ODE\@. Hence, the limit models are 
time-inhomogeneous CTMC (ICTMC). This introduces some additional complexity in the approach, as 
model checking of ICTMC is far more difficult than the homogeneous-time case. To the best of the author's 
knowledge, in fact, there is no known algorithm to solve this problem in general, although related work is 
presented in Section \ref{sec:relatedWork}. We will discuss a general method in Sections \ref{sec:next}, 
\ref{sec:reachability} and \ref{sec:CSLmodelCheking}, based on the solution of variants of the Kolmogorov 
equations, which is expected to work for small state spaces and controlled dynamics of the fluid approximation. 
The main problem with ICTMC model checking is that the truth of a formula can depend on the time at which 
the formula is evaluated. Hence, we need to impose some regularity on the dependency of rates on time to 
control the complexity of time-dependent truth.  We will see that the requirement, piecewise analyticity of rate 
functions, is intimately connected not only with the decidability of the model checking for ICTMC, but also with 
the lifting of convergence results from CTMC to truth values of CSL formulae (Theorems \ref{th:CSLdecidabilityICTMC} 
and  \ref{th:CSLconvergence}).

The paper is organized as follows: in Section \ref{sec:relatedWork} we discuss work related  to our approach. 
In Section \ref{sec:Basics}, we introduce preliminary notions, fixing the class of models considered (Section 
\ref{sec:modelingLanguage}) and presenting fluid limit and fast simulation theorems (Sections \ref{sec:KurtzTheorem} 
and \ref{sec:fastSimulation}). In Section \ref{sec:next} we present the algorithms to compute next-state probability. 
In Section \ref{sec:reachability}, instead, we consider the reachability problem, presenting a method to solve it 
for ICTMC\@. In both cases, we also discuss the convergence of next-state and reachability probabilities for 
increasing population sizes. In Section \ref{sec:CSLmodelCheking}, instead, we focus on the CSL model checking 
problem for ICTMC, exploiting the routines for reachability developed before. We also consider the convergence 
of truth values for formulae about single agent properties.  Finally, in Section \ref{sec:Conc}, we discuss open 
issues and future work. All the proofs of propositions, lemmas, and theorems of the paper are presented in 
Appendix \ref{app:proofs}. A preliminary version of this work has appeared in \cite{My:CONCUR2012:FMC}.

%*************************************

%To see this in practice, consider a model of the client-server system, with 100 clients and 50 servers. Its state space, when we track the evolution of a specific client, has a size on the order of $10^7$, while the state space of a single client, applying theorem \ref{th:fastSimulation}, has size 4!

\section{Related work}
\label{sec:relatedWork}

Model checking (time homogeneous) Continuous Time Markov Chains (CTMC) against Continuous Stochastic 
Logics (CSL) specifications  has a long tradition in computer science \cite{MC:Hermanns:2000:MCofCTMCtransient, 
MC:Aziz:1996:VerifyingCTMC, MC:Kwiatkowska:2004:ProbabilisticModelChecking}. At the core of our approach 
to study time-bounded properties there are similarities to that developed in \cite{MC:Hermanns:2000:MCofCTMCtransient},
because we consider a transient analysis of a Markov chain whose structure has been modified to reflect the formula
under consideration.  But the technical details of the transient analysis, and even the structural modification, differ
to reflect the time-inhomogeneous nature of the process we are studying.

In contrast, the case of time-inhomogeneous CTMCs has received much less attention. To the best of the authors' 
knowledge, there has been no previous proposal of an algorithm to model check CSL formulae on a ICTMC\@. 
Nevertheless model checking of ICTMCs has been considered with respect to other logics.  Specifically, previous
work includes model checking of HML and LTL logics on ICTMC\@.

In \cite{MC:Mereacre:2008:HMLictmc}, Katoen and Mereacre propose a model checking algorithm for Hennessy-Milner 
Logic on ICTMC\@. Their work is based on the assumption of piecewise constant rates (with a finite number of pieces) within
the ICTMC\@. The model checking algorithm is based on the computation of integrals and the solution of algebraic 
equations with exponentials (for which a bound on the number of zeros can be found). 

LTL model checking for ICTMC, instead, has been proposed by Chen \emph{et al.}\ in 
\cite{MC:Mereacre:2009:LTLmcICTMC}. The approach works for time-unbounded formulae by constructing the 
product of the CTMC with a generalized B\"uchi automaton constructed from the LTL formula, and then reducing 
the model checking problem to computation of reachability of bottom strongly connected components in this larger 
(pseudo)-CTMC.    The authors also propose an algorithm for solving time bounded reachability similar to the one 
considered in this paper (for time-constant sets).

Another approach related to the work we present is the verification of CTMC against deterministic time automata 
(DTA) specifications \cite{MC:Mereacre:2011:DTAmc}, in which the verification works by taking the product of the 
CTMC with the DTA, which is then converted into a Piecewise Deterministic Markov Process (PDMP, 
\cite{STOC:Davis:1993:PDMP}), and then solving a reachability problem for the so obtained PDMP.  This extends
earlier work by Baier \emph{et al}.~\cite{asCSL} and Donatelli \emph{et al}.~\cite{CSL-TA}.  These approaches were
limited to considering only a single clock.  This means that they are albe to avoid the consideration of ICTMC, in the
case of \cite{CSL-TA}, through the use of supplementary variables and subordinate CTMCs.

In \cite{MC:Mereacre:2011:MTLmc}, Chen \emph{et al.}\ consider the verification of time-homogenenous CTMC against
formulae in the the metric temporal logic (MTL).  This entails finding the probability of a set timed paths that satisfy the
formula over a fixed, bounded time interval.  The approach taken is one of approximation, based on an estimate
of the maximal number of discrete jumps that will be needed in the CTMC, $N$, and timed constraints over the residence
time within states of a path with up to $N$ steps.  The probabilities are then determined by a multidimensional integral.

Our work is underpinned by the notion of fast simulation, which has previously been applied in a
number of different contexts \cite{STOC:DarlingNorris:2008:DifferentialEquationsCTMC}.  One recent case is a study of 
policies to balance the load between servers in large-scale clusters of heterogeneous processors 
\cite{PA:Gast:2010:workStealing}.  A similar approach is adopted in \cite{GN:Tembine:2009}, in the context of 
Markov games.    These ideas also underlie the work of Hayden \emph{et al.}\ in  \cite{PA:Hayden:2012:passageTimes}.
Here the authors  extend the consideration of transient characteristics as captured by the fluid approximation, to approximation
of first passage times, in the context of models generated from the stochastic process algebra PEPA.  Their approach for
passage times related to individual components is closely related to the fast simulation result and the work presented in this paper.

%*****************************************************************************

\section{Preliminaries}
\label{sec:Basics}

In this section, we will introduce some backgound material needed in the rest of the paper. First of all, we 
introduce a suitable notation to describe the population models we are interested in. This is done in 
Section~\ref{sec:modelingLanguage}. In particular, models will depend parametrically on the (initial) 
population size, so that we are in fact defining a sequence of models. Then, in Section \ref{sec:KurtzTheorem}, 
we present the classic fluid limit theorem, which proves convergence of a sequence of stochastic models to 
the solution of a differential equation. In Section \ref{sec:fastSimulation}, instead, we describe fast simulation, 
a consequence of the fluid limit theorem which connects the system view of the fluid limit to the single agent 
view, providing a description of single agent behaviour in the limit. Finally, in Section \ref{sec:CSL}, we recall 
the basics of Continuous Stochastic Logic (CSL) model checking.

\subsection{Modelling Language}
\label{sec:modelingLanguage}
In the following, we will describe a basic language for CTMC, in order to fix the notation. 
We have in mind population models, where a population of agents, possibly of different kinds, 
interact together through a finite set of possible actions.
To avoid a notational overhead, we assume that the number of agents is  constant during the simulation, 
and equal to $N$. Furthermore, we do not explicitly distinguish between different classes of agents in the notation.

In particular, let $Y\N_i \in S$ represent the state of agent $i$, where $S=\{1,2,\ldots,n \}$ is the state 
space of each agent. Multiple classes of agents can be represented in this way by suitably partitioning 
$S$ into subsets, and allowing state changes only within a single class.  Notice that we made explicit the 
dependence on $N$, the total population size.
\\
A configuration of a system is thus represented by the tuple $(Y\N_1,\ldots,Y\N_N)$. When dealing with 
population models, it is customary to assume that single agents in the same internal state cannot be 
distinguished, hence we can move from the agent representation to the system representation by introducing 
variables counting how many agents are in each state. With this objective, define 
\begin{equation}
X\N_j = \sum_{i=1}^N \vr{1}\{Y\N_i = j\},
\end{equation}
so that the system can be represented by the vector $\X\N = (X\N_1,\ldots,X\N_n)$, whose dimension is 
independent of $N$. The domain of each variable $X\N_j$ is obviously $\{1,\ldots,N\}$.

We will describe the evolution of the system by a set of transition rules at this global level. This simplifies the 
description of synchronous interactions between agents. The evolution from the perspective of a single agent 
will be reconstructed from the system level dynamics. 
In particular, we assume that $\X\N$ is a CTMC (Continuous-Time Markov Chain), with a dynamics described 
by a fixed number of transitions, collected in the set $\calT\N$. 
Each transition $\tau\in\calT\N$ is defined by a \emph{multi-set} of \emph{update rules} $R_\tau$ and by a rate 
function $r\N_\tau$. 
The multi-set\footnote{The fact that  $R_\tau$ is a multi-set, allows us to model events in which agents in the same 
state synchronise.} $R_\tau$ contains update rules $\rho\in R_\tau$ of the form $i \rightarrow j$, where $i,j\in S$. 
Each rule specifies that an agent  changes state from $i$ to $j$. Let $m_{\tau,i\rightarrow j}$ denote the multiplicity 
of the rule $i \rightarrow j$ in $R_\tau$. 
%either $\rho_1,\rho2\in S$ or only one of them equals $\emptyset$. 
% if $\rho_1,\rho_2\neq \emptyset$, or that an agent is created ($\rho_1 = \emptyset$) or destroyed ($\rho_2 = \emptyset$). 
% The assumption that the total number of agents is constant can be enforced by disabling rules containing a $\emptyset$.   
We assume that $R_\tau$ is independent of $N$, so that each transition involves a finite and fixed number of individuals.
Given a multi-set of update rules $R_\tau$, we can define the \emph{update vector} $\v_\tau$ in the following way:
$$\v_{\tau,i} = \sum_{(i\rightarrow j) \in R_\tau} m_{\tau,i\rightarrow j}\vr{1}_j - \sum_{(i\rightarrow j) \in R_\tau} 
m_{\tau,i\rightarrow j}\vr{1}_i,$$
where $\vr{1}_i$ is the vector equal to one in position $i$ and zero elsewhere.
Hence, each transition changes the state from $\X\N$ to $\X\N + \v_\tau$.
The rate function $r\N_\tau(\X)$ depends on the current state of the system, and specifies the speed of the
 corresponding transition. It is assumed to be equal to zero if there are not enough agents available to perform a 
$\tau$ transition. Furthermore, it is required to be \emph{Lipschitz continuous}.  
We indicate such a model by $\calX\N = (\X\N,\calT\N,\x\N_0)$, where $\x\N_0$ is the initial state of the model. 

Given a model $\calX\N$, it is straightforward to construct the CTMC associated with it, exhibiting its infinitesimal generator matrix. First, its state space is $\calD = \{(x_1,\ldots,x_n)~|~x_i\in\{1,\ldots,N\},\sum_i x_i = N\}$. 
The infinitesimal generator matrix $Q$, instead, is the $\calD\times\calD$ matrix defined by $$q_{\x,\x'} = 
\sum\{r_\tau(\x)~|~\tau\in\calT,~\x' = \x +
\vr{v}_\tau\}.$$ We will indicate the state of such a CTMC at time $t$ by $\X(t)$.

\begin{figure}
\begin{center}
\includegraphics[width=.5\textwidth]{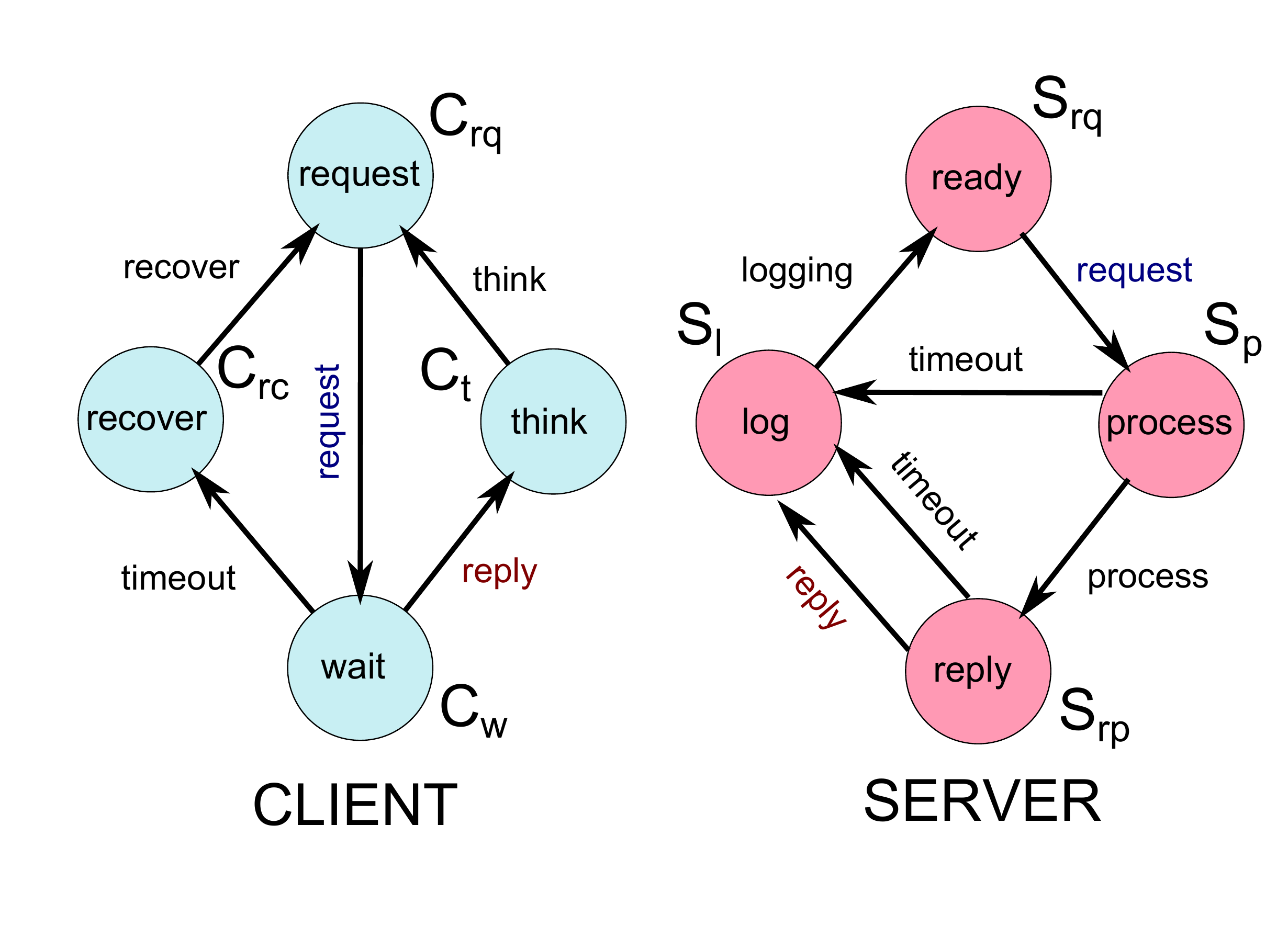}
\end{center}
\caption{Visual representation of the client server system of the running example.} \label{fig:CS}
\end{figure}

\begin{exu}
We introduce now the main running example of the paper: we will consider a model of a simple client-server system, 
in which a pool of clients submits queries to a group of servers, waiting for a reply. In particular, the client asks for 
information from a server and waits for it to reply. It can time-out if too much time passes. The server, instead, after 
receiving a request does some processing and then returns the answer. It can time-out while processing and while it 
is ready to reply. After an action, it always logs data. The client and server agents are visually depicted in 
Figure~\ref{fig:CS}. The global system is described by the following 8 variables:
\begin{itemize}
\item 4 variables for the client states: $C_{rq}$, $C_{w}$, $C_{rc}$, and $C_{t}$. 
\item 4 variables for the server states: $S_{rq}$, $S_{p}$, $S_{rp}$, and $S_{l}$. 
\end{itemize}
Furthermore, there are 9 transitions in total, corresponding to all possible arrows of Figure \ref{fig:CS}. We list them 
below, stressing that synchronization between clients and servers has a rate computed using the minimum, in the 
PEPA style \cite{PA:Hillston:1996:CompositionalPerformanceModeling}. With $\vr{1}_X$ we denote a vector of 
length $n$ which is equal to 1 for component $X$ and zero elsewhere.
\begin{itemize}
\item request: $R_{request} = \{C_{rq} \rightarrow C_w, S_{rq} \rightarrow  S_p\}$, $r_{request} = k_r\cdot \min(C_{rq},S_{rq}))$; 
\item reply: $R_{reply} = \{C_{w} \rightarrow C_{t}, S_{rp} \rightarrow  S_{l}\}$, $r_{reply} = \min(k_w C_w, k_{rp} S_{rp})$; 
\item timeout (client): $R_{timeout1} = \{C_{w} \rightarrow C_{rc}\}$, $r_{timeout1} = k_{to} C_w$; 
\item recover: $R_{recover} = \{C_{rc} \rightarrow C_{rq}\}$, $r_{recover} = k_{rec} C_{rc}$; 
\item think: $R_{think} = \{C_{t} \rightarrow C_{rq}, S_{rp} \rightarrow  S_{l}\}$, $r_{think} = k_t C_t$; 
\item logging: $R_{logging} = \{S_{l} \rightarrow  S_{rq}\}$, $R_{logging} = k_l S_l$; 
\item process: $\v_{process} = \{S_{p} \rightarrow  S_{rp}\}$, $R_{process} = k_p S_p$; 
\item timeout (server processing): $R_{timeout2} = \{S_{p} \rightarrow  S_{l}\}$, $r_{timeout2} = k_{sto} S_p$; 
\item timeout (server replying): $R_{timeout3} = \{S_{rp} \rightarrow  S_{l}\}$, $r_{timeout3} = k_{sto} S_{rp}$; 
\end{itemize}
\end{exu}

\bigskip

The system-level models we have defined depend on the total population $N$ and on the ration between server 
and clients, which is specified by the initial conditions. Increasing  the total population $N$ (keeping fixed the 
client-server ratio), we obtain a sequence of models, and we are interested in their limit behaviour, for $N$ going 
to infinity.

%
%
%
%
%
%The limit results of interest
%describe the behaviour of a CTMC model in the limit of large
%populations, i.e. when the (initial) value of variables $\X$
%diverges. In order to state such theorems, we need to construct a
%sequence $\calXN$ of models depending on $N$, an index related to
%the population level, or \emph{system size}. This notion can be
%instantiated in several ways. For instance, it can represent the
%size of the initial population, like in standard epidemics
%models~\cite{STOC:AndersonBritton:2000:StochasticEpidemics}, or it
%can be a multiplicative factor of the initial population of a basic
%model, like in the fluid semantics of
%PEPA~\cite{PA:Tribastone:2010:FluidPEPA}, or it can also be
%related to a physical notion of size, like the volume for chemical
%reactions~\cite{SB:Gillespie:2006:SimulationBiochemicalKineticsSummary}.
%In this paper, we adopt the following interpretation: the variables
%of the sequence $\calXN$ take integer values, and the size of
%the initial population of the model is proportional to $N$.

In order to compare the models of such a sequence, we will normalize them to the same scale, dividing each 
variable by $N$ and thus introducing the normalized variables $\nXN = \frac{\XN}{N}$. In the case of a constant
 population, normalised variables are usually referred to as the \emph{occupancy measure}, as they represent 
 the fraction of agents in each state.  Update vectors are scaled correspondingly, i.e.\ dividing them by $N$.
Furthermore, we will also require a proper scaling (in the limit) of the rate functions of the normalized models. More precisely, let $\calX\N = (\X\N,\calT\N,\vr{X_0}\N)$ be the $N$-th non-normalized model and $\ncalX\N = (\nX\N,\ncalT\N,\vr{\hat{X}_0}\N)$ the corresponding normalized model. We require that:
\begin{itemize}
  \item initial conditions  scale appropriately: $\vr{\hat{X}_0}\N = \frac{\vr{X_0}\N}{N}$;
  \item for each transition $(\v_\tau,r\N_\tau(\X))$ of the non-normalized model, we let $\hat{r}\N_\tau(\nX)$ be the rate 
  function expressed in the normalised variables (i.e.\ after a change of variables). The corresponding transition in 
  the normalized model is $(R_\tau,\hat{r}\N_\tau(\nX))$, with update vector equal to $\frac{1}{N}\v_\tau$. 
  We assume that there exists a bounded and Lipschitz continuous function $f_\tau(\nX):E\rightarrow\bbR^n$ on normalized variables (where $E$ contains all domains of all $\ncalX\N$), independent of $N$, such that  $\frac{\hat{r}\N_\tau(\x)}{N} \rightarrow f_\tau(\x)$ \emph{uniformly} on $E$.  
\end{itemize}
In accordance with the previous subsection, we will denote the state of the CTMC of the $N$-th non-normalized 
(resp.\ normalized) model at time $t$ as $\X\N(t)$ (resp.\ $\nX\N(t)$).

\begin{exu}
Consider again the running example. If we want to scale the model with respect to the scaling parameter $N$, we 
can increase the initial population of clients and servers by a factor $k$ (hence keeping the client-server ratio 
constant), similarly to \cite{PA:Tribastone:2012:FluidPEPA}. The condition on rates, in this case, automatically holds 
due to their (piecewise) linear nature. For non-linear rate functions, the convergence of rates can usually be enforced 
by properly scaling parameters with respect to the total population $N$. 
\end{exu}

\subsection{Deterministic limit theorem}
\label{sec:KurtzTheorem}

In order to present the ``classic'' deterministic limit theorem, we need to introduce a few more concepts needed 
to construct the limit ODE. Consider a sequence of normalized models $\ncalXN$ and let $\v_\tau$ be the (non-
normalised) update vectors. The drift $\FN(\nX)$ of $\ncalX$ is defined as  
\begin{equation}\label{eqn:drift}
\FN(\nX) = \sum_{\tau\in\ncalT} \frac{1}{N}\v_\tau \hat{r}\N_\tau(\nX) 
\end{equation}
Furthermore, let $f_\tau:E\rightarrow\bbR^n$,
$\tau\in\ncalT$ be the limit rate functions of
transitions of $\ncalXN$. We
define the \emph{limit drift} of the model $\ncalXN$ as
\begin{equation}\label{eqn:drift}
F(\nX) = \sum_{\tau\in\ncalT} \v_\tau f_\tau(\nX) 
\end{equation}
It is easily seen that $\FN(\x)\rightarrow F(\x)$ uniformly.

The limit ODE is $\frac{d\x}{dt} = F(\x)$, with $\x(0) = \vr{x_0}\in
S$. Given that $F$ is Lipschitz in $E$ (as all $f_\tau$ are), the ODE has a unique
solution $\x(t)$ in $E$ starting from $\vr{x_0}$. Then, the following theorem can be proved 
\cite{STOC:Kurtz:1970:ODEandCTMC, STOC:Darling:2002:PracticalFluid}:

\begin{theorem}[Deterministic approximation~\cite{STOC:Kurtz:1970:ODEandCTMC,STOC:Darling:2002:PracticalFluid}]\label{th:Kurtz}
Let the sequence $\nXN(t)$ of Markov processes and $\x(t)$ be
defined as before, and assume that there is some point $\vr{x_0}\in
S$ such that $\nXN(0)\rightarrow\vr{x_0}$ in probability. Then, for
any \emph{finite} time horizon $T<\infty$, it holds that:
$$\bbP\left\{\sup_{0\leq t \leq T}||\nXN(t) -  \x(t)|| >
  \eps\right\} \rightarrow 0.$$
\end{theorem}
Notice that the theorem can be specialised to subsets $E'\subseteq E$, in which case it can also provide an estimate 
of exit times from set $E'$, see \cite{STOC:Darling:2002:PracticalFluid}. Furthermore, if the initial conditions converge 
almost surely, then it also holds that 
$\sup_{0\leq t \leq T}||\nXN(t) -  \x(t)|| \rightarrow 0$ almost surely \cite{STOC:Kurtz:1986:MarkovProcesses}.

\subsection{Fast simulation}
\label{sec:fastSimulation}

We now turn our attention back to a single individual in the population. Even if the system-level dynamics, in the limit 
of a large population, becomes deterministic, the dynamics of a single agent remains a stochastic process. However, 
the fluid limit theorem implies that the dynamics of a single agent, in the limit, becomes essentially dependent on the 
other agents only through the global system state. This asymptotic decoupling allows us to find a simpler Markov 
Chain for the evolution of the single agent. This result is often known in the literature \cite{STOC:DarlingNorris:2008:DifferentialEquationsCTMC} under the name of \emph{fast simulation} \cite{PA:Gast:2010:workStealing}.

To explain this point formally, let us focus on a single individual $Y\N_h$, which is a Markov process on the state 
space $S=\{1,\ldots,n\}$, conditional on the global state of the population $\nX\N(t)$. 
Let $Q\N(\x)$ be the infinitesimal generator matrix of $Y\N_h$, described as a function of the normalized state of the population $\nX\N=\x$, i.e. 
$$\bbP\{Y\N_h(t+dt) = j~|~Y\N_h(t) = i, \,\nX\N(t) = \x\} = q\N_{i,j}(\x)dt.$$
We stress that this is the exact Markov Chain for $Y\N_h$, conditional on $\nX\N(t)$, and that this process is 
\emph{not independent} of $\nX\N(t)$. In fact, without conditioning on $\nX\N$, $Y\N_h(t)$ is not a Markov process. 
This means that in order to capture its evolution in a Markovian setting, one has to consider the Markov chain 
$(Y\N_h(t), \,\nX\N(t))$. 

\begin{exu}
Consider the running example, and suppose we want to construct the CTMC for a single client. For this purpose, 
we have to extract from the specification of global transitions a set of local transitions for the client. 
The state space of a client will consist of four states, $S_c = \{rq,w,t,rc\}$. 

Then, we need to define its rate matrix $Q\N$.  In order to do this, we need to take into account all global transitions 
involving a client, and then extract the rate at which a specific client can perform such a transition.  As a first example,
consider the \texttt{think} transition, changing the state of a client from $t$ to $rq$. Its global rate is 
$r_{think} = k_t C_t$. As we have $C_t$ clients in state $t$, the rate at which a specific one will perform a think 
transition is  $\frac{k_t C_t}{C_t} = k_t$. Hence, we just need to divide the global rate of observing a think transition 
by the total number of clients in state $t$. Notice that, as we are assuming that one specific client is in state $t$, 
then $C_t\geq 1$, hence we are not dividing by zero. 

Consider now a \texttt{reply} transition. In this case, the transition involves a server and a client in state $w$.  The 
global rate is $r_{reply} = min(k_w C_w, k_{rp} S_{rp})$, and $C_w \geq 1$ (in the non-normalized model with total 
population $N$). Dividing this rate by $C_w$, we obtain $\min(k_w, k_{rp} \frac{S_{rp}}{C_w})$, which is defined for 
$C_w > 0$. If we switch to normalised variables, we obtain a similar expression: $\min(k_w, k_{rp} \frac{s_{rp}}{c_w})$, 
which is independent of $N$. 
However, in taking $N$ to the limit we must be careful: even if in the non-normalized model $C_w$ (and hence 
$c_w$) are always non-zero (if a specific agent is in state $w$), this may not be true in the limit: if only one client is 
in state $w$, then the limit fraction of clients in state $w$ is zero (just take the limit of $\frac{1}{N}$). Hence, we need 
to take care of boundary conditions, guaranteeing that the single-agent rate is defined also in these circumstances. 
In this case, we can assume that the rate is zero if $s_{rp}$ is zero (whatever the value of $c_w$), and that the rate 
is $k_w$ if $c_w$ is zero but $s_{rp} > 0$.
\end{exu}

In order to treat the previous set of cases in a homogeneous way, we make the following assumption about rates:

\begin{definition}
\label{def:singleAgentCompatible}
Let $\tau\in\calT$ be a transition such that its update rule set contains the rule $i\rightarrow j$, with multiplicity 
$m_{\tau,i\rightarrow j}$.
The rate $r\N_\tau$ is \emph{single-agent-$i$ compatible} if there exists a Lipschitz continuous function $f^i_\tau(\x)$ 
on normalized variables such that the limit rate on normalized variables $f_\tau(\x)$ can be factorised as 
$f_\tau(\x) = x_i  f^i_\tau(\x)$.
A transition $\tau$ is \emph{single-agent compatible} if and only if it is single-agent-$i$ compatible for any $i$ 
appearing in the left-hand side of an update rule.

Hence, the limit rate of observing a transition from $i$ to $j$ for a specific agent in state $i$ is 
$m_{\tau,i\rightarrow j} f^i_\tau(\x)$, where the factor $m_{\tau,i\rightarrow j}$ comes from the fact that it is one out 
of $m_{\tau,i\rightarrow j}$ agents changing state from $i$ to $j$ due to $\tau$.\footnote{The factor $m$ stems from 
the following simple probabilistic argument: if we choose at random $m$ agents out of $X_i$, then the probability 
to select  a specific agent is $\frac{m}{X_i}$.}
\end{definition} 

Then, assuming all transitions $\tau$ are single-agent compatible, we can define the rate $q\N_{i,j}$ as
$$q\N_{i,j}(\x) = \sum_{\tau\in\calT~|~\{i\rightarrow j\}\subseteq R_\tau}m_{\tau,i\rightarrow j} \frac{r\N_\tau(\x)}{x_i} = \sum_{\tau\in\ncalT~|~\{i\rightarrow j\}\subseteq R_\tau} m_{\tau,i\rightarrow j} \frac{\hat{r}\N_\tau(\nx)}{\hat{x}_i} = q\N_{i,j}(\nx).$$ 
It is then easy to check that $$q\N_{i,j}(\x) \rightarrow q_{i,j}(\x) = \sum_{\tau\in\calT~|~\{i\rightarrow j\}\subseteq R_\tau} m_{\tau,i\rightarrow j} \frac{f_\tau(\x)}{x_i} = \sum_{\tau\in\ncalT~|~\{i\rightarrow j\}\subseteq R_\tau} m_{\tau,i\rightarrow j} f^i_\tau(\x).$$

In the following, we fix an integer $k>0$ and let $Z\N_k = (Y\N_1,\ldots,Y\N_k)$ be the CTMC tracking the state of $k$ 
selected agents among the population, with state space $\calS = S^k$. Notice that $k$ is fixed and independent of $N$, 
so that we will track $k$ individuals embedded in a population that can be very large. 

Let $\x(t)$ be the solution of the fluid ODE, and assume to be under the hypothesis of Theorem \ref{th:Kurtz}.
Consider now $z\N_k(t)$ and $z_k(t)$, the \emph{time-inhomogeneous} CTMCs on $\calS$ defined by the 
following infinitesimal generators (for any $h=1,\ldots,k$):
$$\bbP\{z\N_k(t+dt) = (z_1,\ldots,j,\ldots,z_k)~|~z\N_k(t) = (z_1,\ldots,i,\ldots,z_k)\} = q\N_{i,j}(\x(t))dt,$$
$$\bbP\{z_k(t+dt) = (z_1,\ldots,j,\ldots,z_k)~|~z_k(t) = (z_1,\ldots,i,\ldots,z_k)\} = q_{i,j}(\x(t))dt,$$
Notice that, while $Z\N_k$ describes exactly the evolution of $k$ agents, $z\N_k$ and  $\z_k$ do not. In fact, they 
are CTMCs in which the $k$ agents evolve independently, each one with the same infinitesimal generator, depending 
on the global state of the system via the fluid limit.  

However, the following theorem can be proved \cite{STOC:DarlingNorris:2008:DifferentialEquationsCTMC}:
\begin{theorem}[Fast simulation theorem]
\label{th:fastSimulation}
For any $T< \infty$,\ \
$\bbP\{Z\N_k(t) \neq z\N_k(t),\ \mbox{for some\ } t\leq T\}\rightarrow 0$, and
$\bbP\{Z_k(t) \neq z_k(t),\ \mbox{for some\ } t\leq T\}\rightarrow 0$, as $N\rightarrow \infty$.
\end{theorem}
This theorem states that, in the limit of an infinite population, each fixed set of $k$ agents will behave independently, 
sensing only the mean state of the global system, described by the fluid limit $\x(t)$. Furthermore, those $k$ agents 
will evolve independently, as if there was no synchronisation between them. 
This \emph{asymptotic decoupling} of the system, holding for any set of $k$ agents, is also known in the literature 
under the name of \emph{propagation of chaos} \cite{PA:LeBoudec:2008:MeanFieldContinuousTime}. 
In particular, this holds if we define the rate of the limit CTMC either by the single-agent rates for population $N$ 
($z_k\N$) or by the limit rates ($z_k$). Note that, when the CTMC has density dependent rates 
\cite{STOC:Kurtz:1986:MarkovProcesses}, then $z_k\N(t) = z_k(t)$, as their infinitesimal generators will be the same.

We stress once again that the process $Z\N_k(t)$ is not a Markov process.  It becomes a Markov process when considered together with $\nX\N(t)$. This can be properly understood by observing that it is the projection of the 
Markov process $(Y\N_1(t),\ldots,Y\N_N(t))$ on the first $k$ coordinates, and recalling that a projection of a Markov 
process need not be Markov (intuitively, we can throw away some relevant information about the state of the process). 
However, being the projection of a Markov process, the probability of $Z\N_k(t)$ at each time $t$ is perfectly defined. 
Nevertheless, its non-Markovian nature has consequences for what concerns reachability probabilities and the 
satisfiability of CSL formulae.

\begin{figure}
\begin{center}
\subfigure[$\bbP\{rq\}$] {\label{fig:transientRQ}
\includegraphics[width=.47\textwidth]{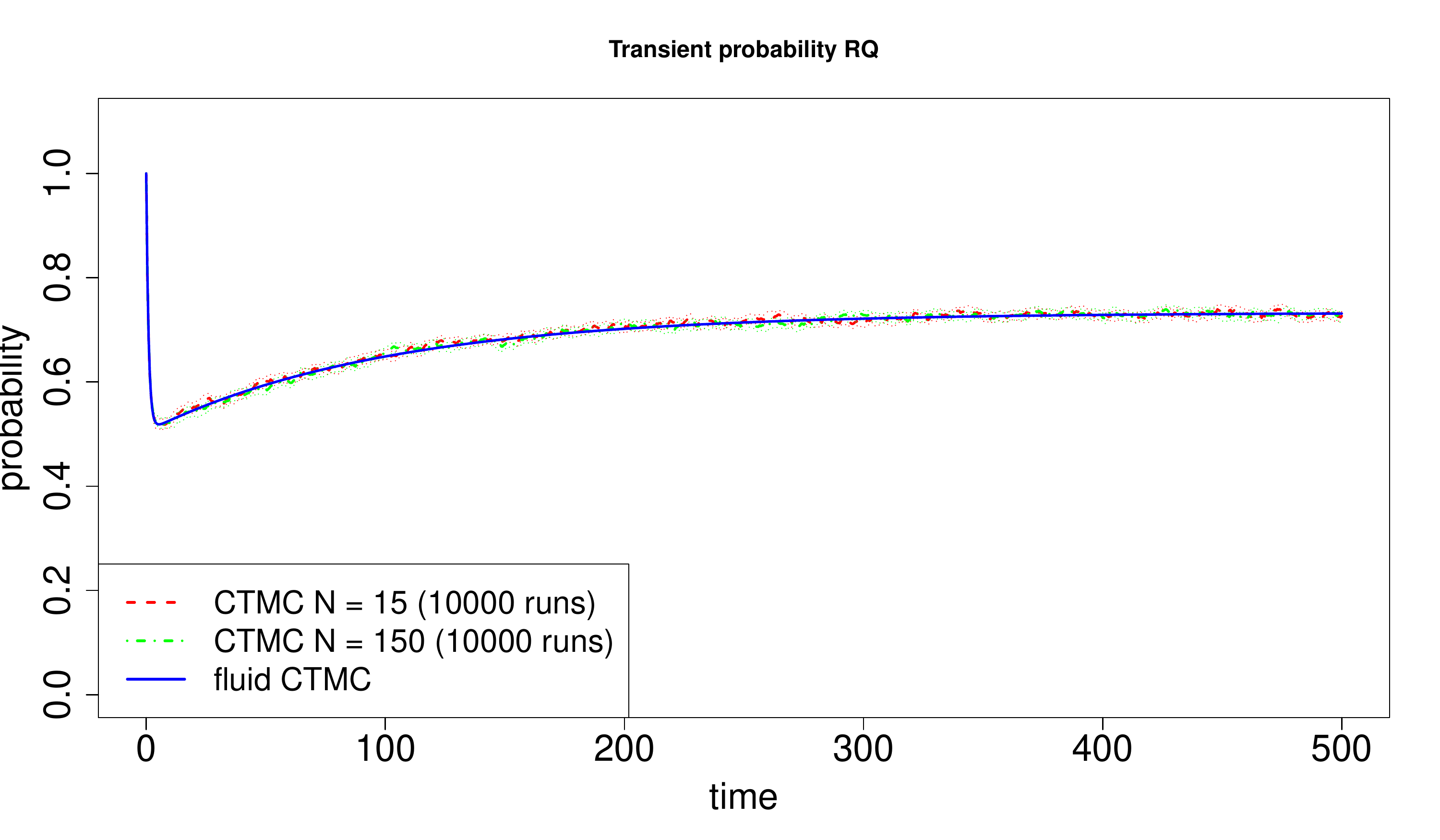} }
\subfigure[$\bbP\{w\}$] {\label{fig:transientW}
\includegraphics[width=.47\textwidth]{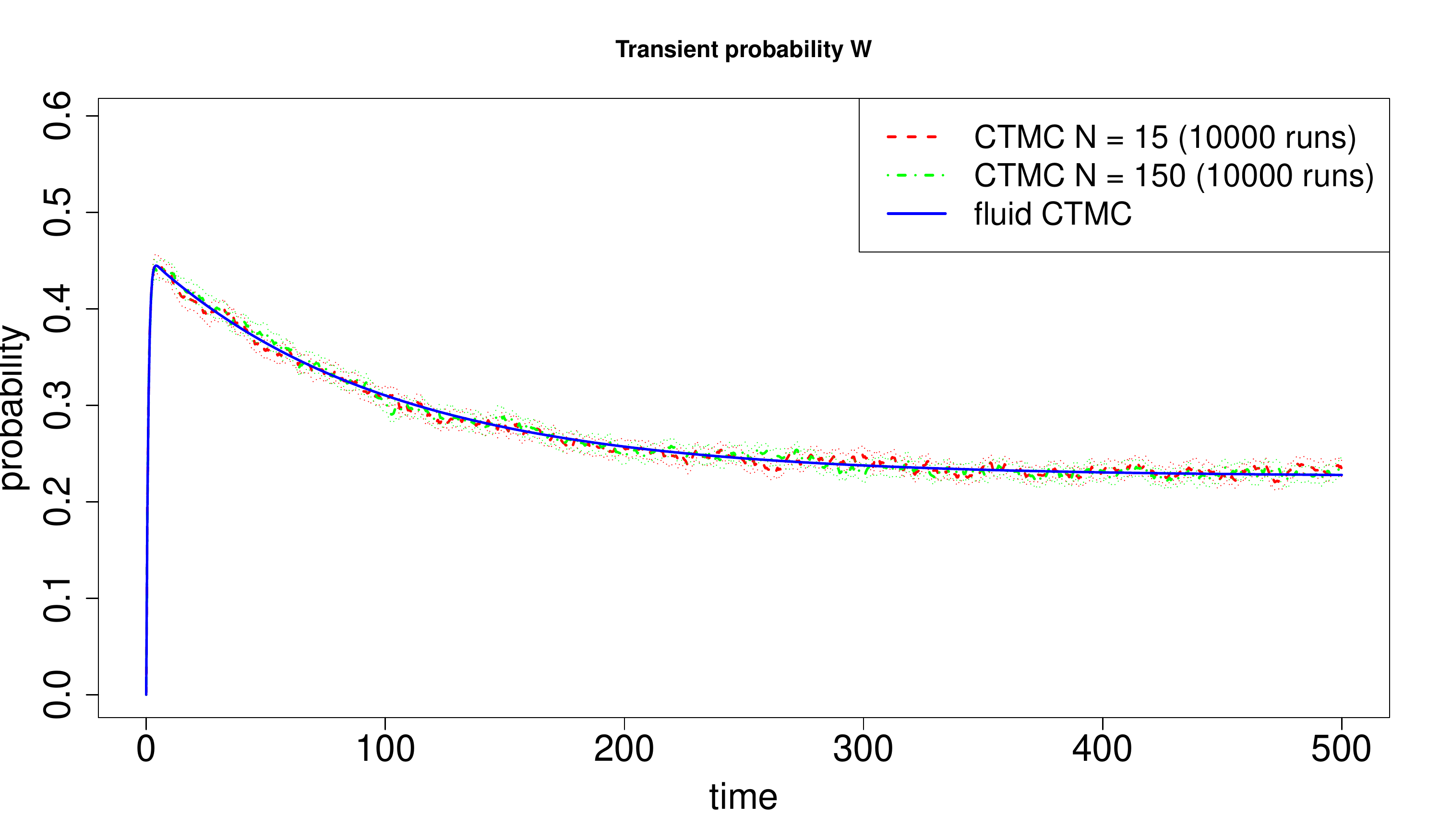} }

\subfigure[$\bbP\{t\}$] {\label{fig:transientT}
\includegraphics[width=.47\textwidth]{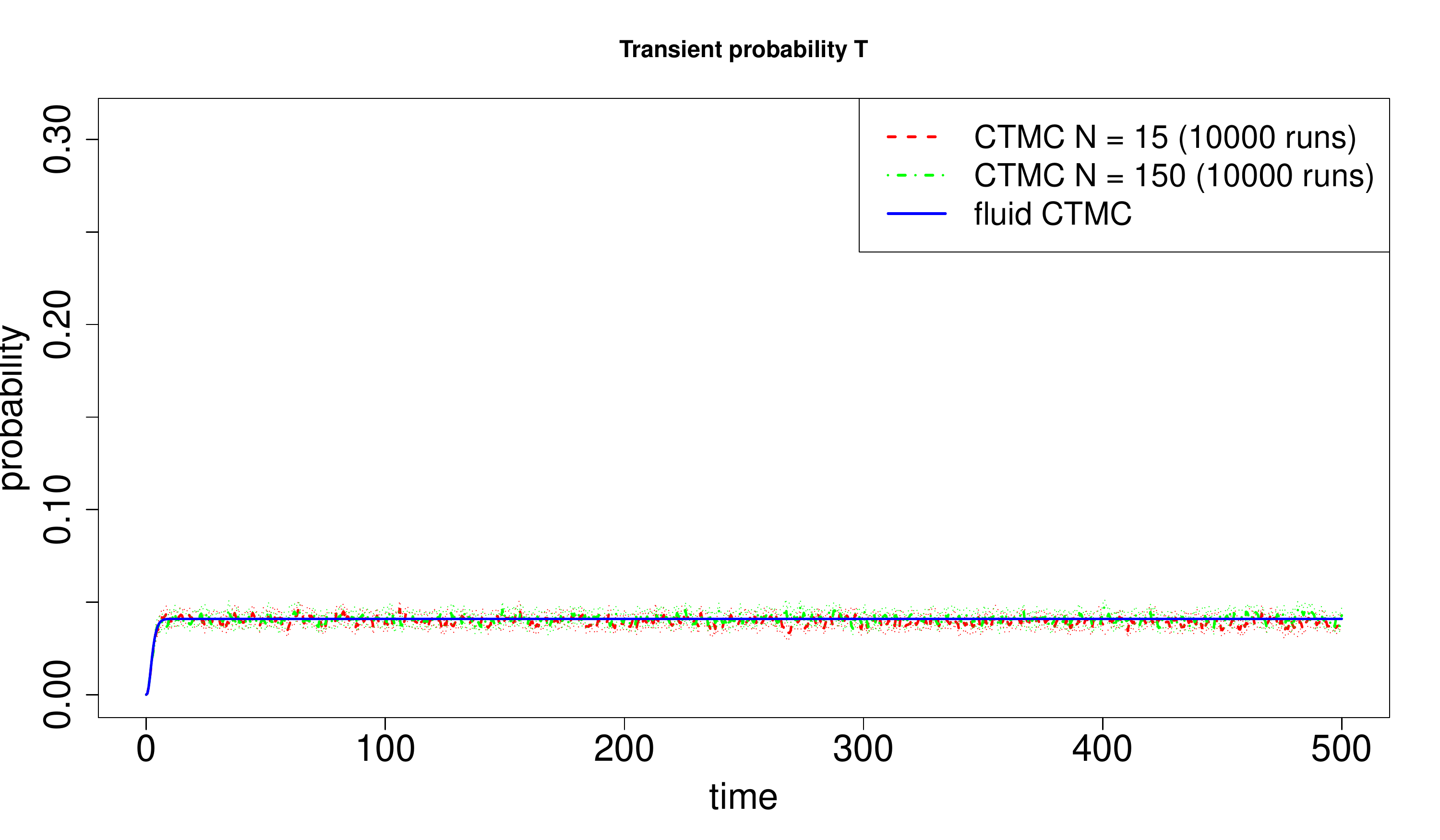} }
\subfigure[$\bbP\{rc\}$] {\label{fig:transientRC}
\includegraphics[width=.47\textwidth]{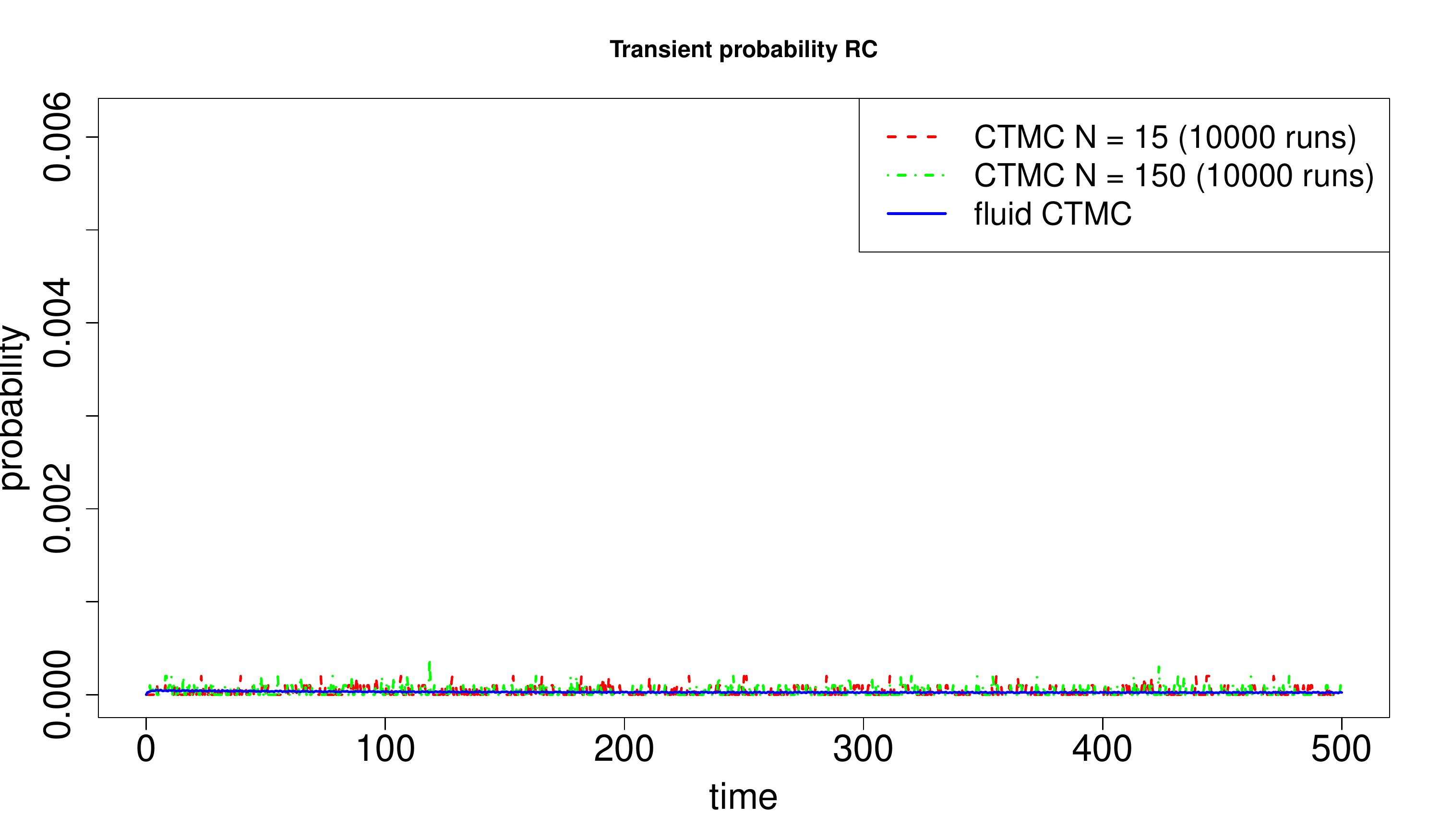} }
\end{center}

\caption{Comparison of the transient probability for all four states of the fluid model of the client server system, 
computed solving the Kolmogorov forward equations, and the transient probability of CTMC models for $N=15$ 
and $N=150$ (2:1 client server ratio). Parameters are $k_r = 1$, $k_w = 100$,  $k_{to} =0.01 $, $k_t = 1$, 
$k_{rc} = 100$ $k_l = 10$, $k_p = 0.1$, $k_{sto} = 0.005$, initial conditions of the full system are $C_{rq} = n$, 
$S_{rq} = m$, while the single client CTMC starts in state $rq$.}
\label{fig:transient}
\end{figure}

\begin{exu}
Consider again the client-server example, and focus on a single client. As said before, its state space is $S_c=\{rq,w,t,rc\}$, and the non-null rates of the infinitesimal generator $Q$ for the process $z_1$ are:
\begin{itemize}
\item $q_{rq,w}(t) = k_r \min\{1,s_{rq}(t)/c_{rq}(t)\}$ (with appropriate boundary conditions);
\item $q_{w,t}(t) = \min\{k_w,k_{rp}s_{rp}(t)/c_w(t)\}$;
\item $q_{w,rc}(t) = k_{to}$;
\item $q_{t,rq}(t) = k_{t}$;
\item $q_{rc,rq}(t) = k_{rc}$.
\end{itemize}

In Figure \ref{fig:transient}, we show a comparison of the transient probabilities for the approximating chain for a 
single client and the true transient probabilities, estimated by Monte Carlo sampling of the CTMC, for different 
population levels $N$. As we can see, the approximation is quite precise already for $N = 15$.
\end{exu}

\begin{remark}
\label{rem:passive}
Single-agent consistency is not a very restrictive condition. However, there are cases in which it is not satisfied. 
One example is passive rates in PEPA \cite{PA:Hillston:1996:CompositionalPerformanceModeling}. In this case, in 
fact, the rate of the synchronization of $P = (\alpha,\top).P1$ and $Q=(\alpha, r).Q1$ is $r X_Q \vr{1}\{X_P>0\}$. 
In particular, the rate is independent of the exact number of $P$ agents. 
If we look at a single $P$-agent rate, it equals $r \frac{X_Q}{X_P} \vr{1}\{X_P>0\}$. 
Normalising variables, we get the rate $r \frac{x_Q}{x_P} \vr{1}\{x_P>0\}$, which approaches infinity as $x_P$ goes 
to zero (for $x_Q$ fixed).  Hence, it cannot be extended to a Lipschitz continuous function.  However, in the case 
$x_P = 0$ and $x_Q > 0$, if we look at a single agent, then the speed at which $P$ changes state is in fact infinite. 
We can see this by letting $X_P = 1$ and $X_Q = Nq$, so that the rate of the transition from the point of view of $P$ 
is $Nq\rightarrow\infty$. Thus, in the limit, the state $P$ becomes vanishing.
\end{remark}

\begin{remark}
\label{rem:birthDeath}
The hypothesis of constant population, i.e.\ the absence of birth and death, can be relaxed. The fluid approximation 
continues to work also in the presence of birth and death events, and so does the fast simulation theorem. 
In our framework, birth and death  can be easily introduced by allowing rules of the form $\emptyset \rightarrow i$ 
(for birth) and $i\rightarrow \emptyset$ (for death).  In terms of a single agent, death can be dealt with by adding a 
single absorbing state to its local state space $\calS$. Birth, instead, means that we can choose the time instant at 
which an agent enters the system (provided that there is a non-null rate for birth transitions at the chosen time).

Another solution would be to assume an infinite pool of agents, among which only finitely many can be alive, and 
the others are an infinite supply of ``fresh souls''.  Even if this is plausible from the point of view of a global model, 
it creates problems in terms of a single agent perspective (what is the rate of birth of a soul?).  A solution can be to 
assume a large but finite pool of agents.  But in this case birth becomes a passive action (and it introduces 
discontinuities in the model, even if in many cases one can guarantee to remain far away the discontinuous boundary), 
hence we face the same issues discussed in Remark \ref{rem:passive}.

%An alternative approach is to keep the system open and condition on the birth time of the agent we are tracking.   
\end{remark}

\subsection{Continuous Stochastic Logic}
\label{sec:CSL}

In this section we consider labelled stochastic processes. A labelled stochastic process is a random process $Z(t)$, 
with state space $\calS$  and 
%infinitesimal generator matrix $Q(t)$ (possibly depending on time), together with 
a labelling function $L:\calS\rightarrow 2^{\calP}$, associating with each state $s\in\calS$ a subset of atomic 
propositions $L(s)\subset\calP=\{a_1,\ldots,a_k\,\ldots\}$ true in that state: each atomic proposition $a_i\in \calP$ is 
true in $s$ if and only if $a_i\in L(s)$. We require that all subsets of paths considered are measurable. This condition 
will be satisfied by all subsets considered in the paper, as $Z(t)$ will always be either a CTMC, defined by an 
infinitesimal generator matrix $Q(t)$ (possibly depending on time), or the projection of a CTMC.
From now on, we always assume we are working with labelled stochastic processes. 

A path of $Z(t)$ is a sequence $\sigma = s_0\lts{t_0}s_1\lts{t_1}\ldots$, such that the probability of going from $s_i$ 
to $s_{i+1}$ at time $t_\sigma[i]  = \sum_{j=0}^i t_j$, is greater than zero.  For CTMCs, this condition is equivalent to  
$q_{s_i,s_{i+1}}(t_\sigma[i])>0$. 
Denote with $\sigma@t$ the state of $\sigma$ at time $t$, with $\sigma[i]$ the i-th state of $\sigma$, and with 
$t_\sigma[i]$ the time of the $i$-th jump in $\sigma$.

A time-bounded CSL formula $\phi$ is defined by the following syntax:
%$$\phi = a \mid \phi_1 \wedge \phi_2 \mid \neg \phi \mid \calP_{\bowtie p}( \next{T_1}{T_2}\phi ) \mid \calP_{\bowtie p}( \phi_1 \until{T_1}{T_2}\phi_2 ).$$ 
$$\phi = a \mid \phi_1 \wedge \phi_2 \mid \neg \phi \mid P_{\bowtie p}( \next{T_1}{T_2}\phi ) \mid P_{\bowtie p}( \phi_1 \until{T_1}{T_2}\phi_2 ).$$ 
The satisfiability relation of $\phi$ with respect to a labelled stochastic process  $Z(t)$ is given by the following rules:
\begin{itemize}
\item $s,t_0\models a$ if and only if $a\in L(s)$;
\item $s,t_0\models \neg\phi$ if and only if $s,t_0\not\models \phi$;
\item $s,t_0\models \phi_1\wedge\phi_2$ if and only if $s,t_0\models \phi_1$ and $s,t_0\models \phi_2$;
%\item $s,t_0\models \calP_{\bowtie p}( \next{T_1}{T_2}\phi )$ if and only if $\bbP\{\sigma~|~\sigma,t_0 \models \next{T_1}{T_2}\phi\} \bowtie p$.
\item $s,t_0\models P_{\bowtie p}( \next{T_1}{T_2}\phi )$ if and only if $\bbP\{\sigma \mid \sigma,\, t_0 \models 
\next{T_1}{T_2}\phi\} \bowtie p$.

%\item $s,t_0\models \calP_{\bowtie p}( \phi_1 \until{T_1}{T_2}\phi_2 )$ if and only if $\bbP\{\sigma~|~\sigma,t_0 \models \phi_1 \until{T_1}{T_2}\phi_2\} \bowtie p$.
\item $s,t_0\models P_{\bowtie p}( \phi_1 \until{T_1}{T_2}\phi_2 )$ if and only if $\bbP\{\sigma \mid \sigma,\, t_0 \models \phi_1 \until{T_1}{T_2}\phi_2\} \bowtie p$.
\item $\sigma,t_0 \models \next{T_1}{T_2}\phi$ if and only if  $t_\sigma[1] \in [T_1,T_2]$ and $\sigma[1],t_0 + t_\sigma[1] \models \phi$.
\item $\sigma,t_0 \models \phi_1 \until{T_1}{T_2}\phi_2$ if and only if  $\exists \bar{t} \in [t_0+T_1,t_0+T_2]$ s.t. $\sigma@\bar{t},\bar{t} \models \phi_2$ and $\forall t_0\leq t < \bar{t}$, $\sigma@t,t \models \phi_1$.
\end{itemize}

Notice that we are considering a fragment of CSL without the steady state operator and allowing only time-bounded 
properties. This last restriction is connected with the nature of convergence theorems \ref{th:Kurtz} and 
\ref{th:fastSimulation}, which hold only on finite time horizons.  However, see Remark~\ref{rem:steadyState} for 
possible relaxations of this restriction.

Model checking of a next CSL formula $P_{\bowtie p}( \next{T_1}{T_2} \phi)$ is usually performed by computing the
 next-state probability via an integral, and then comparing the so-obtained value with the threshold $p$.
Model checking of an until CSL formula $P_{\bowtie p}( \phi_1 \until{T_1}{T_2}\phi_2)$ in a \emph{time-homogeneous 
CTMC} $Z(t)$, instead, can be reduced to the computation of two reachability problems, which themselves can be 
solved by transient analysis \cite{MC:Hermanns:2000:MCofCTMCtransient}.   In particular, consider the sets of states 
$U = \lm\neg\phi_1\rm$ and $G = \lm\phi_2\rm$ and compute the probability of going from state $s_1\not\in U$ to 
a state $s_2\not\in U$ in $T_1$ time units, in the CTMC in which all $U$-states are made absorbing, 
$\pi^1_{s_1,s_2}(T_1)$.  Furthermore, consider the modified CTMC in which all $U$ and $G$ states are made 
absorbing, and denote by $\pi^2_{s_2,s_3}(T_2-T_1)$ the probability of going from a state $s_2\not\in U$ to a state 
$s_3\in G$ in $T_2-T_1$ units of time in such a CTMC\@.
Then the probability of the until formula in state $s$ can be computed as $P_s(\phi) = \sum_{s_3\in G,s_2\not\in U} 
\pi^1_{s_1,s_2}(T_1) \pi^2_{s_2,s_3}(T_2-T_1)$. The probabilities $\pi^1$ and $\pi^2$ can be computed using 
standard methods for transient analysis (e.g.\ by uniformisation \cite{STOC:Jensen:1953:uniformization} or by solving 
the Kolmogorov equations \cite{STOC:Norris:1997:MarkovChains}). Then, to determine the truth value of the formula 
$\phi$ in state $s$, one has just to solve the inequality $P_s(\phi) \bowtie p$. The truth value of a generic CSL formula can therefore be computed recursively on the structure of the formula.

%*****************************************************************************

\section{Next State Probability}
\label{sec:next}

In this section, we start the presentation of the algorithmic procedures that underlie the CSL model checking algorithm. 
We will focus here on next state probabilities for single agents (or a fixed set of agents), in a population of growing 
size. In particular, we will show how to compute the probability that the next state in which the agents jumps  belongs 
to a given set of goal states $G\subseteq \calS$, constraining the jump to happen between time $[t_0+T_1,t_0+T_2]$, 
where $t_0$ is the current time. This is clearly at the basis of the computation of the probability of next path formulae. 
More specifically, we provide algorithms for ICTMC (hence for the limit model $z_k(t)$), focussing on two versions of 
the next state probability: the case in which the set $G$ is constant, and the case in which the set $G$ depends on 
time (i.e.\ a state may belong to $G$ or $U$ depending on time $t$).
\begin{definition}
\label{def:nextState}
Let $Z(t)$ be a CTMC with state space $\calS$ and infinitesimal generator matrix $Q(t)$. 

\begin{enumerate}
\item Let  $G\subseteq \calS$.
The \emph{constant-set next state probability} $P_{next}(Z,t_0,T_1,T_2,G)[s]$ is the probability of the set of 
trajectories of $Z$ jumping into a state in $G$, starting at time $t_0$ in state $s$, within time $[t_0+T_1,t_0+T_2]$. 
$P_{next}(Z,t_0,T_1,T_2,G)$ is the next state probability vector on $\calS$.
\item Let $G:[t_0,t_1]\times\calS\rightarrow\{0,1\}$ be a time-dependent set, identified with its indicator function 
(i.e.\ $G(t$ is the goal set at time $t$). \ifdefined\arxivVersion \\ \fi
 The \emph{time-varying-set next state probability} $P_{next}(Z,t_0,T_1,T_2,
G(t))[s]$ is the probability of the set of trajectories of $Z$ jumping into a state in $G(t)$ at time $t\in [t_0+T_1,t_0+T_2]$, 
starting at time $t_0$ in state $s$.
\end{enumerate}
\end{definition}

The interest in the time-varying sets is intimately connected with CSL model checking. In fact, the truth value of a CSL 
formula in a state $s$ for the (non-Markov) stochastic process $Z\N_k(t)$ depends on the initial time at which we start 
evaluating the formula. This is because $Z\N_k(t)$ depends on time via the state of $\nXN(t)$. Furthermore, time-
dependence of truth values of CSL formulae manifests also for the limit process $z_k(t)$, which is a time-inhomogeneous 
Markov Process. Therefore, in the computation of the probability of the next operator, which can be tackled with the 
methods of this section, we need to consider time-varying sets. As a matter of fact, we will see that dealing with time-varying 
sets (like those obtained by solving the inequality $P_{next}(Z,t_0,T_1,T_2,G(t))\bowtie p$, for $\bowtie\, \in\{<,\leq,>\geq\}$) 
requires us to impose some additional regularity conditions on the rate functions of $Z$ and on the time-dependence of 
goal sets.

In the following sections, we will first deal with the computation of next state probabilities for a generic ICTMC $Z(t)$ and 
then study the relationship between those probabilities for $Z\N_k(t)$ and $z_k(t)$.

\subsection{Computing next-state probability}

Consider a generic ICTMC $Z(t)$, and focus for the moment on a constant set $G\subseteq\calS$. For any fixed $t_0$, 
the probability $P_{next}(Z,t_0,T_1,T_2,G)[s]$ that $Z(t)$'s next jump happens at time $t\in[t_0+T_1,t_0+T_2]$ and ends 
in a state of $G$, given that $Z(t)$ is in state $s$ at time $t_0$, is given by the following integral \cite{MC:vanMoorse:1998:NonHomogUnif, MC:Mereacre:2008:HMLictmc}

\begin{equation}
\label{eqn:intNext}
P_{next}(Z,t_0,T_1,T_2,G)[s] = \int_{t_0+T_1}^{t_0+T_2}
q_{s,G}(t) \cdot e^{-\Lambda(t_0,t)[s]} dt,
\end{equation}
where $\begin{displaystyle}\Lambda(t_0,t)[s] = \int_{t_0}^t -q_{s,s}(\tau)d\tau\end{displaystyle}$,  is the cumulative exit rate of state $s$ from time $t_0$ to time $t$, and \linebreak $q_{s,G}(t) = \sum_{s'\in G} q_{s,s'}(t)$ is the rate of jumping from $s$ to a state $s'\in G$ at time $t$. 

Equation~\ref{eqn:intNext} holds for the following reason. Let $A_{t}$ be the event that we jump into a $G$ state in a time 
$\tau\in [t_0,t]$. Then $A_{t_1}\subseteq A_{t_2}$ for $t_1\leq t_2$, and $\begin{displaystyle}\bbP\{A_t\} =\int_{t_0}^{t}
q_{s,G}(t) \cdot e^{-\Lambda(t_0,t)[s]} dt\end{displaystyle}$. We are interested in the probability of the event $A = A_{t_0+T_2}\setminus A_{t_0 + T_1}$, which has probability 
\[ \bbP\{A\} = \bbP\{A_{t_0 + T_2}\} - \bbP\{A_{t_0 + T_1}\} =  \int_{t_0+T_1}^{t_0+T_2} q_{s,G}(t) \cdot e^{-\Lambda(t_0,t)[s]} dt. \]

In order to compute $P_{next}(Z,t_0,T_1,T_2,G)[s]$ for a given $t_0$, we can numerically compute the integral, or 
transform it into a differential equation, and integrate the so-obtained ODE with standard numerical methods. This 
simplifies the treatment of the nested integral $\Lambda(t_0,t)[s]$ involved in the computation of  $P_{next}$. More 
specifically, we can introduce two variables, $P$ and $L$, initialise $P(t_0+T_1) = 0$ and $L(t_0+T_1) = \Lambda(t_0,t_0+T_1)$, and then integrate the following two ODEs from time $t_0+T_1$ to time $t_0+T_2$:
\begin{equation}
\label{eqn:nextFixedTime}
\left\{\begin{array}{l}
\begin{displaystyle}\frac{d}{dt} P(t) = q_{s,G}(t)\cdot e^{-L(t)} \end{displaystyle}\\[6pt]
\begin{displaystyle}\frac{d}{dt} L(t) = -q_{s,s}(t) \end{displaystyle}
\end{array} \right.
\end{equation}

However, for CSL model checking purposes, we need to compute\\ $P_{next}(Z,t_0,T_1,T_2,G)[s]$ as a function of $t_0$: 
$\bP_s(t_0) = P_{next}(Z,t_0,T_1,T_2,G)[s]$. One way of doing this is to compute the integral (\ref{eqn:intNext}) for any 
$t_0$. A better approach is to use the differential formulation of the problem, and define a set of ODEs with the initial time $t_0$ as independent variable. First, observe that the derivative of $\bP_s(t_0)$ with respect to $t_0$ is
\[
\begin{split}
\frac{d}{dt_0} \bP_s(t_0) = & \;q_{s,G}(t_0+T_2)\cdot e^{-\Lambda(t_0,t_0+T_2)} - q_{s,G}(t_0+T_1)\cdot e^{-\Lambda(t_0,t_0+T_1)}\\ & + \int_{t_0+T_1}^{t_0+T_2} \frac{\partial}{\partial t_0}  q_{s,G}(t)\cdot e^{-\Lambda(t_0,t)}dt \\[4pt]
= &\; q_{s,G}(t_0+T_2)\cdot e^{-\Lambda(t_0,t_0+T_2)} - q_{s,G}(t_0+T_1)\cdot e^{-\Lambda(t_0,t_0+T_1)}\\
& -q_{s,s}(t_0)\bP_s(t_0) 
\end{split}
\]  
Consequently, we can compute the next-state probability as a function of $t_0$ by solving the following set of ODEs:
\begin{equation}
\label{eqn:nextTimeVarying}
\left\{\begin{array}{l}
\frac{d}{d t} \bP_s(t) = q_{s,G}(t+T_2)\cdot e^{-L_2(t)} - q_{s,G}(t+T_1)\cdot e^{-L_1(t)} -q_{s,s}(t)\bP_s(t)\\[4pt]
\frac{d}{d t} L_1(t) =-q_{s,s}(t) + q_{s,s}(t+T_1)\\[4pt]
\frac{d}{d t} L_2(t) =-q_{s,s}(t) + q_{s,s}(t+T_2)\\
\end{array}\right.
\end{equation}
where $L_1(t) = \Lambda(t,t+T_1)$ and $L_2(t) = \Lambda(t,t+T_2)$.\\
Initial conditions are $P_s(t_0) = P_{next}(Z,t_0,T_1,T_2,G)[s]$, $L_1(t_0) = \Lambda(t_0,t_0+T_1)$, and $L_2(t_0) = \Lambda(t_0,t_0+T_2)$, and are computed solving the equations \ref{eqn:nextFixedTime}.  The algorithm is sketched in Figure \ref{algo:nextState}.

\begin{figure}[!t]
\begin{algorithmic}
\Function{next-state-probability}{$Z$, $G$, $T_1$, $T_2$, $t_0$, $t_1$}  

\ForAll{$s\in\calS$}

\State Compute $\bP_s(t_0)$ by solving ODE \ref{eqn:nextFixedTime}

\State Compute $\bP_s(t)$ for $t\in[t_0,t_1]$ by solving ODE \ref{eqn:nextTimeVarying}

\EndFor

\Return  $\bP(t)$, $t\in[t_0,t_1]$
\EndFunction
\end{algorithmic}
\caption{Algorithm for the computation of next-state probability $\bP(t)$, for any state $s$ and $t\in[t_0,t_1]$. Other input parameters are as in the text.}
\label{algo:nextState}
\end{figure}

We turn now to discuss the case of a time-varying next-state set $G(t)$. In this case, the only difference with respect to 
the constant-set case is that the function $q_{\cdot,G(t)}$ is piecewise continuous, rather than continuous. In fact, each time a 
state  $s'$ gains or loses membership of $G(t)$, the range of the sum defining $q_{\cdot,G(t)}$ changes, and a 
discontinuity can be introduced. However, as long as these discontinuities constitute a set of measure zero (for instance, 
they are finite in number), this is not a problem: the integral (\ref{eqn:intNext}) is defined and absolutely continuous, and so is the 
solution of the set of ODEs (\ref{eqn:nextTimeVarying}) (because the functions involved are discontinuous with respect to 
time). It follows that the method for computing the next-state probability for constant sets works also for time-varying sets. 

Now, if we want to use this algorithm in a model checking routine, we need to be able also to solve the equation 
$\bP_s(t) = p$, for $p\in[0,1]$ and each $s\in\calS$.  In particular, for obvious computability reasons, we want the number of 
solutions to this equation to be finite. 
This is unfortunately not true in general, as even a smooth function can be equal to zero on an uncountable and nowhere 
dense set of Lebesgue measure 0 (for instance, on the Cantor set \cite{THMAT:Rudin:1976:analysis}). 

Consequently, we have to introduce some restrictions on the class of functions that we can use. In particular, we will require 
that the rate functions of $z_k$ and of $Z\N_k$ are \emph{piecewise real analytic functions}.

\subsubsection{Piecewise Real analytic functions}

A function $f : I\rightarrow \bbR$, $I$ an open subset of $\bbR$, is said to be analytic 
\cite{THMAT:Kranz:2002:RealAnalyticFunctions} in $I$ if and only if for each point $t_0$ of $I$ there is an open 
neighbourhood of $I$ in which $f$ coincides with its Taylor series expansion around $t_0$. Hence, $f$ is locally a power 
series. For a piecewise analytic function, we intend a function from $I\rightarrow \bbR$, $I$ interval, such that there exists 
$I_1,\ldots,I_k$ disjoint open intervals, with $I=\bigcup_j \bar{I}_j$, such that $f$ is analytic in each $I_j$. A similar definition 
holds for functions from $\bbR^n$ to $\bbR$, considering their multi-dimensional Taylor expansion.

Analytic functions are a class of functions closed by addition, product, composition, division (for non-zero analytic functions), 
differentiation and integration. Piecewise analytic functions also satisfy these closure properties, by considering the 
intersections of their analytic sub-domains.  Many functions are analytic: polynomials, the exponential, logarithm, sine, cosine. 
Using the previous closure properties, one can show that most of the functions we work with in practice are analytic. 

Analytic functions have two additional properties that make them particularly suitable in this context:
\begin{enumerate}
\item The zeros of an analytic function $f$ in $I$, different from the constant function zero, are isolated. In particular, 
if $I$ is bounded, then the number of zeros is finite. This is true also for the derivatives of any order of the function $f$.
\item If $f$ is analytic in a set $E$, then the solution $\x$ of $\frac{d\x}{dt} = f(\x)$ in $E$ is also analytic (this is a 
consequence of the Cauchy-Kowalevski theorem \cite{THMAT:Folland:1995:PDE}).
\end{enumerate}

This second property, in particular, guarantees that if the rate functions of $z_k$ and $Z\N_k$ are piecewise analytic, then all 
the probability functions computed solving the differential equations, like those introduced in the previous section, are 
also piecewise analytic. 

In the following, we will need the following straightforward property of piecewise analytic functions:
\begin{proposition}
\label{prop:measureSimpleZeroSetsPiecewiseAnalytic}
Let $f:I\rightarrow \bbR$ be a piecewise analytic function, with $I\subseteq \bbR$ a compact interval.  
Let $E_f = \{x\in \bbR~|~\mu_\ell(f^{-1}(\{x\}))=0\}$ be the set of all values $x$ such that $f$ is not locally constantly equal to $x$, where $\mu_{\ell}$ is the Lebesgue measure.
Furthermore, let $Z_x = f^{-1}(\{x\})$ be the set of solutions of $f(t) = x$ and let 
$DZ_f =\{x\in \bbR~|~\forall t\in Z_x, f'(t)\neq 0\}$. Then
\begin{enumerate}
\item $\forall x\in E_f$, $Z_x$ is \emph{finite}.
\item $\mu_\ell(E_f\cap DZ_f) = 1$
\end{enumerate}
\end{proposition}

%\proof Point 1 follows from basic properties of the piecewise analytic function $(f - x)$: in any analytic piece, either the function is constantly equal to zero, or it has only a finite number of zeros. Point 2, instead, follows from the fact that the derivative $f'(t)$ of $t$ is piecewise analytic, hence has only a finite number of zeros (in the analytic pieces in which $f$ is not constant).
%\qed

\subsection{Convergence of next-state probability}
We consider now the problem of relating the next-state probabilities for the limit single agent process $z_k(t)$ and the 
sequence of single agent processes $Z_k\N(t)$ in a population of size $N$. In particular, we want to show that the 
probability $\bP_s\N(t) = P_{next}(Z_k\N,t_0,T_1,T_2,G)[s]$ converges to $\bP_s(t)=P_{next}(z_k,t_0,T_1,T_2,G)[s]$ 
uniformly for $t\in[t_0,t_1]$, as $N$ goes to infinity. We will prove this result in a general setting. More specifically, we will 
consider time-varying sets that can depend on $N$, and that converge to a limit time-varying set in a suitable sense. This 
is needed because the time-varying sets we need to consider are obtained by solving (for each $s\in\calS$) equations of 
the form $\bP_s\N(t) -p=0$ or $\bP_s(t) - p = 0$, which are generally different, but intuitively converge (as $\bP_s\N(t)$ 
converges to $\bP_s(t)$). 

\subsubsection{Robust time-varying sets}

We first introduce a notion of robustness for \emph{time-varying} sets, which will be enforced on limit sets:

\begin{definition}
\label{def:robustSet}
A time-dependent subset $V(t)$ of $\calS$, $t\in I$, is \emph{robust} if and only if there is a piecewise analytic function $h_V:\calS\times I \rightarrow \bbR$ and an operator $\bowtie\in\{<,\leq,\geq,>\}$, such that for each $s\in\calS$, the indicator function 
$V_s:I\rightarrow\{0,1\}$ of $s$ is given by $\vr{1}\{h_V(s,t)\bowtie 0\}$, and it further satisfies:
\begin{enumerate}
\item the number of discontinuity points $Disc(V) = \{(s,\bar{t})~|~V_s(\bar{t}^-\neq V_s(\bar{t}^+) \}$ is finite;
\item if $h_V$ is analytic in $(s,t)$ and $h_V(s,t) = 0$, then $\frac{d}{dt} h_V(s,t) \neq 0$ (zeros of $h_V$ are simple);
\item if $h_V$ is \emph{not} analytic in $(s,t)$, then $h_V(s,t^-) \neq 0$ and $h_V(s,t^+) \neq 0$.\footnote{This condition states that, if $h_V$ is continuous but not analytic in $(s,t)$, then it cannot be equal to zero in those points, implying that first order derivatives exist and are non-null in all continuity points in which $h_V$ crosses zero. Moreover, if  $h_V(s,t^-) \neq h_V(s,t^+)$, $h_V$ can cross zero in $(s,t)$ only if the jump contains zero,  meaning that $\min\{h_V(s,t^-), h_V(s,t^+) \} < 0 < \max\{h_V(s,t^-), h_V(s,t^+) \}$ .} 
\end{enumerate}
%has only a finite number of discontinuities and it is either right or left continuous in those 
%discontinuity points.  
\end{definition} 

In the following, we will usually indicate with $V(t)$ both a time dependent set $V$ and its indicator function (with values in 
$\{0,1\}^m$, $m=|\calS|$), and with $h_V$ the piecewise analytic function defining it.\\ 
%Moreover, denote by $Disc(V) = \{\bar{t} \mid V_s(\bar{t})\ \mbox{is discontinuous for some\ } s\in\calS\}$, the set of 
%discontinuity points of $V$.  
As we will see later on, the notion of robustness is closely related to the computability of the 
reachability probability for time-varying sets and with the decidability of the model checking algorithm for ICTMCs, both 
discussed in Section \ref{sec:CSLmodelCheking}. 

Furthermore, we need the following notion of convergence for time-varying sets:

\begin{definition}
\label{sec:robustConvergence}
A sequence of time-varying sets $V\N(t)$, $t\in I$, converges \emph{robustly} to a \emph{robust} time-varying set $V(t)$, $t\in I$, if and 
only if, for each $s\in \calS$ and each open neighbourhood $U$ of $Disc(V_s)$ (i.e.\ the set of discontinuity points of $V_s$), $V_s\N(t)\rightarrow V_s(t)$ 
\emph{uniformly} in $I\setminus U$.
\end{definition}

Connecting the notions of robust set and robust convergence, we have the following:
\begin{proposition}
\label{prop:convergenceToRobustSet}
Let $V\N(t)$ be a sequence of time varying sets converging robustly to a robust set $V(t)$, $t\in I$. Let 
$D\N_V = \{t \mid V\N(t)\neq V(t)\}$. Then $\mu_\ell(D\N_V)\rightarrow 0$, where $\mu_{\ell}$ is the Lebesgue measure on 
$\bbR$.
\end{proposition}

%\proof A straightforward consequence of the definition of robust convergence is that, for each open neighbourhood $U$ of $Disc(v)$, there exists an $N_0$ such that, for all $N\geq N_0$, $V\N(t) = V(t)$ for $t\in I\setminus U$. Now, as $V$ is robust, then $|Disc(V)|=m<\infty$. 
%Fix $\eps>0$ and define $U_\eps = \bigcup_{\bar{t}\in Disc(V)}B(\bar{t},\eps)$, where $B(\bar{t},\eps)$ is the open ball centred in $\bar{t}$ of radius $\eps$. Then $\mu_\ell(U_\eps)\leq 2m\eps$. Now, fix $\eps_k\rightarrow 0$. 
%For each $k$, there is an $N_k$ such that, for all $N\geq N_k$ $V\N(t) = V(t)$ for $t\in I\setminus U_{\eps_k}$, and therefore $D\N_V\subseteq U_{\eps_k}$.  \qed

\subsubsection{Convergence results}

We are now ready to state the convergence result for next-state probabilities. We will assume that the limit time-varying set 
is robust. The following lemma will be one of the key ingredients to prove the inductive step in the convergence for truth of 
CSL formulae, see Section \ref{sec:convergenceCSL }.

\begin{lemma}
\label{lemma:timeDepNextProb}
Let $\calXN$ be a sequence of CTMC models, as defined in Section \ref{sec:modelingLanguage}, and 
let $Z\N_k$ and $z_k$ be defined from $\calXN$ as in Section \ref{sec:fastSimulation}, with piecewise real analytic rates, in 
a compact interval $[0,T']$, for $T'>t_1+T_b$. \\
Let $G(t)$, $t\in[t_0,t_1+T_b]$ be a \emph{robust} time-varying set, and let $G\N(t)$ be a sequence of time-varying sets 
converging robustly to $G$. \\ 
Furthermore, let $\bP(t) = P_{next}(z_k,t,T_a,T_b,G)$ and\\ $\bP\N(t) = P_{next}(Z_k\N,t,T_a,T_b,G)$, $t\in[t_0,t_1]$.\\
Finally, fix $p\in[0,1]$, $\bowtie\in\{\leq,<,>,\geq\}$, and let $V_p(t) = \vr{1}\{\bP(t)\bowtie p\}$, $V\N_p(t) = \vr{1}\{\bP\N(t)\bowtie p\}$. 
Then
\begin{enumerate}
\item $\bP\N(t)\rightarrow \bP(t)$, uniformly in  $t\in [t_0,t_1]$. 
\item For almost every $p\in[0,1]$, $V_p$ is robust and the sequence $V\N_p$ converges robustly to $V_p$.
\end{enumerate}
\end{lemma}

\section{Reachability}
\label{sec:reachability}

In this section, we will focus on the computation of reachability probabilities of a single agent (or a fixed set of agents), in 
a population of increasing size. Essentially, we want to compute the probability of the set of traces reaching some goal 
state $G\subseteq \calS$ within $T$ units of time, starting at time $t_0$ and avoiding unsafe states in $U\subseteq \calS$. 
The key point is that the reachability probability of the limit CTMC $z_k(t)$ obtained by Theorem \ref{th:fastSimulation} 
approximates the reachability probability of a single agent in a large population of size $N$, i.e. the reachability probability 
for $Z\N_k(t)$. 

Similarly to Section \ref{sec:next}, we will consider two versions of the reachability problem: one for constant goal and 
unsafe sets, and one in which $G$ and $U$ depend on time. We will state these problems for a generic ICTMC $Z(t)$ 
on state space $\calS$:

\begin{definition}
\label{def:reachability}
Let $Z(t)$ be an ICTMC with state space $\calS$ and infinitesimal generator matrix $Q(t)$. 

\begin{enumerate}
\item Let  $U,G\subseteq \calS$.
The \emph{constant-set reachability} $P_{reach}(Z,t_0,T,G,U)[s]$ is the probability of the set of trajectories of $Z$ reaching 
a state in $G$ without passing through a state in $U$, within $T$ time units, starting at time $t_0$ in state $s$. 
$P_{reach}(Z,t_0,T,G,U)$ is the reachability probability vector on $\calS$.
\item Let $U,G:[t_0,t_1]\times\calS\rightarrow\{0,1\}$ be time-dependent sets, identified with their indicator function (i.e.\ 
$G(t),U(t)$ are the goal and the unsafe sets at time $t$).
The \emph{time-varying-set reachability} $P_{reach}(Z,t_0,T,G(t),U(t))[s]$ is the probability of the set of trajectories of $Z$ 
reaching a state in $G(t)$ at time $t\in [t_0,t_0+T]$ without passing through a state in $U(t')$, for $t'\in[t_0,t]$, starting at 
time $t_0$ in state $s$.
\end{enumerate}
\end{definition}

%The interest in the time-varying set reachability is intimately connected with CSL model checking. In fact, the truth value of a CSL formula in a state $s$ for the process $Z\N_k(t)$, where it is a non-Markov process,  depends on the initial time at which we start evaluating the formula. This is because $Z\N_k(t)$ depends on time via the state of $\nXN(t)$. Furthermore, time-dependence of truth values of CSL formulae manifests also for the limit process $z_k(t)$, which is a time-inhomogeneous Markov Process. Therefore, in the computation of the probability of an until formula, which can be reduced to the computation of two reachability problems, we need to consider time-varying-sets. As a matter of fact, we will see that dealing with time varying sets (like those obtained by solving the inequality $P_{reach}(Z,t_0,T,G(t),U(t))\bowtie p$, for $\bowtie\in\{<,\leq,>\geq\}$) is much harder, and we will need to impose some additional regularity on the rate functions of $Z$ and on the time-dependence of goal and unsafe sets.

In the following sections, we will first deal with the specific reachability problem for a generic ICTMC $Z(t)$, presenting 
an effective way of computing such probability, and then studying the relationship between the reachability probabilities of 
$Z\N_k(t)$ and $z_k(t)$. 

\subsection{Constant-set reachability}
\label{sec:constantReach}

We consider constant-set reachability, according to Definition \ref{def:reachability}. For the rest of this section let $Z(t)$ be 
an ICTMC on $\calS$, with rate matrix $Q(t)$ and initial state $Z(0) = Z_0 \in \calS$. We will solve the reachability problem 
in a standard way, by reducing it to the computation of  transient probabilities in a modified ICTMC 
\cite{MC:Hermanns:2000:MCofCTMCtransient}. The solution is similar to the one proposed in 
\cite{MC:Mereacre:2009:LTLmcICTMC}.

Let $\Pi(t_1,t_2)$ be the probability matrix of $Z(t)$, in which entry $\pi_{s_1,s_2}(t_1,t_2)$ gives the probability of being in state $s_2$ at time $t_2$, given that we were in state $s_1$ at time $t_1$. 
The \emph{Kolmogorov forward and backward equations} describe the time evolution of $\Pi(t_1,t_2)$ as a function of 
$t_2$ and $t_1$, respectively. More precisely, the forward equation is $\frac{\partial\Pi(t_1,t_2)}{\partial t_2} = 
\Pi(t_1,t_2)Q(t_2)$, while the backward equation is $\frac{\partial\Pi(t_1,t_2)}{\partial t_1} = -Q(t_1)\Pi(t_1,t_2)$.

The constant-set reachability problem, for a given initial time $t_0$, can be solved by integration of the forward Kolmogorov 
equation (with initial value given by the identity matrix) in the modified ICTMC $Z'(t)$, with infinitesimal generator matrix 
$Q'(t)$, in which all unsafe states and goal states are made absorbing \cite{MC:Hermanns:2000:MCofCTMCtransient} 
(i.e.\ $q'_{s_1,s_2}(t) = 0$, for each $s_1\in G\cup U$). 
In particular, $P_{reach}(Z,t_0,T,G,U) = \Pi'(t_0,t_0+T) \vr{e}_G$, where $\vr{e}_G$ is an $n\times 1$ vector equal to $1$ 
if $s\in G$ and 0 otherwise, and $\Pi'$ is the probability matrix of the modified ICTMC $Z'$.\footnote{Clearly, alternative ways 
of computing the transient probability, like uniformization for ICTMC \cite{PA:Wolf:2011:uniformizationICTMC}, could also be 
used.  However, we stick to the ODE formulation in order to deal with dependency on the initial time $t_0$.} 
We emphasise that, in order for the initial value problem defined by the Kolmogorov forward equation to be well posed, the 
infinitesimal generator matrix $Q(t)$ has to be sufficiently regular (e.g.\ bounded and integrable).

As already remarked,  in contrast with time-homogeneous CTMC, the reachability probability for ICTMC can depend on the 
initial time $t_0$ at which we start the process. 
Consider now the problem of computing $P(t) = P_{reach}(Z,t,T,G,U)$ as a function of $t\in [t_0,t_1]$. To this end, we 
can solve the forward equation for $t_0$ and then use the chain rule to define a differential equation for $\Pi(t,t+T)$, solving 
it using $\Pi(t_0,t_0+T)$ as the initial condition, i.e.
\begin{equation}
\label{eqn:rechForwBack}
\begin{split}
\frac{d\Pi(t,t+T)}{dt} & = \frac{\partial \Pi(t,t+T)}{\partial t} + \frac{\partial \Pi(t,t+T)}{\partial (t+T)}\frac{d(t+T)}{dt} \\ & =   -Q(t)\Pi(t,t+T) + \Pi(t,t+T)Q(t+T).
\end{split}
\end{equation}

Using a numerical solver for the ODE, this gives an effective algorithm  (Figure \ref{algo:reachabilityConstantSet}) to 
compute the probability of interest (for any fixed error bound).  Furthermore, if we can guarantee that the number of zeros 
of the equation $P(t) - p$ is finite, then we also have an effective procedure to compute the truth value of $P(t) \bowtie p$, 
for $\bowtie\in\{<,\leq,\geq,>\}$ (provided we can find those zeros, as will be discussed in Section~\ref{sec:CSLmodelCheking}). 

\begin{figure}[!t]
\begin{algorithmic}
\Function{reachability-constant-set}{$Z$, $T$, $G$, $U$, $t_0$, $t_1$}  

\State Construct the CTMC in which $G$ and $U$ states are absorbing, with rate matrix $Q'(t)$.

\State Compute $\Pi'(t_0,t_0+T)$ by solving the forward Kolmogorov ODE for the modified CTMC.

\State Compute $\Pi'(t,t+T)$ for $t\in[t_0,t_1]$ by solving ODE (\ref{eqn:rechForwBack}) for the modified CTMC with initial conditions $\Pi'(t_0,t_0+T)$.

\Return  $P(t) = \Pi'(t_0,t_0+T) \vr{e}_G$, $t\in[t_0,t_1]$.
\EndFunction
\end{algorithmic}
\caption{Algorithm for the computation of reachability probability $P(t)$ for $t\in[t_0,t_1]$ and constant goal and unsafe sets $G$ and $U$. Other input parameters are as in the text.}
\label{algo:reachabilityConstantSet}
\end{figure}

\bigskip

Consider now the sequence of processes $Z\N_k$ defined in Section \ref{sec:fastSimulation}. We are interested in the asymptotic behaviour of $P_{reach}(Z\N_k,t,T,G,U)$. 
The following result is a consequence of Theorem \ref{th:fastSimulation}:

\begin{proposition}
\label{prop:reachability}
Let $\calXN$ be a sequence of CTMC models, as defined in Section \ref{sec:modelingLanguage}, and 
let $Z\N_k$ and $z_k$ be defined from $\calXN$ as in Section \ref{sec:fastSimulation}. Assume that the infinitesimal 
generator matrix $Q(t)$ of $z_k$ is bounded and integrable in every compact interval $[0,T]$. Then 
\[ P_{reach}(Z\N_k,t,T,G,U) \rightarrow P_{reach}(z_k,t,T,G,U), \text{ uniformly in } [t_0,t_1], \text{ as } N\rightarrow\infty \] 
i.e.\ $\sup_{t\in [t_0,t_1]}\|P_{reach}(Z\N_k,t,T,G,U) - P_{reach}(z_k,t,T,G,U) \|\rightarrow 0$. 

\end{proposition}

The previous proposition shows that the reachability probability for $Z\N_k$ converges to the reachability probability for 
$z_k$, hence for large $N$ we can approximate the former with the latter. 

It is interesting to observe how the reachability probability for $Z\N_k(t)$ depends on the initial time. 
As previously remarked, $Z\N_k(t)$ is not a Markov-process, but $(Z\N_k,\nXN)(t)$ is. 
Furthermore, we can obtain $Z\N_k(t)$ by projecting on the first component of $(Z\N_k,\nXN)(t)$. 
The reachability probability for $Z\N_k(t)$ can be obtained from that of $(Z\N_k,\nXN)(t)$ in the following way: 
compute the reachability probability $P_{U,G}(s,x,T)$ for each state $(s,\x)$ of $(Z\N_k,\nXN)$ with time horizon $T$. 
As $(Z\N_k,\nXN)$ is a time-homogeneous CTMC, this probability is independent of the initial time. 
Fix a state $s\in\calS$ of $Z\N_k$, and consider the probability $P_{s,\x}(t|s) = \bbP\{(Z\N_k,\nXN)(t) = (s,x) \mid Z\N_k(t)
=s\}$ of being in $(s,x)$ at time $t$, conditional on being in $s$, i.e.\ $P_{s,\x}(t|s) = \bbP\{(Z\N_k,\nXN)(t) = (s,x)\}/\sum_{\x} 
\bbP\{(Z\N_k,\nXN)(t) = (s,x)\}$ (when the denominator is non-zero). 
Then, this is the initial distribution of $(Z\N_k,\nXN)$ that we have to take into account when computing the reachability 
probability $P_{reach}(Z\N_k,t,T,G,U)[s]$, starting at time $t$.  It follows that 
\begin{equation}
\label{eqn:probPN}
P_{reach}(Z\N_k,t,T,G,U)[s] = \sum_{\x\in \hat{\calD}} P_{s,\x}(t|s) P_{U,G}(s,x,T),
\end{equation}
which depends on $t$ via $P_{s,\x}(t|s)$.\\ 
As a consequence, the answer to a question like  $P_{reach}(Z\N_k,t,T,G,U)[s] > p$, $p\in[0,1]$, for $Z\N_k$ depends on 
the initial time $t$: \emph{truth is time-dependent in $Z\N_k$}.

%
%, even if $Z\N_k$ is obtained from the time-homogeneous CTMC $\nXN(t)$. In fact, the process $Z\N_k(t)$ is not a CTMC by itself. To make it a CTMC, we have to considered it together with $\nXN(t)$. Now, although $(Z\N_k(t),\nXN(t))$ is a time-homogeneous CTMC, the reachability probability  of $Z\N_k(t)$ is obtained from that of  $(Z\N_k(t),\nXN(t))$ by marginalising on $Z\N_k(t)$. Hence, fixing an initial state of $(Z\N_k,\nXN)$ at time 0 (satisfying Kurtz theorem assumptions), then the reachability probability for $Z\N_k$ does depend on the time $t$ via the state of the global system $\nXN(t)$. 

\begin{figure}
\begin{center}
\subfigure[$T$ varying, $n=10$, $m=5$] {\label{fig:timeoutonlyS1end}
\includegraphics[width=.47\textwidth]{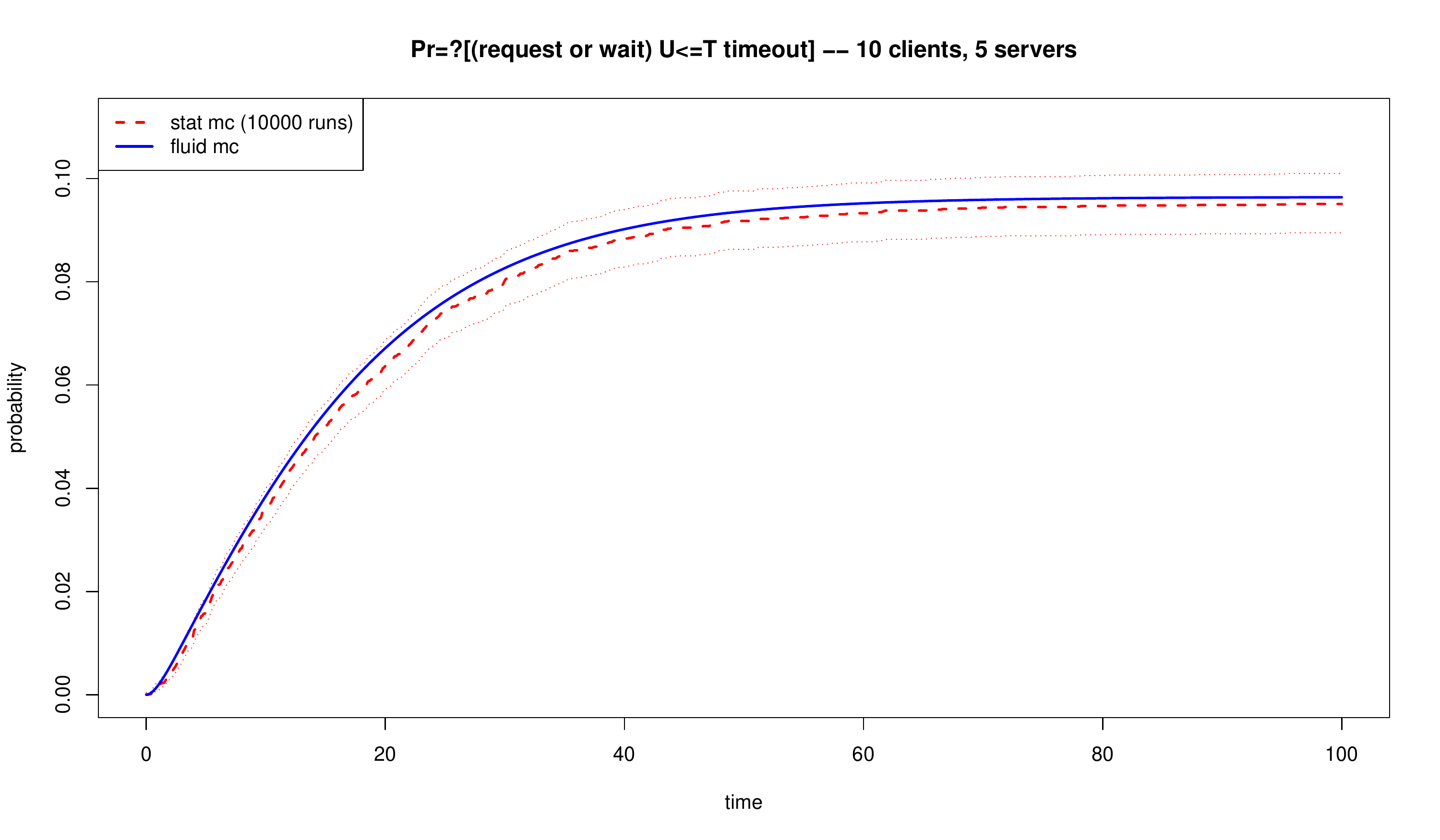} }
\subfigure[$T$ varying, $n=100$, $m=50$] {\label{fig:timeoutonlyS10end}
\includegraphics[width=.47\textwidth]{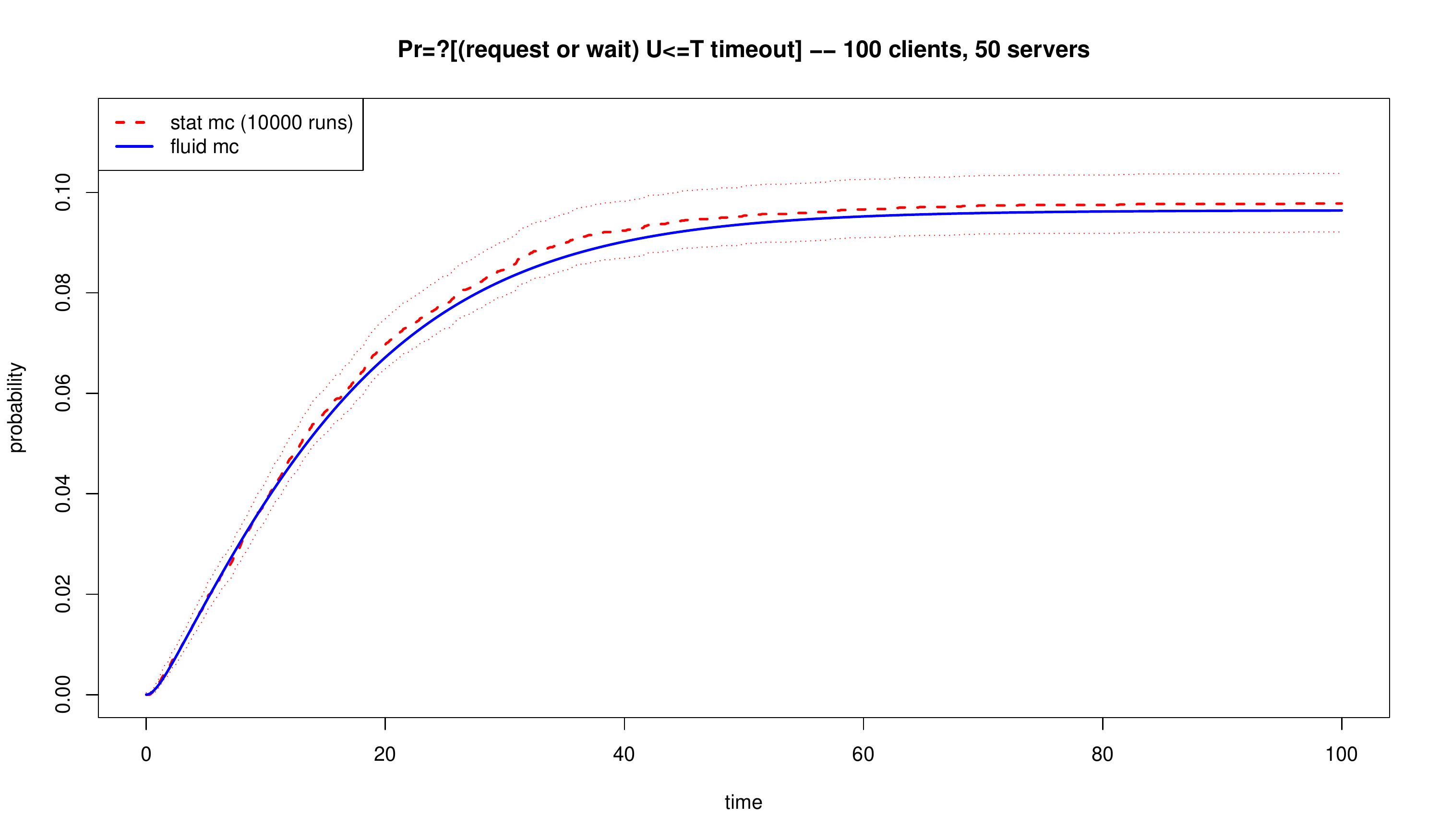} }

\subfigure[$t_0$ varying, $n=10$, $m=5$] {\label{fig:timeoutonlyS1tv}
\includegraphics[width=.47\textwidth]{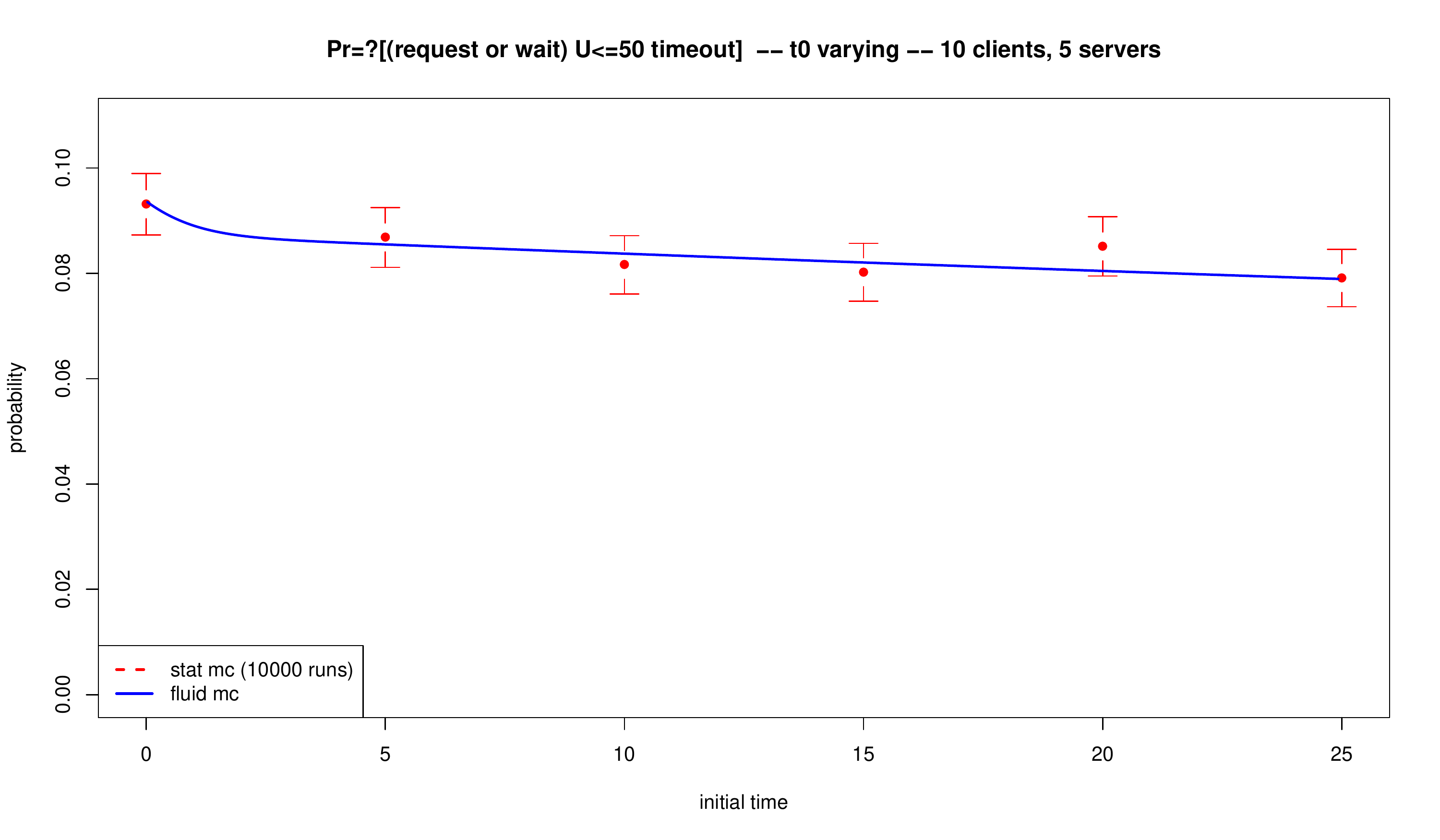} }
\subfigure[$t_0$ varying, $n=100$,
$m=50$] {\label{fig:timeoutonlyS10tv}
\includegraphics[width=.47\textwidth]{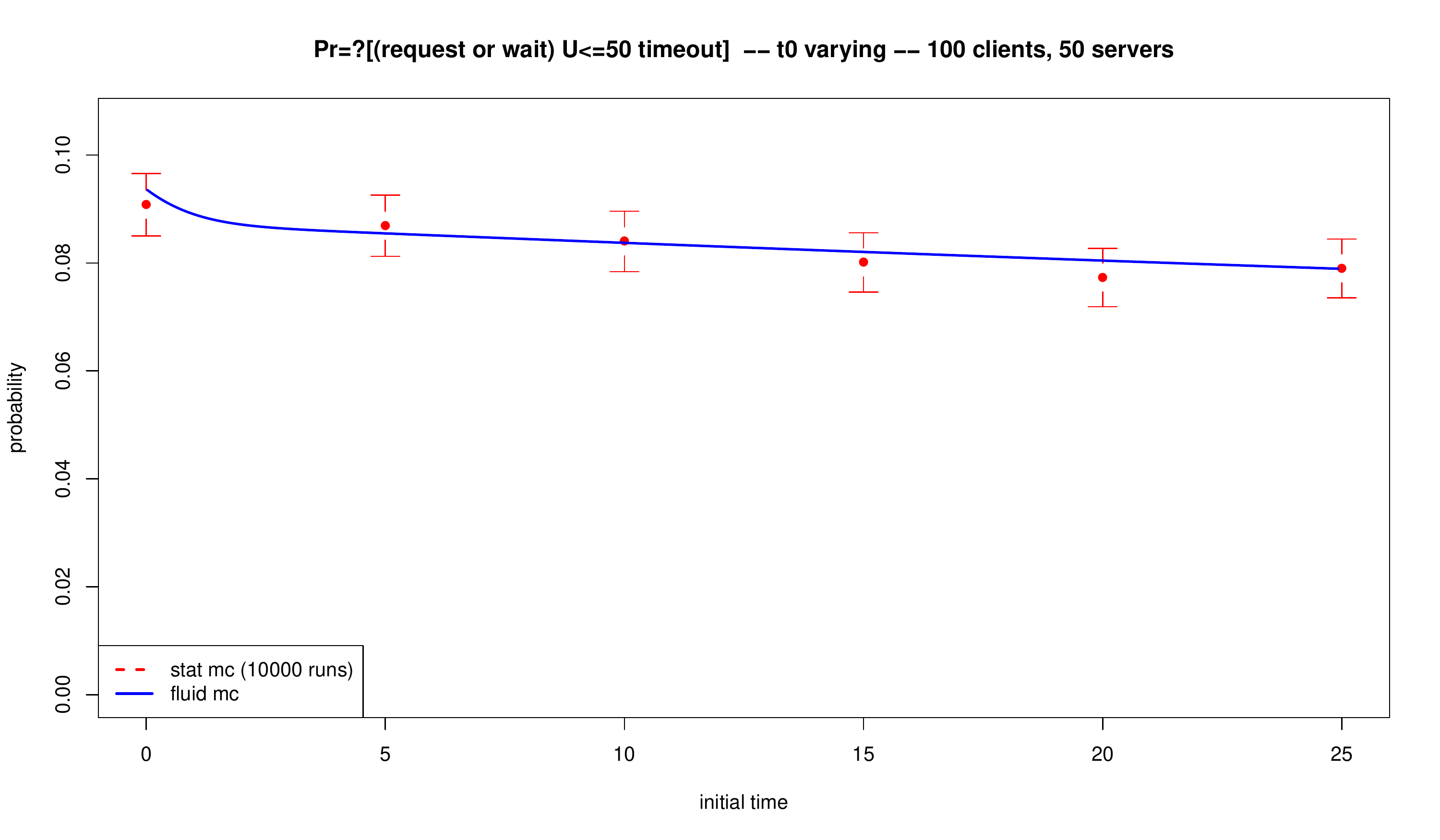} }

\subfigure[$n$, $m$ varying] {\label{fig:timeoutonlyS1nm}
\includegraphics[width=.47\textwidth]{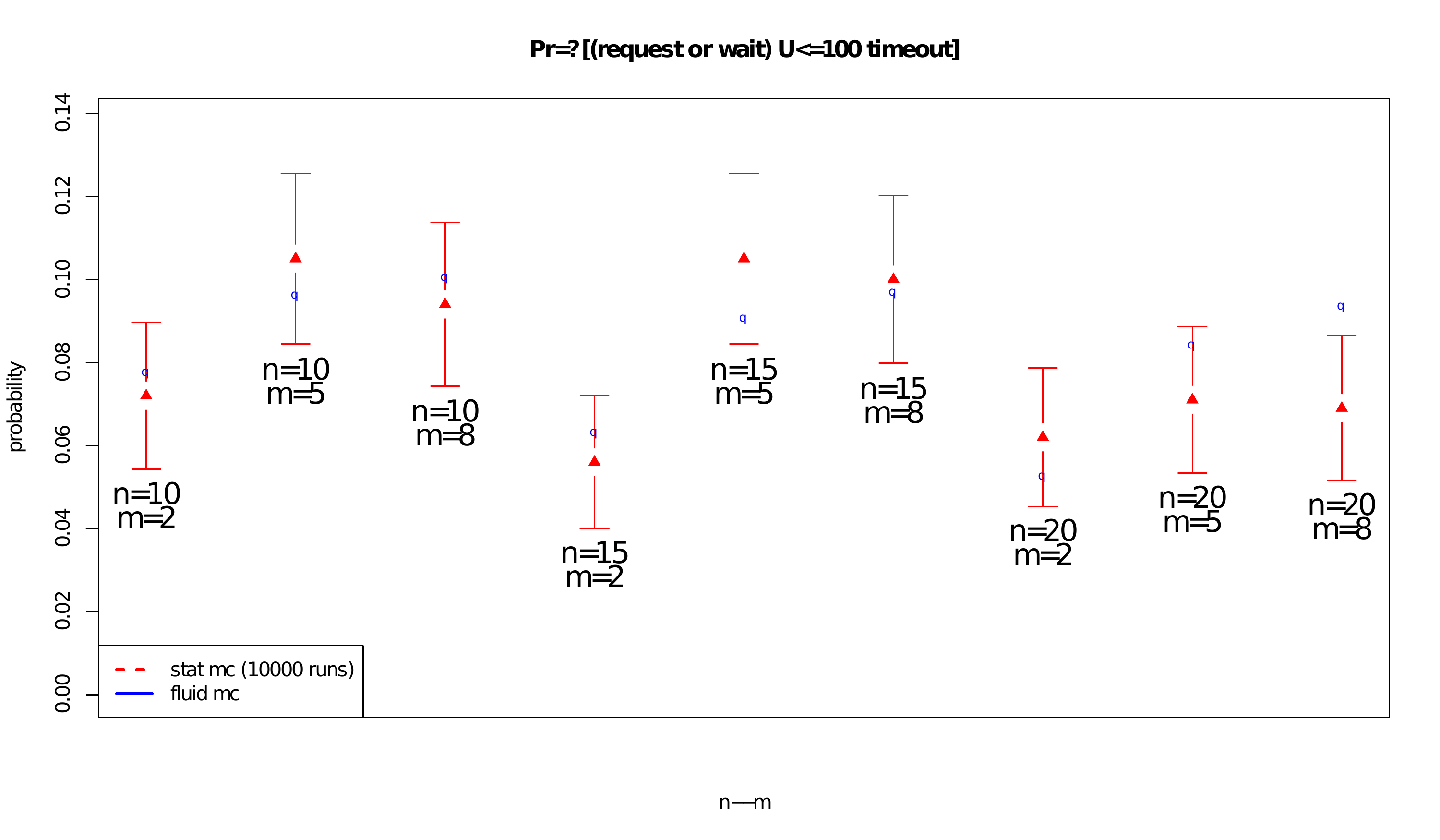} }
\subfigure[$n$, $m$ varying] {\label{fig:timeoutonlyS10nm}
\includegraphics[width=.47\textwidth]{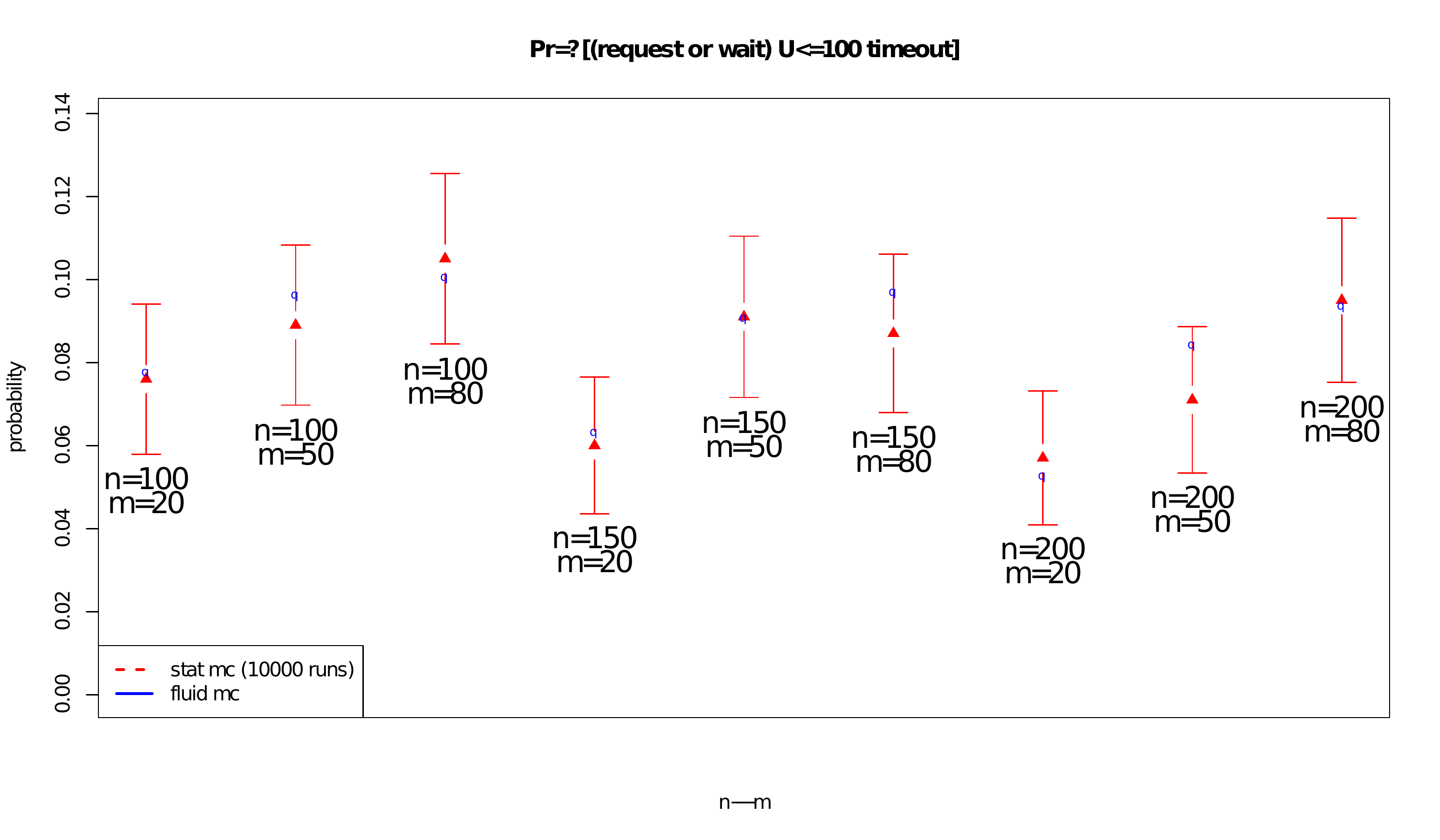} }
\end{center}

\caption{Client-Server model of Section \ref{sec:modelingLanguage}, single client CTMC. First line: comparison of time-out 
before being served  probability (point 1) for fluid and CTMC models as a function of time horizon $T$. Second line: 
comparison of time-out before being served  probability (point 1)  for fixed time horizon $T=50$ and variable initial time $t_0$. 
Third line: time-out before being served  probability (point 1) at time $T=250$, and variable number of client and servers.}
\label{fig:timeoutonly}
\end{figure}

\begin{figure}
\begin{center}
\subfigure[$T$ varying, $n=10$, $m=5$] {\label{fig:timeoutS1end}
\includegraphics[width=.47\textwidth]{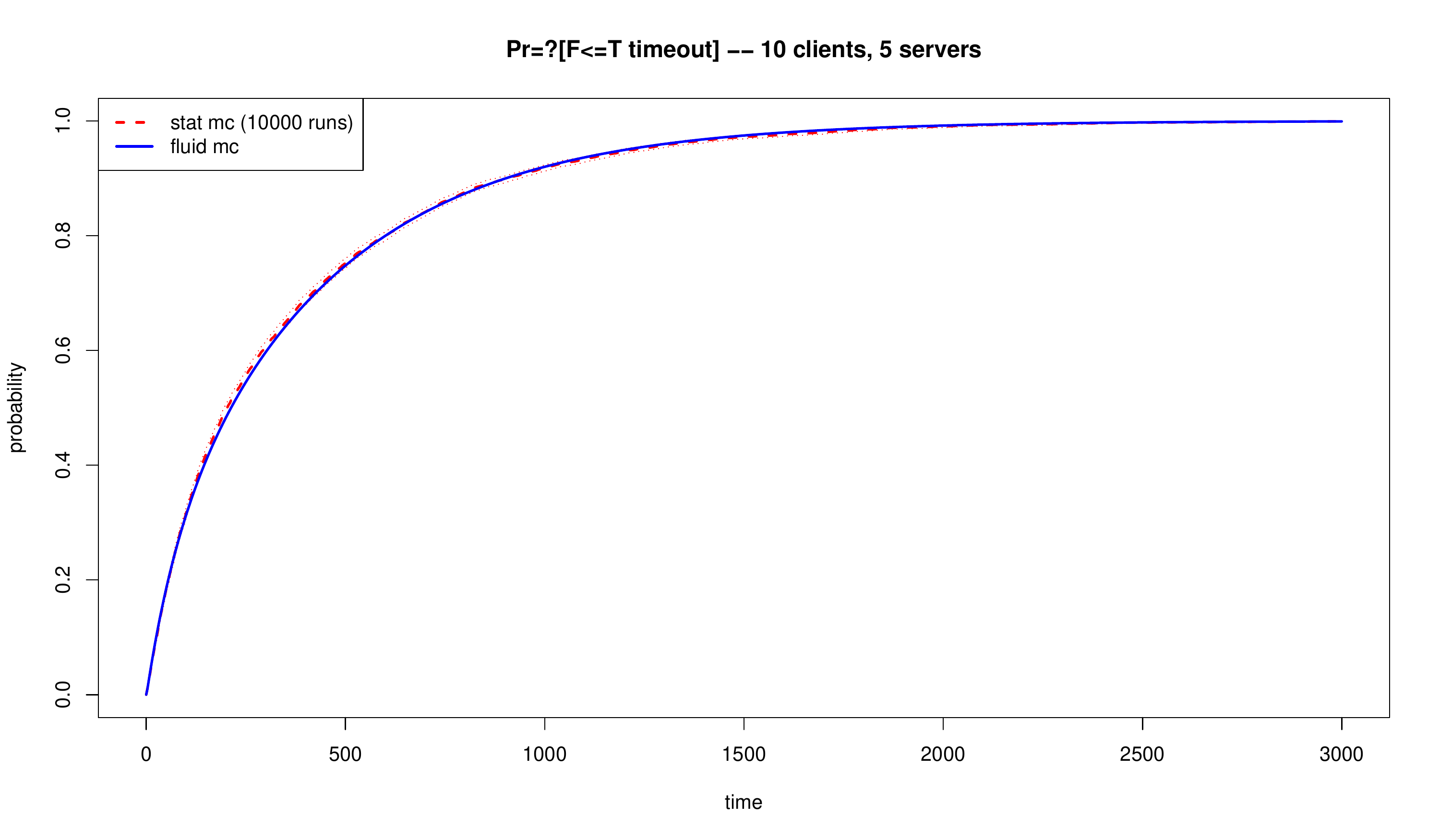} }
\subfigure[$T$ varying, $n=100$, $m=50$] {\label{fig:timeoutS10end}
\includegraphics[width=.47\textwidth]{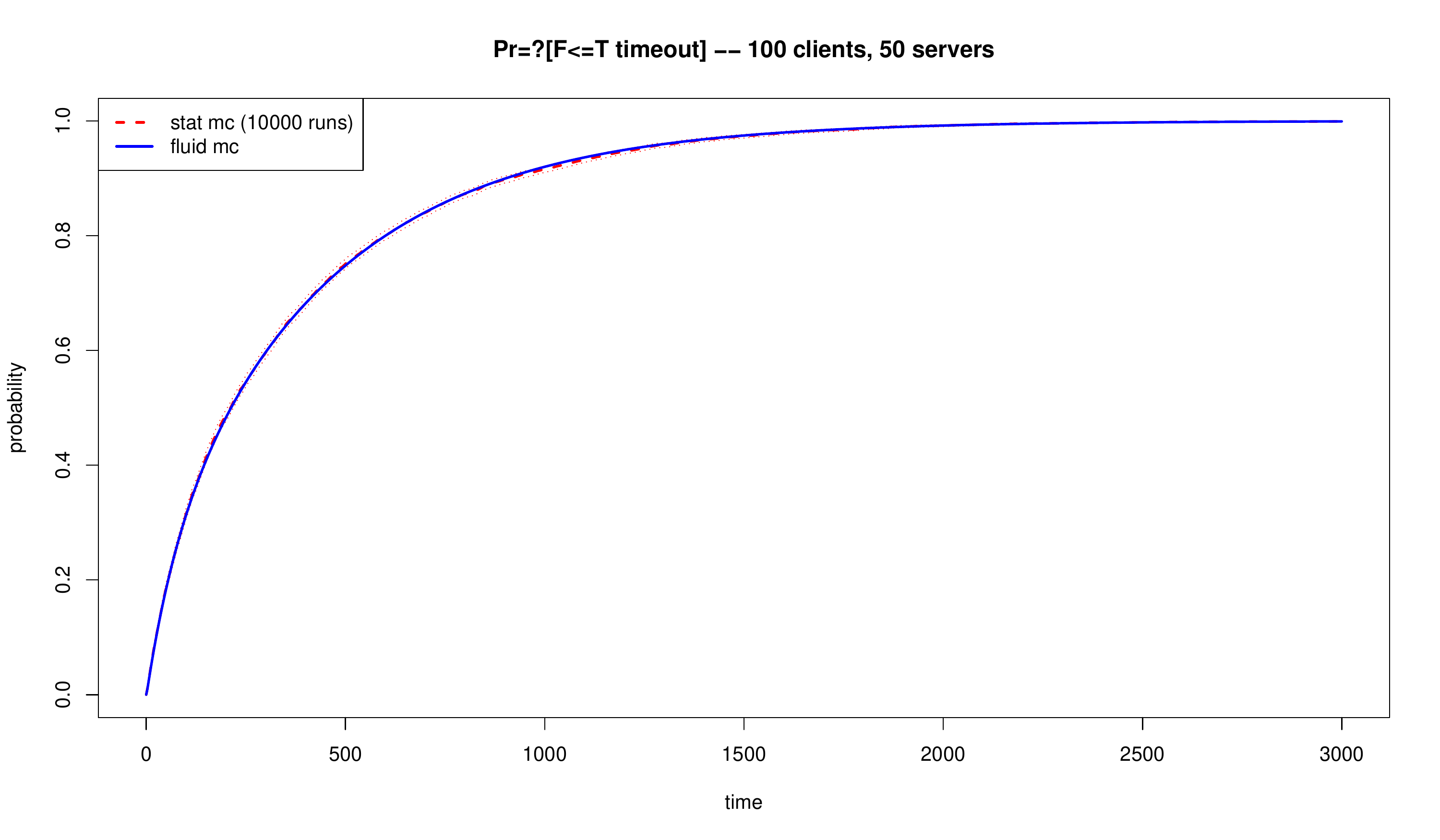} }

\subfigure[$t_0$ varying, $n=10$, $m=5$] {\label{fig:timeoutS1tv}
\includegraphics[width=.47\textwidth]{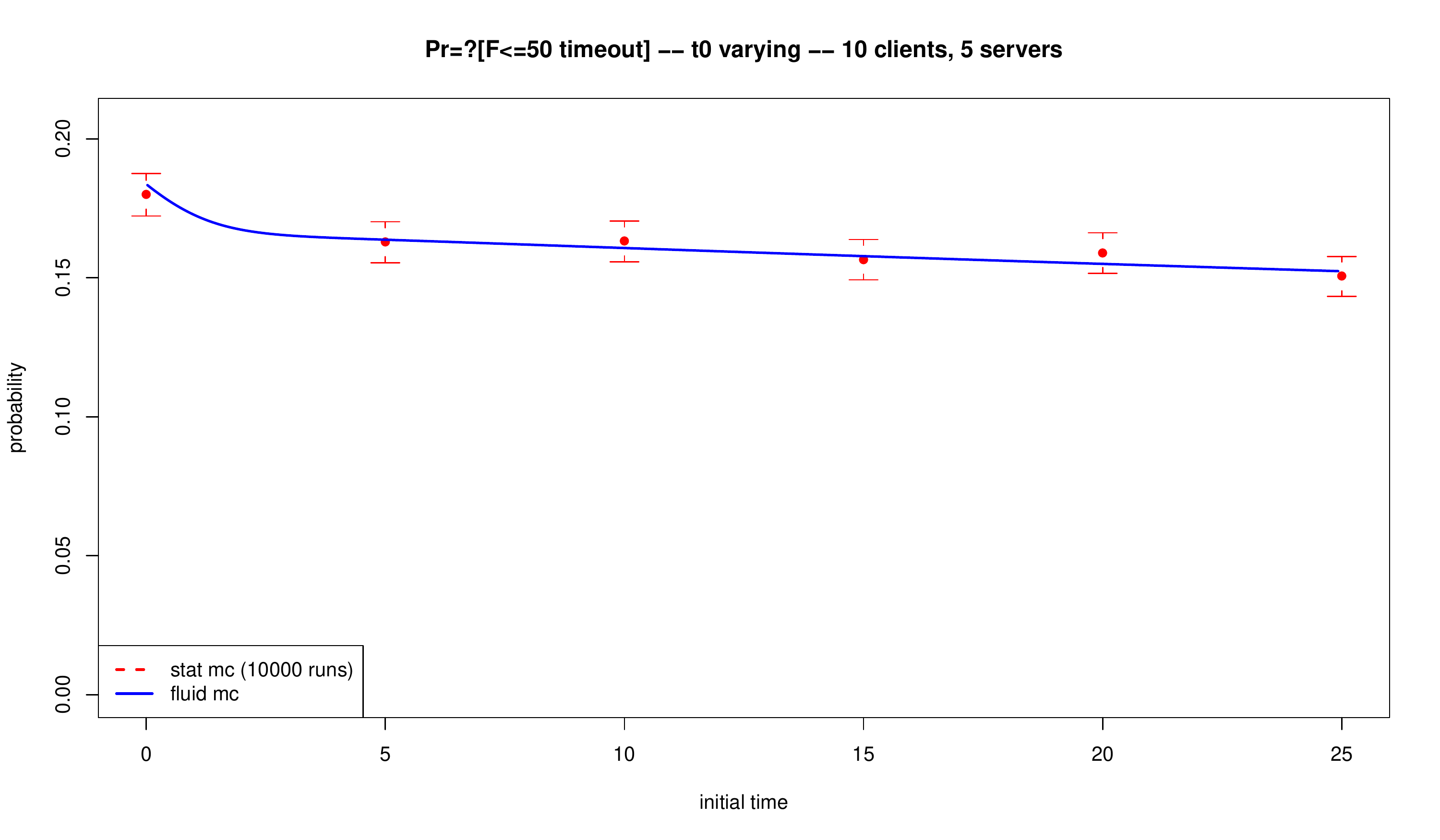} }
\subfigure[$t_0$ varying, $n=100$, $m=50$] {\label{fig:timeoutS10tv}
\includegraphics[width=.47\textwidth]{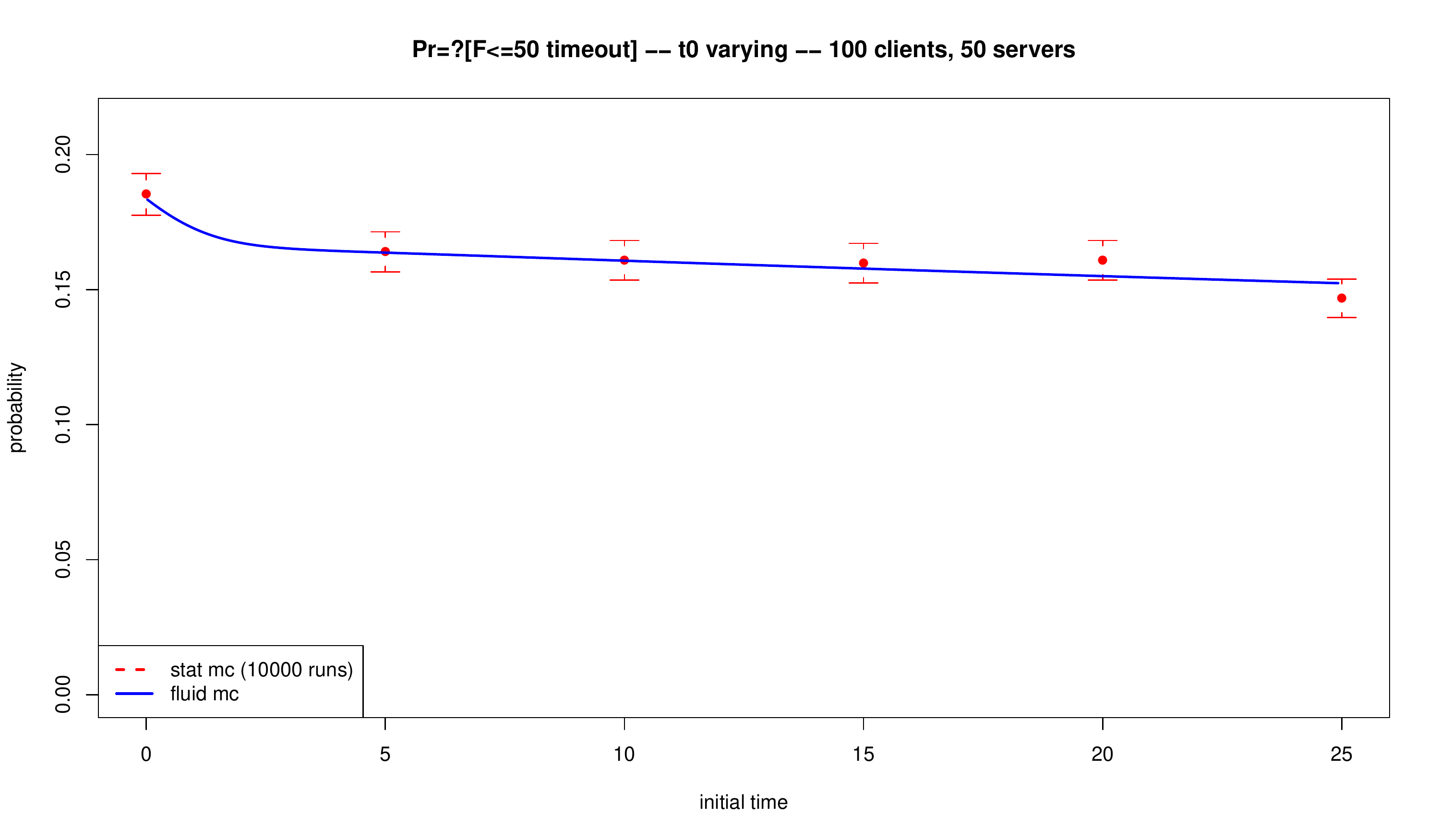} }

\subfigure[$n$, $m$ varying] {\label{fig:timeoutS1nm}
\includegraphics[width=.47\textwidth]{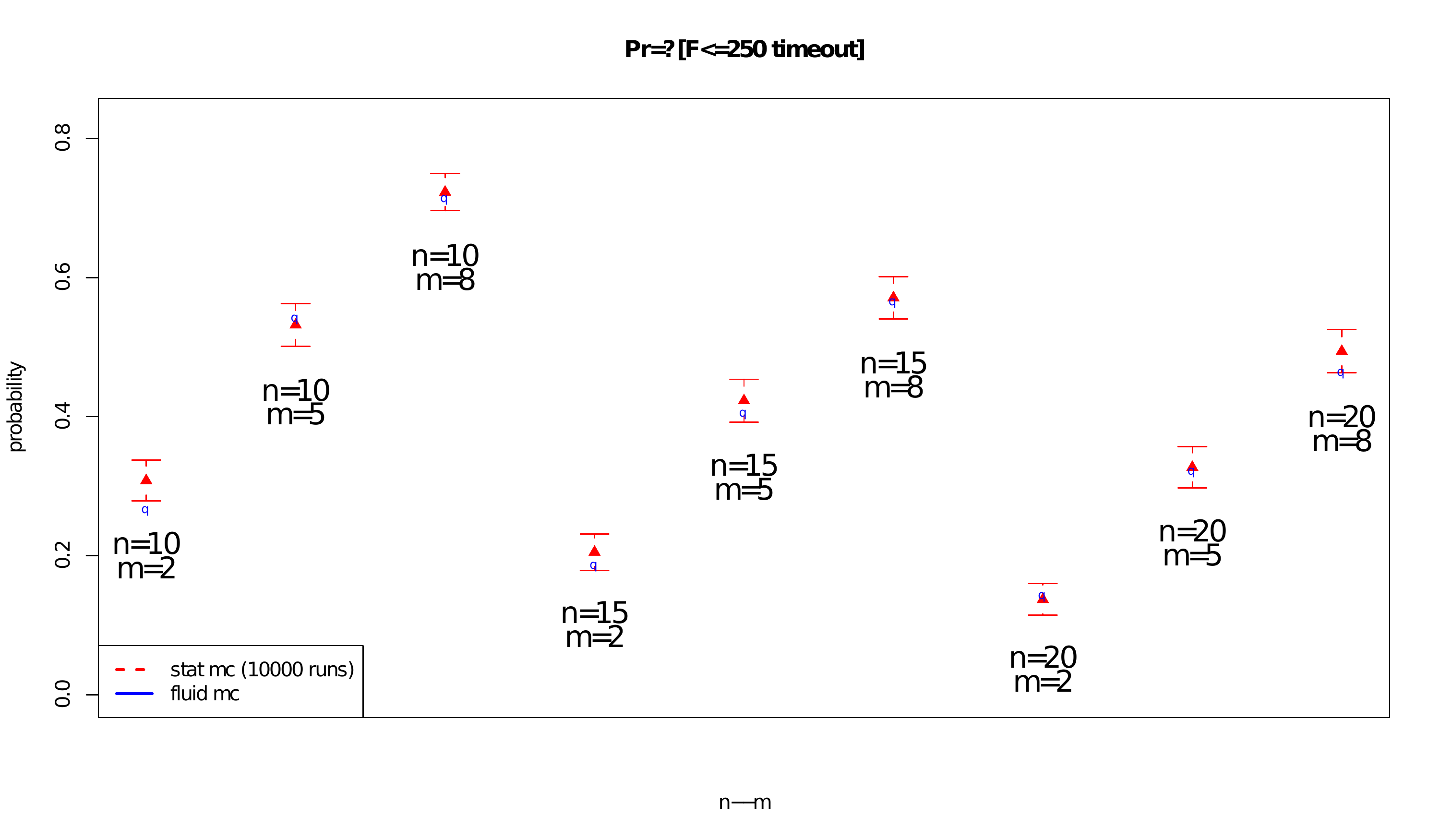} }
\subfigure[$n$, $m$ varying] {\label{fig:timeoutS10nm}
\includegraphics[width=.47\textwidth]{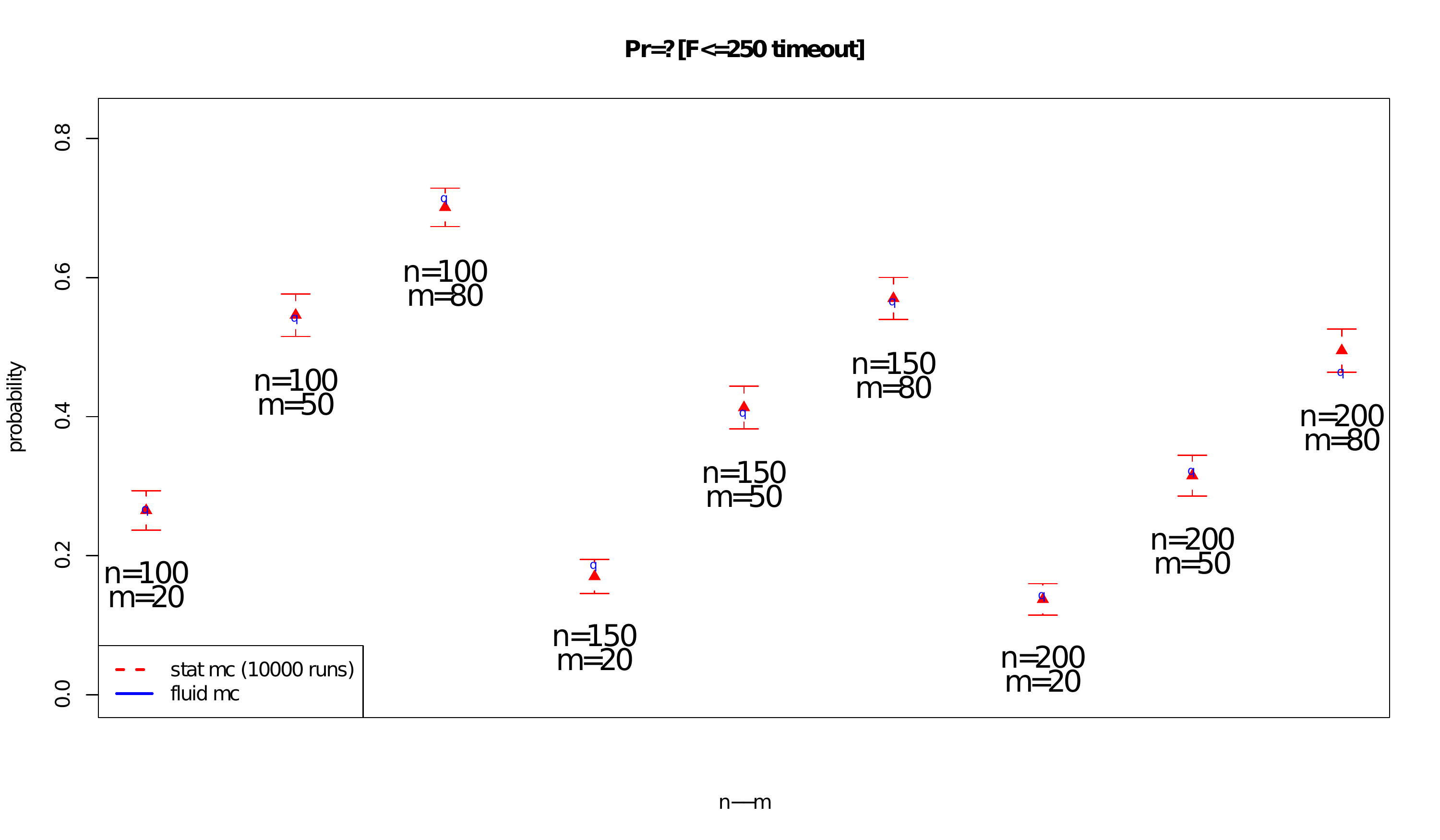} }
\end{center}

\caption{Client-Server model of Section \ref{sec:modelingLanguage}, single client CTMC. First line: comparison of time-out 
probability (point 2) for fluid and CTMC models as a function of time horizon $T$. Second line: comparison of time-out 
probability (point 2) for fixed time horizon $T=50$ and variable initial time $t_0$. Third line: comparison of time-out probability 
(point 2) at time $T=250$, and variable number of client and servers.}
\label{fig:timeout}
\end{figure}

\begin{exu}
We consider again the client-server example of Section \ref{sec:modelingLanguage}, and focus on  two reachability probabilities for a single client:
\begin{enumerate}
\item The probability of observing a time-out before  being served for the first time within time $T$. This is a reachability 
problem with goal set $G=\{rc\}$ and unsafe set $U=\{rq,w\}$.
\item The probability of observing a timeout within time $T$. This is a reachability problem with goal set $G=\{rc\}$ and 
unsafe set $U=\emptyset$.\footnote{In fact, this is a first passage time problem.}
\end{enumerate}
In Figures \ref{fig:timeoutonlyS1end}, \ref{fig:timeoutonlyS10end}, \ref{fig:timeoutS1end}  and \ref{fig:timeoutS10end} we 
can observe a comparison between the values computed for the limit ICTMC $z$ and the exact ICTMC $Z\N$, 
for $N = 15$ or $N=150$ (with a client-server ratio of 2:1), as a function of the time horizon $T$. As can be seen, the 
probability for $z$ is in very good agreement with that of $Z\N$ (computed using a statistical approach, from a sample of 
10000 traces) even for $N$ relatively small. As far as running time is concerned, the fluid model checking is 100 times faster 
for $N=15$, and 1000 times faster for $N=150$, than the stochastic simulation. What is even more important is that the 
complexity of the fluid approach is independent of $N$, hence its computational cost (on the order of 200 milliseconds for all 
cases considered here) can scale to much larger systems. 
Furthermore, another advantage of the fluid approach is that, by solving a set of differential equations, we are computing the 
reachability probability for each $t\in[0,T]$ (or better for any finite grid of points in $[0,T]$), while a method based on 
uniformisation (as in PRISM \cite{MC:Kwiatkowska:2004:PRISM}) has to deal with each time point separately.

In Figures  \ref{fig:timeoutonlyS1tv}, \ref{fig:timeoutonlyS10tv}, \ref{fig:timeoutS1tv}  and \ref{fig:timeoutS10tv}, instead, 
we focus on the reachability probability for both problem 1 and 2 for $T = 50$ as a function of the initial time $t_0\in [0,25]$. 
The value for the fluid model is compared with the probability of $Z\N_k$ obtained by simulating the full CTMC up to time 
$t_0$ and then focussing attention on a specific client in state $request$ and starting the computation of the reachability 
probability.\footnote{This is done by using two indicator variables $X_G$ and $X_U$ that are set equal to one when a 
trajectory reaches a goal or an unsafe set, respectively. Then, we estimate the reachability probability by the sample mean of 
$X_G$ at the desired time.} 
As we can see, the agreement is good also in this case. 

Finally, in Figures \ref{fig:timeoutonlyS1nm}, \ref{fig:timeoutonlyS10nm}, \ref{fig:timeoutS1nm}  and \ref{fig:timeoutS10nm}, 
we compare the reachability probability for $T=100$ (reachability problem 1) or $T=250$ (reachability problem 2) of the 
ICTMC for different populations $N$ and different proportions of clients ($n$) and servers ($m$), with the fluid limit. 
This data confirms that the agreement is good also for small populations for this model.  
\end{exu}

\subsection{Time-varying set reachability}
\label{sec:timeVaryingReach}

Now we turn our attention to the reachability problem for time-varying sets. First, we will focus on solving the problem for a 
generic ICTMC $Z(t)$, considering then the limit behaviour of $Z\N_k$. 

In order to deal with the reachability problem for time varying sets, the main difficulty is that, at each time $T_i$ in which the 
goal or the unsafe set changes, also the modified Markov chain that we need to consider to compute the reachability 
probability changes structure. This can have the effect of introducing a discontinuity in the probability matrix.

In particular, if at time $T_i$ a state $s$ becomes a goal state, then the probability $\pi_{s_1,s}(t,T_i)$ suddenly needs to 
be added to the reachability probability from state $s_1$. Therefore, a change in the goal set at time $T_i$ introduces a 
discontinuity in the reachability probability at time $T_i$. 
Similarly, if  a state $s$ was safe and then becomes unsafe, we have to discard the probability of trajectories that are in that
state at time $T_i$, as those trajectories become suddenly unsafe. 

In the following, let  $G(t)$ and $U(t)$ be the goal and unsafe sets, and assume that the set of time points in which $G$ 
or $U$ change value (at least in one state) is finite and equal to $T_1\leq T_2 \ldots \leq T_k$. This can be enforced by 
requiring that rate functions are piecewise analytic. 
Let $T_0 = t$ and $T_{k+1} = t+T$. 

In order to compute the reachability probability, we can exploit the semi-group property of the Markov process, stating that 
$\Pi(T_0,T_{k+1}) =  \prod_{i=0}^k \Pi(T_i,T_{i+1})$. Then, we also need to deal appropriately with the discontinuity effects 
at each time $T_i$, mentioned above. We proceed in the following way:

\begin{enumerate}
\item We double the state space, letting $\tilde{\calS} = \calS \cup \bar{\calS}$, where a state $\bar{s}\in\bar{\calS}$ 
represents state $s$ when it is a goal state. 
Hence, in the probability matrix $\tilde{\Pi}$, $\tilde{\pi}_{s_1,\bar{s_2}}$ is the probability of having reached $s_2$ avoiding unsafe states, while $s_2$ was a goal state. 
%This essentially corresponds to doubling the state space into accepting and non accepting states. 
%[CITATION?]
\item Consider a discontinuity time $T_i$ and let $t_1 \in [T_{i-1},T_i)$ and $t_2 \in (T_i,T_{i+1}]$.
Define $W(t) = S\setminus(G(t)\cup U(t))$.
Then, for $s_1\in W(t_1)$ and $s_2\in W(t_2)$, the probability of being in $s_2$ at time $t_2$, given that we were in 
$s_1$ at time $t_1$ and avoiding both unsafe and goal sets, can be written as $\tilde{\pi}_{s_1,s_2}(t_1,t_2) = \sum_{s \in 
W(t_1)\cap W(t_2)} \tilde{\pi}_{s_1,s}(t_1,T_i)\tilde{\pi}_{s,s_2}(T_i,t_2)$.
Hence, we have to appropriately restrict the summation set at time $T_i$, to account for changes in $W$.  
\item Consider again a discontinuity time $T_i$ and let $t_1 \in [T_{i-1},T_i)$ and $t_2 \in (T_i,T_{i+1}]$. Suppose 
$s_2\in W(t_1)$ and $s_2\in G(t_2)$. Then, the probability of reaching the goal state $s_2$ at time $t_2$, given that at 
time $t_1$ we were in $s_1$, can be written as  \ifdefined\arxivVersion \[  \else $ \fi  \tilde{\pi}_{s_1,s_2}(t_1,T_i) + \sum_{s \in W(t_1)\cap 
W(t_2)}\tilde{\pi}_{s_1,s}(t_1,T_i)\tilde{\pi}_{s,\bar{s}_2}(T_i,t_2) \ifdefined\arxivVersion .\] \else $. \fi  The first term is needed because all safe trajectories 
that are in state $s_2$ at time $T_i$ suddenly become trajectories satisfying the reachability problem, hence we have to 
add them to compute the reachability probability.
\end{enumerate}

All the previous remarks can be formally incorporated into the semi-group expansion of $\tilde{\Pi}(t,t+T)$ by multiplying 
on the right each term $\tilde{\Pi}(T_i,T_{i+1})$ by a suitable 0/1 matrix, depending only on the structural changes at time 
$T_{i+1}$. 
Let $|\calS| =  n$ and let $\zeta_W(T_i)$ be the $n\times n$ matrix equal to 1 only on the diagonal elements corresponding 
to states $s_j$ belonging to both $W(T_i^-)$ and $W(T_i^+)$ (i.e.\ states that are safe and not goals both before and after 
$T_i$), and equal to 0 elsewhere. 
Furthermore, let $\zeta_G(T_i)$ be the $n\times n$  matrix equal to 1 in the diagonal elements corresponding to states 
$s_j$ belonging to $W(T_i^-)\cap G(T_i^+)$, and zero elsewhere. 
Finally, let $\zeta(T_i)$ be the $2n\times 2n$ matrix defined by:
$$\zeta(T_i) = \left(\begin{array}{cc} 
\zeta_W(T_i) & \zeta_G(T_i)\\
0 & I
\end{array} \right).$$

Consider now the following ICTMC $\tilde{Z}$ on $\tilde{\calS}$, with rate matrix $\tilde{Q}(t)$, where 
\begin{enumerate}
\item for $\bar{s}_1 \in\bar{\calS}$ and any $s_2\in \tilde{\calS}$, $\tilde{q}_{\bar{s}_1,s_2}(t) = 0$;  
\item for $s_1\not\in W(t)$ and all $s_2\in\tilde{\calS}$,  $\tilde{q}_{s_1,s_2}(t) = 0$
\item for $s_1\in W(t)$ and $s_2 \in S \setminus G(t)$, $\tilde{q}_{s_1,s_2}(t) = q_{s_1,s_2}(t)$, while $\tilde{q}_{s_1,\bar{s}_2}(t) = 0$;
\item for $s_1\in W(t)$ and $s_2 \in G(t)$, $\tilde{q}_{s_1,\bar{s}_2}(t) = q_{s_1,s_2}(t)$, while $\tilde{q}_{s_1,s_2}(t) = 0$.
\end{enumerate}

In the previous chain, all unsafe and goal states are absorbing, while transitions leading from a safe state $s$ to a goal 
state are readdressed to the copy $\bar{s}$ of $s$. States in $\bar{\calS}$ are absorbing, too. 

Now let $\tilde{\Pi}(t_1,t_2)$ be the probability matrix associated with the ICTCM $\tilde{Q}(t)$. Given the interval $I = [t,t+T]$, we indicate with $T_1,\ldots,T_{k_I}$ the ordered sequence of discontinuity points of goal and unsafe sets internal to $I$. Let 
\begin{equation}
\label{eqn:upsilonDef}
\Upsilon(t,t+T) = \tilde{\Pi}(t,T_1)\zeta(T_1)\tilde{\Pi}(T_1,T_2)\zeta(T_2)\cdots \zeta(T_{k_I})\tilde{\Pi}(T_{k_I},t+T).
\end{equation}
Then, we have that
\begin{equation}
\label{eqn:reachabilityProb}
P_s(t) = P_{reach}(Z,t,T,G,U)[s] = \sum_{\bar{s}_1\in\bar{\calS}} \Upsilon_{s,\bar{s}_1}(t,t+T) + \vr{1}\{s\in G(t)\},
\end{equation}
where the first term takes into account the probability of reaching a goal state starting from a non-goal state, while the second term is needed to properly account for states $s\in G(t)$, for which $P_s(t)$ has to be equal to 1 (a formal proof can be given by induction on the number of discontinuity points).
%
%Furthermore, let $\vr{\bar{e}}$ be the $2n\times 1$ vector equal to 1 for states $\bar{s}\in\bar{S}$, and zero elsewhere, and $\vr{e}_{G}(t)$ be the vector equal to 1 for $s\in G(t)$ and zero elsewhere. Then, from the previous discussion, it should be clear that $P(t) = P_{reach}(t,T,U,G) = \Upsilon(t,t+T)\vr{\bar{e}} + \vr{e}_{G}(t)$ (projected on the first $n$ coordinates): 
$\Upsilon(t,t+T)$ can be obtained by computing each $\tilde{\Pi}(T_i,T_{i+1})$ solving the associated forward Kolmogorov equation and then multiplying those matrices and the appropriate $\zeta$ ones, according to the definition of $\Upsilon$. 

If we want to compute $P(t)$ as a function of $t$, instead, we need a way to compute $\Upsilon(t,t+T)$ as a function of $t$. 
This can be done by observing that $\Upsilon$ depends on $t$ only from the first and last factors in the multiplication. 
Defining $\Gamma(T_1,T_k) = \zeta(T_1)\tilde{\Pi}(T_1,T_2)\zeta(T_2)\cdots \tilde{\Pi}(T_{k-1},T_k)\zeta(T_k)$, writing 
$\Upsilon(t,t+T) = \tilde{\Pi}(t,T_1)\Gamma(T_1,T_k)\tilde{\Pi}(T_k,t+T)$, differentiating with respect to $t$ and applying 
the forward or backward equation for $\tilde{\Pi}$, we find the following differential equation for $\Upsilon$:
\begin{equation}
\label{eqn:upsilonODE}
\frac{d\Upsilon(t,t+T)}{dt} = -\tilde{Q}(t)\Upsilon(t,t+T) + \Upsilon(t,t+T)\tilde{Q}(t+T).
\end{equation}
This equation holds until either $t$ or $t+T$ becomes equal to a discontinuity point. When this happens, the integration 
has to be stopped and restarted, recomputing $\Upsilon$ accordingly.

%%%%[MAYBE CAN BE DROPPED IN A CONFERENCE PAPER]

Practically, to solve this problem we can proceed as follows:
\begin{enumerate}
\item Given an interval $[t_0,t_1]$ of interest for the initial time of the reachability, find all discontinuity points of the sets 
$G$ and $U$ contained in $[t_0,t_1+T]$, and let them be $t_0 = T_0 < T_1 < \ldots < T_k < T_{k+1} = t_1 + T$.
Furthermore, let $T_i' = T_i + T$ for $i=0,\ldots, k$,  let $pre(t)$ be the greatest $T_j$ preceding $t$, and $post(t)$ the 
smallest $T_j$ following $t$.
\item Compute $\tilde{\Pi}(T_i,T_{i+1})$ and $\tilde{\Pi}(pre(T_i'),T_i')$ for $i\leq k$, using the forward Kolmogorov equations
\footnote{Notice, that, if $T_j = pre(T_i')$, then $\tilde{\Pi}(T_j,T_i')$ and $\tilde{\Pi}(T_j,T_{j+1})$ can be computed during 
the same numerical integration of the forward equation.}. Compute also each $\zeta(T_i)$. 
\item Compute $\Upsilon(t_0,t_0+T)$ and integrate until time $t = \min\{T_1,T_{j+1} - T\}$, where $t_0 + T \in [T_j,T_{j+1}]$.
\item If $t+T = T_{j+1}$, multiply $\Upsilon$ on the right by $\zeta(T_{j+1})$ and continue the integration. If $t = T_1$, then 
recompute $\Upsilon$ as $\tilde{\Pi}(T_1,T_2) \Gamma(T_2,T_j)\tilde{\Pi}(T_j,T_1+T)$, where $\tilde{\Pi}(T_j,T_1+T) = \tilde
{\Pi}(pre(T_1'),T_1')$.
\item Integrate piecewise using the previous rules until time $t_1$.
\end{enumerate}

A more detailed algorithmic presentation of the procedure is given in Figure \ref{algo:reachability}.
Notice that, if the infinitesimal generator matrix $Q(t)$ of $Z$ is sufficiently well-behaved (for instance, Lipschitz continuous), 
then the function $P(t)$ will be at least piecewise continuous, with a finite number of discontinuity points at instants $T_i$ 
and $T_i'$.

\begin{remark}
\label{rem:discontPoints}
The precise behaviour of $G$ and $U$ functions at their discontinuity points (i.e.\ if they are left-continuous or right-
continuous) is irrelevant for the computation of $\Upsilon$: the set of trajectories of $Z$ differing in those time points has 
probability 0.
\end{remark}

\begin{remark}
\label{rem:complexity}
In the previous method, we need to integrate repeatedly a set $4n^2$ differential equations. However, most of these variables are redundant. In fact, we only need $n^2$ variables for the probability transition matrix $\Pi$ on $\calS$ and an additional $n$ variables to store the reachability probability vector. The method presented above can be easily reconfigured to this restricted set of variables.
\end{remark}

\begin{figure}[!t]
\begin{algorithmic}
\Function{reachability}{$Z$, $T$, $G$, $U$, $t_0$, $t_1$}  

\State Construct the CTMC on the modified state space $\tilde{\calS}$, according to the recipe in the text. 

\State Let $t_0 = T_0, T_1, \ldots, T_k, T_{k+1} = t_1 + T$ be the time instants at which $G$ or $U$ has a discontinuity. %Let $T_i' = T_i+T$.

\For{$i = 0$ to $k$} 

\State Compute $\tilde{\Pi}(T_i,T_{i+1})$ and $\tilde{\Pi}(pre(T_i + T),T_i + T)$ using the forward Kolmogorov equations 

\EndFor  

\State Compute $\Upsilon(t_0,t_0+T)$ according to equation (\ref{eqn:upsilonDef}) and $P(t_0)$ according to equation (\ref{eqn:reachabilityProb})

\State $t \leftarrow t_0$

\Repeat  

\State $T_a \leftarrow post(t)$

\State $T_b \leftarrow post(t+T)$

\State $\bar{t} \leftarrow \min\{T_a,T_b-T\}$

\State Compute $\Upsilon$ and $P$ from $t$ to $\bar{t}$, according to ODE (\ref{eqn:upsilonODE}) and equation 
(\ref{eqn:reachabilityProb}), with initial conditions $\Upsilon(t,t+T)$ (previously computed).

\If{$\bar{t} + T = T_b$} 

\State $\Upsilon(\bar{t},\bar{t}+T) \leftarrow \Upsilon(\bar{t},\bar{t}+T) \zeta(T_b)$

\ElsIf{$\bar{t} = T_a$} 

\State $\Upsilon(\bar{t},\bar{t}+T) \leftarrow  \tilde{\Pi}(T_a,post(T_a)) \Gamma(post(T_a),pre(T_b))\tilde{\Pi}(pre(T_b),T_b+T)$

\EndIf

\State $t \leftarrow \bar{t}$

\Until{$t \geq t_1$}

\Return  $[\Upsilon(t,t+T),P(t)]$, $t\in[t_0,t_1]$.
\EndFunction
\end{algorithmic}
\caption{Algorithm for the computation of reachability probability $P(t)$ for $t\in[t_0,t_1]$ and time-varying goal and unsafe sets $G(t)$ and $U(t)$, with a finite number of discontinuities. Other input parameters are as in the text.}
\label{algo:reachability}
\end{figure}

\subsubsection*{Limit behaviour}

We consider now the limit behaviour of time-varying reachability probability for $Z\N_k$, proving that it converges 
(almost everywhere) to that of $z_k$. As in Section \ref{sec:next},  we state this result in a more general form, assuming 
that also the goal and unsafe sets depend on $N$, and converge robustly to some robust limit sets $G$ and $U$. 
% compatibility condition
Furthermore, we require that $G$ and $U$ are \emph{compatible} in the sense that they do not have a discontinuity at the same time for the same state $s$: $\forall s\in \calS$, $Disc(G_s)\cap Disc(U_s) = \emptyset$. 
The following lemma, which is also the basic inductive step to prove convergence for CSL model checking formulae, 
relies on the functions involved being piecewise analytic.

\begin{lemma}
\label{lemma:tameTimeDepReachab}
Let $\calXN$ be a sequence of CTMC models, as defined in Section \ref{sec:modelingLanguage}, and 
let $Z\N_k$ and $z_k$ be defined from $\calXN$ as in Section \ref{sec:fastSimulation}, with piecewise analytic rates, 
in a compact interval $[0,T']$, for $T'$ sufficiently large. \\
Let $G(t)$, $U(t)$, $t\in[t_0,t_1+T]$ be \emph{compatible} and \emph{robust} time-varying sets, and let $G\N(t)$, $U\N(t)$ be sequences of 
time-varying sets converging robustly to $G$ and $U$, respectively. \\ 
Furthermore, let $P(t) = P_{reach}(z_k,t,T,G,U)$ and\\ $P\N(t) = P_{reach}(Z\N_k,t,T,G\N,U\N)$, $t\in[t_0,t_1]$.\\
Finally, fix $p\in[0,1]$, $\bowtie\in\{\leq,<,>,\geq\}$, and let $V_p(t) = I\{P(t)\bowtie p\}$, $V\N_p(t) = I\{P\N(t)\bowtie p\}$. Then
\begin{enumerate}
\item For all but finitely many $t\in [t_0,t_1]$, $P\N(t)\rightarrow P(t)$, with uniform speed (i.e. independently of $t$). 
\item For almost every $p\in[0,1]$, $V_p$ is robust and the sequence $V\N_p$ converges robustly to $V_p$.
\end{enumerate}
\end{lemma}

\begin{exu}
If we consider our running example, then it is easy to check that the rate functions defining the infinitesimal generator 
matrices of interest are piecewise analytic. In fact, even if the vector field of the fluid ODE is not analytic, due to the 
minimum function, the two functions $g_1(\x)$ and $g_2(\x)$ of which we take the minimum are analytic. Piecewise 
analyticity follows from the fact that the solutions of the associated ODE cross the surface $g_1(\x)-g_2(\x)=0$ (where 
the minimum is not analytic) only a finite number of times. 
\end{exu}

%*****************************************************************************

\section{CSL Model Checking}
\label{sec:CSLmodelCheking}

We turn now to consider the model checking of CSL formulae and the relationship between the truth of formulae for 
$Z\N_k$ and $z_k$. 

Consider an until CSL formula $\phi = \calP_{\bowtie p}(\phi_1\until{0}{T} \phi_2)$, where $\phi_1$ and $\phi_2$ are boolean 
combinations of atomic propositions. The major consequence of the time-inhomogeneity of $z_k$ is that the truth value of 
$\phi$ in a state $s$ depends on the time $t$ at which we evaluate such a formula.  In particular, $\phi$ may be true in state 
$s$ at time $t_1$, but false at a different time $t_2$. Consequently, the set of states that satisfy a CSL formula $\phi$ can be 
time dependent, and this introduces an additional layer of complexity to the analysis of $z_k$. Indeed, this requires the 
computation of next-state and reachability probabilities for time-varying sets. There is a similar problem with next formulae of 
the form $\phi = \calP_{\bowtie p}(\next{T_a}{T_b}\phi_1)$, as the next-state probability also depends on the evaluation time 
$t$.  Notice that we have the same issue about time-dependence also for  model checking CSL formulae against $Z\N_k(t)$.

%Indeed, the problem of model checking CSL formulae on time-inhomogeneous CTMC is difficult, and to the authors' knowledge, there is no general method in literature. An exception is \cite{HML ICTMC}, in which the authors show a model checking algorithm for the Hennessy-Milner logics, under the hypothesis of piecewise constant rates. 

The method we put forward in the previous sections can cope with these issues, but in general may require a large 
computational effort (for until formulae, the solution of systems of ODE quadratic in the size of the state space of the ICTMC, 
and it depends on the number of discontinuity points of the sets $U$ and $G$). However, in our setting we are interested in 
$z_k$, which is an abstract and approximate model of the behaviour of a single agent. Usually, a single agent has a very small 
state space, hence the given approaches to compute next-state probability and reachability of time-varying sets should be 
feasible in practice.

An orthogonal issue is the asymptotic correctness of CSL model checking, when considering the sequence $Z\N_k$ and the 
limit $z_k$. As boolean operators pose no real problem, we only need to concentrate on next formulae $\phi = \next{T_a}{T_b}\phi_2$ and on until formulae  $\phi = \calP_{\bowtie p}(\phi_1\until{0}{T} \phi_2)$, with time varying sets satisfying $\phi_1$ 
and $\phi_2$.

In particular, we can reduce this problem to the computation of the next-state probabilities $\bP\N(t) = P_{next}(Z\N_k,t,T_a,
T_b,G\N)$ and $\bP(t) = P_{next}(z_k,\ifdefined\arxivVersion  \else \linebreak \fi  t,T_a,T_b,G)$ (for next formulae) or to reachability probabilities
$P\N(t) = P_{reach}(Z\N_k,t,T,G\N,U\N)$ and $P(t) = P_{reach}(z_k,t,T,G,U)$ (for until formulae), 
where $G\N(t)$ ($U\N(t)$) is the set of states satisfying $\phi_2$ ($\neg\phi_1$) at time $t$ for $Z\N_k$, while $G$ and $U$ 
are defined similarly for $z_k$.\footnote{We can restrict our attention to until formulae with time between $[0,T]$, as intervals 
$[T_a,T_b]$ can be dealt with by essentially solving two reachability problems of this kind and combining their solution (or 
better, by computing two transient probabilities and then combining the two so obtained probabilities, see \cite{MC:Hermanns:2000:MCofCTMCtransient}).} 
Then, we may resort to Lemmas \ref{lemma:timeDepNextProb} and \ref{lemma:tameTimeDepReachab} to prove convergence 
of $\bar{P}\N(t)$ to $\bar{P}(t)$ and of $P\N(t)$ to $P(t)$.

However, in CSL model checking we are interested in truth values rather than in probabilities, and lifting the previous convergence 
to truth values is not so straightforward. Consider the path formula $\phi_1\until{0}{T} \phi_2$. The problem is that we have to 
compute its probability $P(t)$ (depending on the initial time $t$) for $z_k$ and then solve the algebraic equation $P_s(t) - p = 0$ 
for each state $s$, to identify for which time instants state $s$ satisfies the formula. Now, the point is that, even in case 
$P\N(t) \rightarrow P(t)$ uniformly, we are not guaranteed that $P\N(t)\bowtie p \rightarrow P(t)\bowtie p$. 
For instance, if $P(t) = p$, and $\bowtie$ is $\leq$, then if $P\N(t)$ converges to $P(t)$ from above, it never satisfies 
$P\N(t)\bowtie p$ for any $N$, hence convergence of $P\N(t)\bowtie p$ to $P(t)\bowtie p$ does not hold. However, things can 
go wrong only when $P(t) = p$, and the main point of the convergence theorem is to prove that this happens sufficiently 
``rarely'' not to impact on the computation of probabilities of a next or of an until formula in which $\phi$ is a sub-formula.

In the following, we first outline an algorithm for CSL model checking of ICTMC, and then discuss convergence in more detail. 
Finally, at the end of the section, we will compare in more detail the CSL model checking problem for $Z\N_k$ and 
$(Z\N_k,\nXN)$. 

\subsection{Model Checking CSL for ICTMC}
\label{sec:CSLforICTMC}

The computation of next-state probabilities for time-varying target sets can be done by the method presented in 
Section~\ref{sec:next}, in particular the algorithm in Figure \ref{algo:nextState}. 

The algorithm of Section \ref{sec:timeVaryingReach} for computing reachability in the presence of piecewise constant goal 
and update sets, instead, is the core procedure to compute the probability of an until formula. 
In fact, consider the path formula $\phi_1 \until{T_a}{T_b} \phi_2$. 
To compute its probability for initial time in $[t_0,t_1]$,\footnote{The appropriate value of $t_0$ and $t_1$ are to be deduced 
from $\phi_1$, $\phi_2$ and the superformula of the until, in a standard way \cite{SB:Zuliani:2009:StatMC}} 
we solve two reachability problems separately and then combine the results.

The first reachability problem is for unsafe set $U_1=\lm\neg\phi_1\rm$ and empty goal set $G(t+T_a) = \emptyset$. Let 
$\Upsilon^1(t,t+T_a)$ be the probability matrix of this reachability problem. In order for the computation of the until probability 
to work, we must then discard the probability of being in an unsafe state, essentially multiplying $\Upsilon^1(t,t+T_a)$ by 
$\zeta^1(t+T_a)$ on the right (see Section \ref{sec:timeVaryingReach}).\footnote{In fact, this reachability problem can be solved 
in a simpler way: it just requires trajectories not to enter an unsafe state, and then collects the probability to be in a safe state 
at the time $t+T$.  In particular, we can get rid of the copy $\bar{\calS}$ of the state space, and define a simplified $\Upsilon$ 
function using $\zeta_W$ matrices instead of $\zeta$ ones.} 

The second reachability problem is for unsafe set $U=\lm\neg\phi_1\rm$ and goal set $G = \lm \phi_2 \rm$, and is solved for 
initial time $t\in[t_0 + T_a, t_1+T_a]$, and time horizon $T_b-T_a$. Let $\Upsilon^2(t+T_a,t+T_b)$ be the function computed 
by the algorithm in Section \ref{sec:timeVaryingReach} for this second problem.  Then, for each  state $s$, safe at time $t$, we 
compute $P(t) = \Upsilon^1(t,t+T_a) \zeta^1(t+T_a) \Upsilon^2(t+T_a,t+T_b) \vr{e}_{\bar{\calS}}$, where $\vr{e}_{\bar{\calS}}$ is 
the vector equal to 1 for states $\bar{s}\in\bar{\calS}$ and zero elsewhere.  $P_s(t)$ contains the probability of the until formula 
in state $s$.  Then, we can determine if state $s$ at time $t$ satisfies $\calP_{\bowtie p} (\phi_1 \until{T_a}{T_b} \phi_2)$ by  
solving the inequality $P_s(t) \bowtie p$. 

This  provides an algorithm to approximately solve the CSL model checking for ICTMC recursively on the structure of the 
formula, provided that the number of discontinuities of sets satisfying a formula is finite and that we are able to find all the 
zeros of the computed probability functions, to construct the proper time-dependent satisfiability sets (or approximations
thereof). The full procedure is sketched in Figure \ref{algo:CSLMC}.

\begin{figure}[!t]
\begin{algorithmic}
\Function{CSL\_MC}{$Z$, $\phi$, $t_0$, $t_1$}  
\Comment Computes $V_s(t) = \vr{I}\{s,t\models \phi\}$ for $s\in\calS$ and $t\in[t_0,t_1]$.

\If{ $\phi = p$ }

\State $V_s(t) \leftarrow \vr{I}\{p\in L(s)\}$, $s\in\calS$

\ElsIf{ $\phi = \neg \phi_1$ }

\State $V_1 \leftarrow \mbox{\textsc{CSL\_MC}}(Z,\phi_1,t_0,t_1)$

\State $V(t) \leftarrow 1-V_1(t)$

\ElsIf{ $\phi = \phi_1 \wedge \phi_2$ }

\State $V_1 \leftarrow \mbox{\textsc{CSL\_MC}}(Z,\phi_1,t_0,t_1)$

\State $V_2 \leftarrow \mbox{\textsc{CSL\_MC}}(Z,\phi_2,t_0,t_1)$

\State $V(t) \leftarrow \min\{V_1(t),V_2(t)\}$

\ElsIf{ $\phi = \calP_{\bowtie p}(\next{T_a}{T_b}\phi_1)$ }

\State $V_1 \leftarrow \mbox{\textsc{CSL\_MC}}(Z,\phi_1,t_0,t_1)$

\State $\bP \leftarrow \mbox{\textsc{next-state-probability}}(Z,V_1,T_a,T_b,t_0,t_1)$

\State $V(t) \leftarrow \vr{I}\{\bar{P}(t)\bowtie p \}$

\ElsIf{ $\phi = \calP_{\bowtie p}(\phi_1 \until{T_a}{T_b}\phi_2)$ }

\State $V_1 \leftarrow \mbox{\textsc{CSL\_MC}}(Z,\neg \phi_1,t_0,t_1+T_b)$

\State $V_2 \leftarrow \mbox{\textsc{CSL\_MC}}(Z,\phi_2,t_0,t_1+T_b)$

\State $\Upsilon^1 \leftarrow \mbox{\textsc{reachability}}(Z,T_a,\emptyset,V_1,t_0,t_1+T_a)[1]$
\Comment Returns $\Upsilon$ component of \textsc{reachability}.

\State $\Upsilon^2 \leftarrow \mbox{\textsc{reachability}}(Z,T_b-T_a,V_2,V_1,t_0+T_a,t_1+T_b)[1]$

\State $P(t) = \Upsilon^1(t,t+T_a) \zeta^1(t+T_a) \Upsilon^2(t+T_a,t+T_b) \vr{e}_{\bar{\calS}}$

\State $V(t) \leftarrow \vr{I}\{P(t)\bowtie p \}$

\EndIf

\Return  $V$
\EndFunction
\end{algorithmic}
\caption{Core algorithm for solving the CSL model checking problem, by computing the satisfiability of a CSL-formula $\phi$ 
as a function of the time $t\in[t_0,t_1]$ at which it is evaluated. The truth value of $\phi$ is then the value it has in $t_0$, which
is usually 0. }
\label{algo:CSLMC}
\end{figure}

Below we will consider this algorithm in more detail, focussing particularly on correctness and termination.  In this consideration 
we will make the following assumption about the numerical algorithms that it uses.

\begin{assumption}
\label{ass:intervalcomputability}
There are interval arithmetic routines that can compute bounding sets for the rate functions of $z_k$ and $Z\N_k$, in such a 
way that the approximation error can be made arbitrary small. We call such functions \emph{interval computable}.
\end{assumption}
Notice that this assumption is not very restrictive. It applies to all the standard functions, and also to solutions of ODEs of functions 
which satisfy it, to derivatives of these functions and to their integrals \cite{THMAT:Neumeier:1990:IntervalAnalysis, 
THMAT:Alefeld:2000:IntervalAnalysis}. In particular, if the rate functions are interval computable, then so will be all the 
probabilities computed by solving reachability problems. 

The approach presented above relies, in addition on the solution of ODEs, also on two other key numerical operations: given 
a computable real number $p$, determine if $p$ is zero and and given an analytic function $f$, find all the zeros of such a 
function (or better an interval approximation of these zeros of arbitrary accuracy). 
However, it is not clear if these two operations can be carried out effectively for any input that we can generate, see 
Remark~\ref{rem:undecidableSources} for further comments.   Therefore, we need some further assumptions. Instead of 
restricting the class of functions (which seems a difficult problem since we have to consider the solution of differential 
equations), we will follow the approach of \cite{COMP:Franek:2011:quasiDecRealEquations}, introducing a notion of 
\emph{robust CSL formula} and proving decidability for this subset of formulae. 
This will not solve the decidability problem in theory, but makes it quasi-decidable \cite{COMP:Franek:2011:quasiDecRealEquations}, which may be enough in practice. 
As we will see, the set of CSL formulae which is not robust has measure zero (see Theorem \ref{th:satisfiabilityAE}).

In order to introduce the concept of robust CSL formula, consider a CSL formula $\phi$ and let $p_1,\ldots,p_k$ be the 
constants appearing in the $\calP_{\bowtie p}$ operators of next and until sub-formulae of $\phi$.  We will treat  
$\phi = \phi(p_1,\ldots,p_k)$ as a function of those $p_1,\ldots,p_k$. Furthermore, we will call the next or until sub-formulae 
of $\phi$ \emph{top next sub-formulae} or \emph{top until sub-formulae} if they are not sub-formulae of other next or until 
formulae. The other next or until formulae will be called \emph{dependent}. Finally, given two robust time-varying sets 
$V_1$ and $V_2$, we recall that $V_1$ and $V_2$ are \emph{compatible} if they do not have discontinuities for the same state $s$ happening at the same time instant $t$.

\begin{definition}
\label{def:robustCSL}
A CSL formula $\phi = \phi(\vr{p})$, $\vr{p}\in[0,1]^k$ is robust if and only if 
\begin{enumerate}
\item  there is an open neighbourhood $W$ of $\vr{p}$ in $[0,1]^k$ such that for each $\vr{p_1}\in W$, $$s,0 \models \phi(\vr{p}) \Leftrightarrow s,0 \models \phi(\vr{p_1}).$$
\item The time-varying sets of any dependent next or until sub-formula of $\phi$ holds are \emph{robust}.
\item The time-varying sets of sub-formulae of $\phi$ that are part of an until formula or of a conjunction/disjunction are \emph{compatible} among them.
\end{enumerate} 
\end{definition}

We now prove the following theorem, which states that the CSL model checking algorithm we put forward works at least for 
robust formulae:
\begin{theorem}
\label{th:CSLdecidabilityICTMC}
The CSL model checking for ICTMC, for piecewise analytic interval computable rate functions, is decidable for a robust CSL formula $\phi(p_1,\ldots,p_k)$. 
\end{theorem}

The following corollary is a straightforward consequences of the proof of the previous theorem:
\begin{corollary}
\label{cor:correctness}
The algorithm for CSL model checking presented in this section is correct for robust CSL formulae. \qed
\end{corollary}

We turn now to characterise the set of robust formulae from a topological and measure-theoretic point on view. 
We have the following

\begin{theorem}
\label{th:satisfiabilityAE}
Given a CSL formula $\phi(\vr{p})$, with $\vr{p}\in[0,1]^k$, then the set $\{\vr{p}~|~\phi(\vr{p})\ \mbox{is robust}\}$ is relatively 
open\footnote{A set $U\subset V$ is relatively open in $V\subset W$, where $W$ is a topological space, if it is open in the 
subspace topology, i.e.\ if there exists an open subset $U_1 \subseteq W$ such that $U = V \cap U_1$.} in $[0,1]^k$ and 
has Lebesgue measure 1. 
\end{theorem}

The openness of the set of robust thresholds for a formula allows us to prove the following corollary about quasi-decidability. 
In this paper, we consider a notion of quasi-decidability, which is slightly different than the one defined in 
\cite{COMP:Franek:2011:quasiDecRealEquations}. In fact, we take advantage of the fact that out input values belong to a 
compact subset $K\subseteq \bbR^n$, for which a standard notion of measure exists. 

\begin{definition}
\label{def:quasidec}
A problem with inputs in a compact subset $K\subseteq \bbR^n$ of Lebesgue measure $\mu_\ell(K)>0$, is 
\emph{quasi-decidable} if there is an algorithm that solves it correctly for an open subset $U\subset K$, with 
$\mu_\ell(U)/\mu_\ell(K)$ = 1. 
\end{definition}

Combining Theorems \ref{th:CSLdecidabilityICTMC} and \ref{th:satisfiabilityAE}, we obtain the following:
\begin{corollary}
\label{cor:quasiDec}
The CSL model checking for ICTMC, for piecewise analytic interval computable rate functions, is quasi-decidable for any 
formula $\phi(\vr{p})$. \qed
\end{corollary}

\begin{remark}
\label{rem:robustnessInPractice}
The notions of robustness and quasi-decidability have a practical side.  First, the openness property of the set of robust 
thresholds for a formula $\phi(\vr{p})$ guarantees that if we ``perturb'' a formula (by varying the set $\vr{p}$ of threshold 
constants of the path probability operators), then the formula remains robust. Furthermore, by the definition of robustness, 
also its truth value remains the same (as the notion of quasi-decidability of \cite{COMP:Franek:2011:quasiDecRealEquations} 
requires). This explains the use of the terminology ``robust''. 

Secondly, the characterisation of the set $R$ of robust thresholds for a formula $\phi$ provided in 
Theorem~\ref{th:satisfiabilityAE},  implies that if we choose thresholds at ``random'', we are likely to select a robust formula. 
In fact, consider the grid of rational numbers with $\frac{1}{n}$ in $[0,1]$, i.e.  $GR_n = \{\frac{m}{n}~|~m<n,m,n\in\bbN\}$, 
and take the Cartesian product $GR_n^k \subset [0,1]^k$. Let $\mu_n$ be the uniform distribution in $GR_n^k$, then 
$\mu_n\rightarrow \mu$, the uniform distribution on $[0,1]^k$ (which coincides with the Lebesgue measure on Borel sets). 
Now, as $R$ is open and has Lebesgue measure 1, then $\mu(R) = 1$ and $\mu(\partial R) = 0$, hence $R$ is a continuity 
set for $\mu$. Therefore, $\mu_n(R) \rightarrow \mu(R) = 1$ by the Portmanteau theorem 
\cite{STOC:Billingsley:1979:ProbabilityTheory}.  This means that, fixing $\eps>0$, if we choose the thresholds of the until 
sub-formulas from the set $GR_n^k$, for $n$ large enough, the probability of choosing a bad set of thresholds, for which 
the formula is not robust and the CSL model checking algorithm may not terminate, will be less than $\eps$. 
\end{remark}

\begin{remark}
\label{rem:undecidableSources}
The semi-decidability result presented here is in contrast with the decidability result of model checking for time-homogeneous 
CTMC.  However, in that case the result follows because  $P_s(0)$ has a special form allowing the application of 
Lindeman-Weierstass theorem for transcendental numbers (together with zero testing procedures for algebraic numbers) 
\cite{MC:Aziz:2000:modelCheckingCTMC}. This, in turn, is a consequence of having constant (rational) rates.
In our case, instead, rates are piecewise analytic functions, and we cannot rely on the method of \cite{MC:Aziz:2000:modelCheckingCTMC} anymore.  In fact, in the algorithm for computing the probability, there are two numerical 
operations that are potential sources of undecidability:
\begin{enumerate}
\item Given a number $p$, which is the analytic image of a rational, decide if it is zero. This is  a classical problem whose decidability is not known, even restricting to expressions made up by polynomials and exponentials only 
\cite{COMP:Richardson:2007:zeroTest, COMP:RIchardson:1991:ZeroFinding}.  Indeed, its decidability is connected with the 
truth of the Schanuel conjecture  \cite{COMP:Richardson:2007:zeroTest, COMP:RIchardson:1991:ZeroFinding}, which is in 
turn connected with decidability of the theory of reals extended by the exponential. However, even in case the Schanuel 
conjecture holds, it is not clear if the zero problem will be decidable for any analytic function. 
\item Detecting the zeros of an analytic function with arbitrary precision.
In this case the problem is caused by non-simple zeros, i.e. points in which the function and some of its derivatives are zero. 
The method sketched in a footnote of the proof of theorem \ref{th:CSLdecidabilityICTMC} does not work, as it relies on the 
fact that we can bound the derivative away from zero on null points of the function. 
Furthermore, in the presence of non-simple zeros, detecting if a compact interval  is bounded away from zero is 
semi-decidable (the decision procedure fails if the interval contains a non-simple zero). Whether there is a decidable 
algorithm for this problem is not known to the authors (even assuming the Schanuel conjecture is true). 
It may be possible, however, to find algorithms for some subclass of analytic functions large enough for practical purposes. 
For instance, if we know a lower bound on the radius of convergence of power series in each analytic point, we can effectively 
extend the real analytic function to an open ball in the complex plane, and then use methods developed for complex analytic 
functions \cite{THMAT:Johnson:2009:zerosAnalyticFunctions} which can effectively compute the number of zeros in any 
sufficiently simple open set, by integrating a function on its boundary with interval arithmetic routines 
\cite{THMAT:Ahlfors:1953:ComplexAnalysis, THMAT:Johnson:2009:zerosAnalyticFunctions}. 
\end{enumerate}

Our conjecture is that the model checking problem for time-inhomogeneous CTMC is not decidable in general,
although it may be decidable for some restricted subclass of rate functions if the Schaunel conjecture is true. Further 
investigations on this issue are required.  
\end{remark}

Finding an upper bound on the complexity of the approximation algorithm, when it converges, requires us to find an upper 
bound on the number of zeros of the solution of a linear differential equation with piecewise analytic rates. This is a non trivial 
problem. However, we can rely on a result for linear systems with bounded analytic rate functions 
\cite{THMAT:Novikov:2000:zerosLinearODEsystems}, which gives an upper bound $\Psi$ on the number of zeros, expressible 
as an elementary function of the upper bound on coefficients of the ODE. For piecewise analytic rate functions, simply multiply 
this bound for the number of analytic pieces. The number of analytic pieces is $K_Q + K$, where $K_Q$ comes from the 
piecewise analytic nature of rate functions, and $K$ from the number of structural changes of $\lm\neg\phi_1\rm$ and 
$\lm\phi_2\rm$ sets.  Hence, the number of zeros of $P(t) - p$ can be bounded by $(K+K_Q)\Psi$. By induction, if $h$ is the 
degree of a formula $\phi$ and $h_u$ is the number of nested next or until subformulae,\footnote{Notice that, for next or 
until formulae not containing any other next or until subformula, $K=0$.} then the total  complexity is bounded above by 
$C(n,T_{max},\eps)(2^{h}\Psi^{h_u}K_Q)$, where the constant $C(n,T_{max},\eps)$ hides the cost of integrating ODEs and 
finding roots in each analytic piece. It is proportional to $n^3$ (matrix multiplication), time $T_{max}$ for which the ODEs have 
to be solved, and the precision $\eps$ of root finding and numerical integration\footnote{The precision $\eps$ depends on the specific analytic functions considered. However, we can imagine a procedure which takes an $\eps$ as input and does not 
provide an answer if the precision is not small enough.}.

However, this is a theoretical upper bound, and we will not expect such a complexity in practice. 

%In fact, in most applications, the rate functions (fluid equations) will converge to a stationary value, usually without damped oscillations, hence 

\subsection{Convergence for CSL formulae}
\label{sec:convergenceCSL }

We are now ready to state a convergence result for CSL model checking. Also in this case, we will restrict our attention to 
robust CSL formulae. This is reasonable, as we want to use Lemmas \ref{lemma:timeDepNextProb} and
 \ref{lemma:tameTimeDepReachab}, which require robustness of time-varying sets. 

%
%In the following, given a CSL formula $\phi$, we will denote by $p(\phi)$ the set of all $p_j$ constants appearing in until subformulae of $\phi$, and treat  $\phi = \phi(p_1,\ldots,p_k)$ as a function of those $p_1,\ldots,p_k$. 

\begin{theorem}
\label{th:CSLconvergence}
Let $\calXN$ be a sequence of CTMC models, as defined in Section \ref{sec:modelingLanguage}, and 
let $Z\N_k$ and $z_k$ be defined from $\calXN$ as in Section \ref{sec:fastSimulation}.\\
Assume that $Z\N_k$, $z_k$ have piecewise analytic infinitesimal generator matrices.\\ 
Let $\phi(p_1,\ldots,p_k)$ be a robust CSL formula. 
Then, there exists an $N_0$ such that, for $N\geq N_0$ and each $s\in\calS$
$$s,0 \vDash_{Z\N_k} \phi \Leftrightarrow  s,0 \vDash_{z_k} \phi.$$
\end{theorem}

\begin{corollary}
\label{cor:convergenceAE}
Given a CSL formula $\phi(\vr{p})$, with $\vr{p}\in[0,1]^k$, then the subset of $[0,1]^k$ in which convergence holds has 
Lebesgue measure 1 and is open in $[0,1]^k$. \qed
\end{corollary}

The previous theorem shows that the results that we obtain abstracting a single agent in a population of size $N$ with the fluid 
approximation is consistent. However, the theorem excludes the sets of constants $\vr{p}$ for which the formula is not robust. 
Interestingly, this is the same condition required for decidability of the model checking problem for ICTMC, a fact that shows how 
these two aspects  are intimately connected. Notice that, contrary to decidability, this limitation is unavoidable and is present also 
in the case of sequences of processes converging to a time-homogeneous CTMC\@. In this case, in fact, the next-state and 
reachability probabilities are constant with respect to the initial time, and their value $p$ (in the limit model) can cause 
convergence of truth values to fail. 

However, notice that the constants $p$ appearing in a formula that can make convergence fail depend only on the limit CTMC 
$z_k$, hence we can detect potentially dangerous situations while solving the CSL model checking for the limit process (in
 these cases the model checking algorithm may fail to provide an answer).

%Furthermore, notice that convergence of truth values will work better if in a given point the value of $P(t)$ is far away from the threshold $p$, in the sense that in this case we expect to observe coincidence for a smaller value of $N$. 

%
%\begin{remark}
%\label{rem:ominimality}
%The restriction on the class of functions that guarantee  the convergence to hold is intimately related with finite representability of properties over real numbers. In fact, the theory of real numbers extended with analytic functions restricted on a compact set is o-minimal \cite{THMAT:vanDenDries:1998:ominimal}, meaning that any set definable in such a theory is a finite union of points and intervals. It is precisely the finiteness of zeros of the probability functions we obtain that allows us to prove convergence and to set up an approximate effective procedure to compute the probability of a formula and hence semi-decide CSL formulae. 
%\end{remark}

%\begin{remark}
%\label{rem:next}
%The version of CSL considered in this paper lacks the (time bounded) next operator. However, it can be easily included in our framework. As for the model checking of ICTMC, time bounded next operator $\vr{X}^I \phi_1$ can be dealt with by computing an integral of the rate functions (taking into account discontinuities of $\lm\phi_1\rm$). Finally, to compute its probability as a function of time, we can obtain a set of ODEs by taking the derivative of the integrals.
%The convergence of probabilities and truth values of the next operator  follows from arguments similar to the ones used in this paper.  
%\end{remark}

\begin{remark}
\label{rem:steadyState}

In this paper, we are considering only time bounded operators. This limitation is a consequence of the very nature of the 
approximation theorem \ref{th:fastSimulation}, which holds only for a finite time horizon. However, there are situations in which 
we can extend the validity of the theorem to the whole time domain, but this extension depends on properties of the phase 
space of the fluid ODE \cite{STOC:Benaim:2003:MeanFieldGames, STOC:Benaim:1998:UrnProcesses, 
STOC:BenaimLeBoudec:2011:stationaryConvergence}.

In those cases, we can prove convergence of the steady state behaviour of $Z\N_k$ to that of $z_k$, hence we can incorporate 
also operators dealing with steady state properties. 

In order to deal with time unbounded operators, instead, convergence to steady state is not enough. We also need to ensure that the equation $P(t) - p$ has a finite number of zeros on the whole positive time axis. Piecewise analyticity is not sufficient in this case (think about sine and cosine), and stronger conditions have to be required. However, for periodic functions, we may reason similarly to \cite{MC:Mereacre:2009:LTLmcICTMC}, if we can prove that periodicity of rate functions implies periodicity in the reachability probabilities as a function of initial time.
\end{remark}

\begin{figure}[!h]

\begin{center}
\subfigure[$\textit{true}\, \until{0}{50}\, \textit{timeout}$ -- $rq$] {\label{fig:TVR}
\includegraphics[width=.47\textwidth]{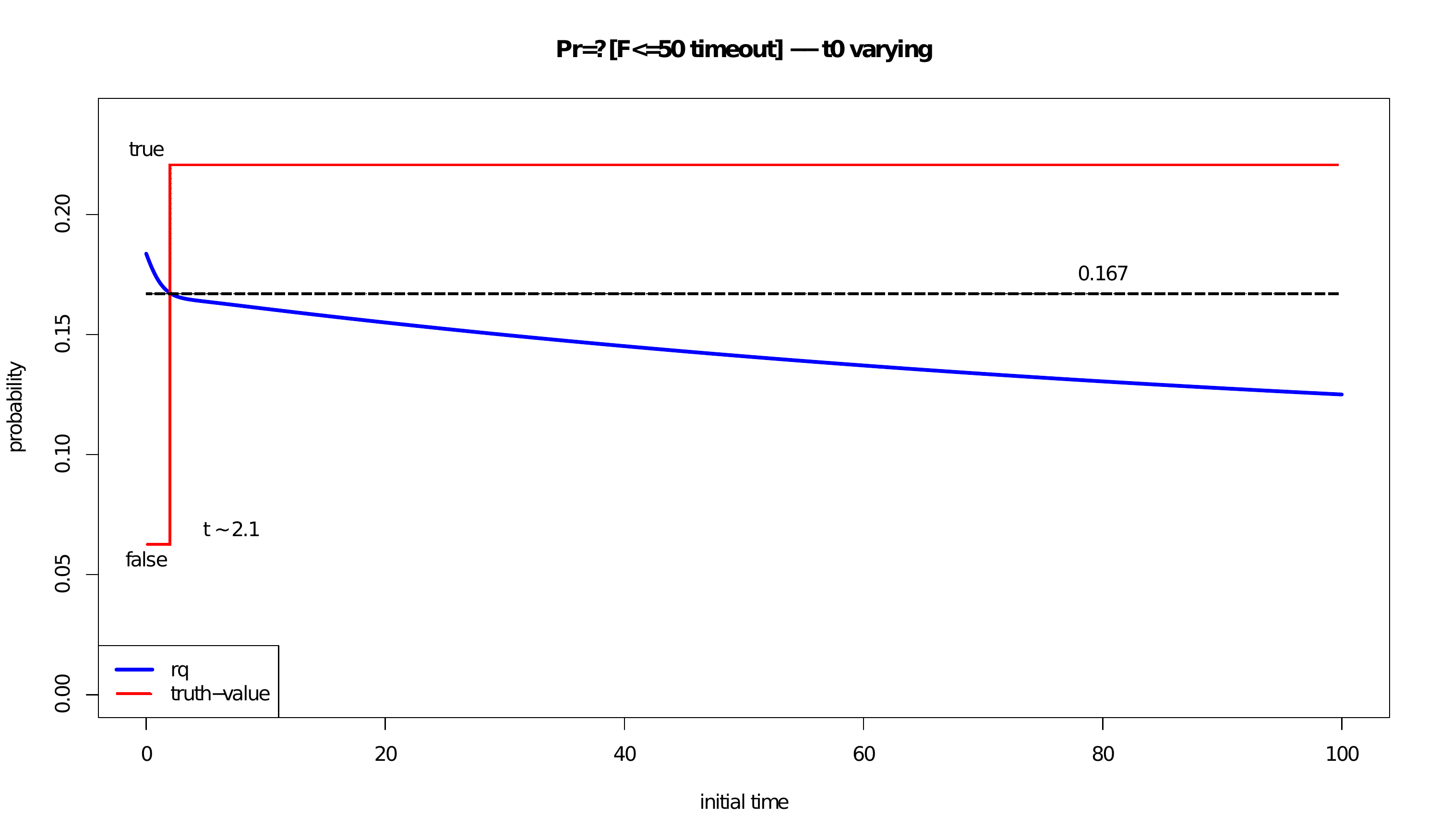} }
\subfigure[$\textit{true}\, \until{0}{50}\, \textit{timeout}$ -- $w$] {\label{fig:TVW}
\includegraphics[width=.47\textwidth]{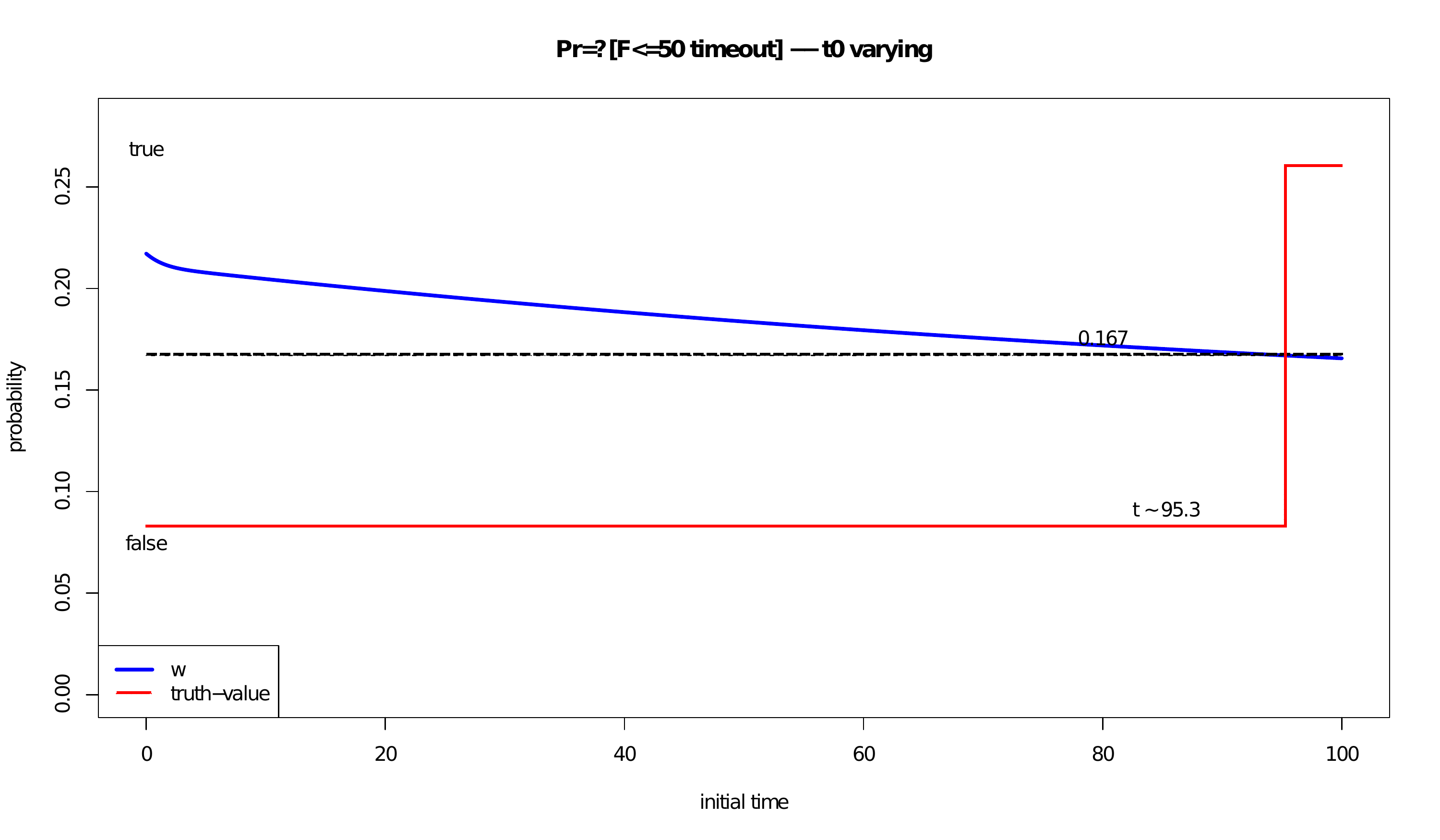} }

\subfigure[$\textit{true} \until{0}{50}\, \textit{timeout}$ -- $t$] {\label{fig:TVT}
\includegraphics[width=.47\textwidth]{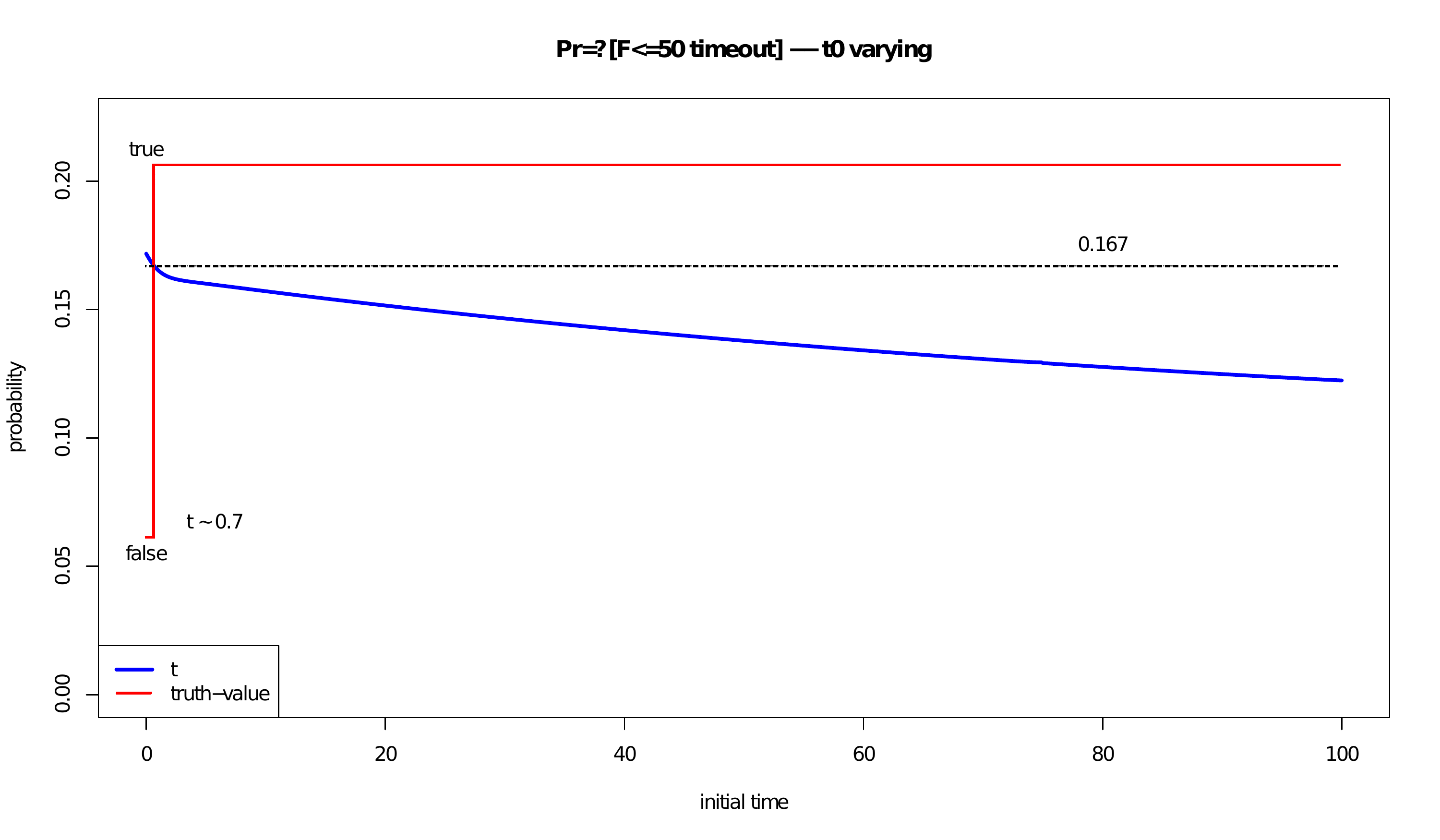} }
\subfigure[$\textit{true}\,\until{0}{T} ( \calP_{<0.167}(\textit{true}\, \until{0}{50}\, \textit{timeout})) $] {\label{fig:TVMC}
\includegraphics[width=.47\textwidth]{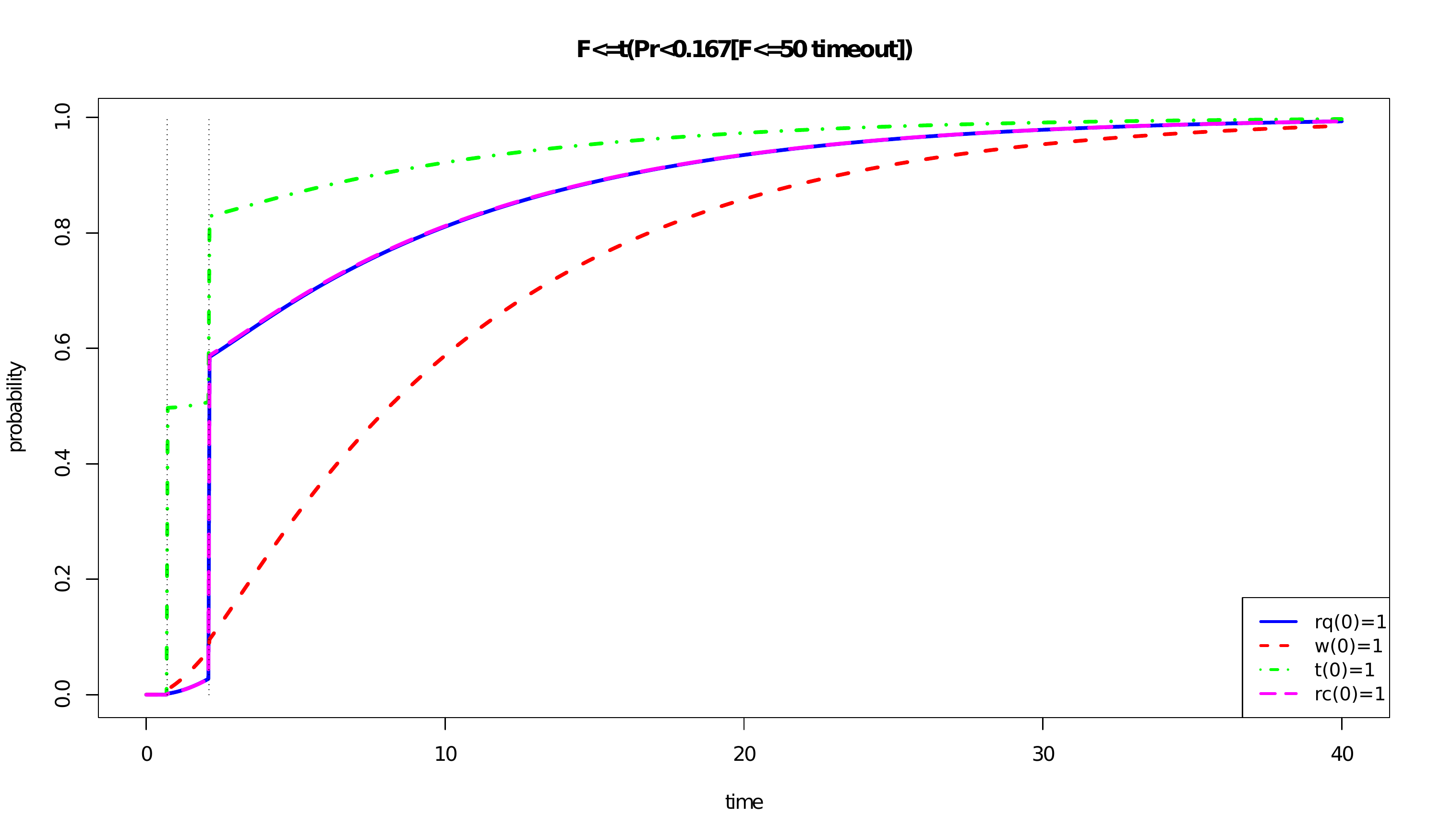} }
\end{center}

\caption{ Figures \ref{fig:TVR}, \ref{fig:TVW}, and \ref{fig:TVT}. Probability of the formula $\textit{true}\, \until{0}{50}\, 
\textit{timeout}$, for varying initial time, and different initial states ($rq$, $w$, and $t$ respectively). The dotted line shows the 
time varying truth function for the CSL formula $\calP_{<0.167}(true \until{0}{50} timeout))$, which is obtained by finding the 
zeros of the initial-time dependent probability. 
Figure \ref{fig:TVMC}. Probability of the until path formula $\textit{true}\, \until{0}{T} ( \calP_{<0.167}(\textit{true}\, \until{0}{50}\, 
\textit{timeout})) $, as a function of time bound $T$. Vertical dotted lines show the discontinuity points time-dependent truth 
of the until sub-formula. 
 }
\label{fig:mc}
\end{figure}

\begin{exu}
Going back to the running example, consider the until path formula $\textit{true}\, U^{[0,50]} \,\textit{timeout}$, where 
$\textit{timeout}$ is true only in state $rc$. Its  probability, as a function of the initial time, is shown in Figures \ref{fig:TVR}, 
\ref{fig:TVW}, and \ref{fig:TVT}, for the states $rq$, $w$, and $t$, respectively. In the same figures, we also show the time-
dependent truth of the CSL formula $\calP_{<0.167}(\textit{true} \,U^{[0,50]}\, \textit{timeout})$, which is obtained by solving 
the inequality $P_s(t) < 0.167$, where $P_s(t)$ is one of the previous time-dependent probability functions. In this case, we 
can observe that for time $t_0\in[0,100]$, there is only one solution, as the probability is monotone. This depends on the 
solution of the fluid equations. In this case, in fact, they converge to a steady state, hence we do expect that also the time 
dependent truth value of CSL until formulae stabilises (when the fluid ODE are close to steady state, the rates of the ICTMC 
are practically constant). This suggests that in many practical cases, the number of changes of truth value of until formulae 
will be very small, as in the running example.
Notice that in the case of the running example, if we had chosen a threshold bigger, say, than 0.25, then the time-dependent 
truth formulae would have been a constant function. 

In Figure \ref{fig:mc}, instead, we show the probability of the path formula $$\textit{true}\, \until{0}{T} 
( \calP_{<0.167}(\textit{true}\, U^{[0,50]}\, \textit{timeout})), $$ as a function of the time horizon $T$. In the plot, it is evident 
how this probability has discontinuities in those time instants in which the truth values function of its until sub-formula change. 
These discontinuities differentiate the model checking of ICTMC  from that for time-homogeneous CTMC.
\end{exu}

\subsection{Comparison of CSL model checking for $Z\N_k$ and $(Z\N_k,\nXN)$}
\label{sec:spaceVsTime}

In this paper we have considered two possible descriptions of a single agent at a fixed population level $N$, i.e. $Z\N_k(t)$ 
and $(Z\N_k(t),\nXN(t))$. From the discussion in Sections \ref{sec:fastSimulation}, \ref{sec:next}, and \ref{sec:reachability}, 
we already know that, while $(Z\N_k(t),\nXN(t))$ is a CTMC with finite (but extremely large) state space, $Z\N_k(t)$ has a 
much smaller state space but it is not a Markov process. Furthermore, its behaviour is time dependent. 
The non-Markovian nature of $Z\N_k(t)$ has consequences for its reachability probability (see Section \ref{sec:timeVaryingReach}), meaning that its value is dependent on the initial time at which we compute it. This implies that 
the satisfiability of a CSL formula (with the truth value of atomic propositions depending only on $\calS$) for $Z\N_k(t)$ can 
depend on the time at which we evaluate it. Hence we need to consider time-dependent sets to compute the probabilities of 
next or until path formulae.  But time-dependent sets can introduce discontinuities in such probabilities, as discussed in 
Section \ref{sec:timeVaryingReach}. 
On the other hand, $(Z\N_k(t),\nXN(t))$ is a time-homogeneous CTMC, hence its next-state and reachability probabilities do 
not depend on time and no time-dependent notion of satisfaction has to be considered in this case. In particular, when 
considering $(Z\N_k(t),\nXN(t))$, its reachability probability is always a continuous function. 
This implies that the truth value of a CSL formula containing nested next or until sub-formulae, can be different if we consider 
its satisfiability with respect to $Z\N_k(t)$ or $(Z\N_k(t),\nXN(t))$.

However, despite this discrepancy for finite $N$, we will prove that the satisfiability for $Z\N_k(t)$ and  $(Z\N_k(t),\nXN(t))$ 
is asymptotically the same, at least if we restrict to robust CSL formulae.
In order to show this, we will combine the convergence results of the previous sections with additional results relative to 
$(Z\N_k(t),\nXN(t))$ and $(z(t),\x(t))$.

\begin{exu}
If we observe Figures \ref{fig:timeoutS1tv} and \ref{fig:timeoutS10tv}, we can easily convince ourselves that the reachability 
probability for $Z\N_k$ in the running example for the formula $\phi_1 = \textit{true}\, \until{0}{50}\, \textit{timeout}$ depends 
on the initial time.  Hence it gives rise to a time-dependent set for the satisfiability of the formula $\phi_2 = \calP_{<0.167}(\phi_1)$. 
This implies that for $Z\N_k$, the probability of the formula $\phi = \textit{true}\,\until{0}{T}\, \phi_2$ will have discontinuities as 
a function of $T$, similarly to the case for $z_k$. 
However, if we compute the reachability probability for $\phi$ in $(Z\N_k,\nXN)$, in a state $s,\x_0$, this will be a continuous function of $T$, hence the two probabilities are different. 
\end{exu}

We will now prove the convergence of the standard CSL model checking for $\Y\N(t) = (Z\N_k(t),$ $\nXN(t))$ in state 
$s,\x_0$, to the equivalent CSL model checking procedure for $\y(t) = (z(t),\x(t))$. This procedure requires us to compute, 
given a next formula $\phi = \next{T_a}{T_b}\phi_1$ or  an until formula $\phi = \phi_1 \until{T_a}{T_b}\phi_2$, its probability 
$P(s,\x)$ starting from time 0, in each point $(s,\x)$  of the state space $\calS\times E$ of $\y(t)$, and then solve the inequality 
$P(s,\x)\bowtie p$, to determine the truth of $\calP_{\bowtie p} (\phi)$ in $(s,\x)$. 
This defines a subset of $\calS\times E$ where $\calP_{\bowtie p}(\phi)$ is true.

The intuition behind the proof is that the truth value of an until formula in a state $(s,\x_0)$ for $\y(t)$ does not depend on the whole state space $\calS\times E$, but only on the points of $E$ intersected by the solution of the fluid ODE starting in $\x_0$, 
i.e.\ on $\calS \times \flow{[0,T]}{\x_0}$, where $\flow{t}{\x_0}$ is the flow of the differential equation\footnote{The solution of the fluid ODE at time $t$ starting in $\x_0$ at time 0}.
Furthermore, the convergence of $\nXN(t)$ to $\x(t)$ allows us to restrict the attention to an arbitrary small neighbourhood of $\flow{[0,T]}{\x_0}$, in order to solve the model checking problem for $\Y\N(t)$, for $N$ large enough. 

In the following, we need some additional concepts and definitions. 

Consider the domain $\hat{\calD}\N\subset E$ of $\nXN$. With each point $\x\in E$, we associate a point $\prjN{\x}\in 
\hat{\calD}\N$, such that $\|\x-\prjN{\x}\|<\frac{n}{N}$. The existence of such a point is guaranteed by the definition of $E$. 
Now, we further assume that, given a point $(s,\x)\in E$, the initial state $\Y\N(0)$ is $(s,\prjN{\x})$, so that $\Y\N(0)$ converges 
to $(s,\x)$ uniformly in space.
This choice of $\Y\N(0)$ guarantees uniform bounds in space for Kurtz theorem and the fast simulation theorem,  for 
convergence in probability.\footnote{The speed of convergence to the fluid limit depends on  the initial conditions only through $\|\nXN(0) - \x(0)\|$; the choice of $\prjN{\x}$ guarantees the uniform convergence of this quantity with respect to $\x$.}.

Now, consider the fluid limit differential equation, and let $\flow{t}{\x_0}$ be its flow. We assume that  $\flow{t}{\x_0}$  is a 
piecewise analytic function with respect to $t$ and $\x$. 
%i.e. the point $\x(t)\in E$, for the solution $\x(t)$ with initial conditions $\x(0) = \x_0$.  
The  \emph{$T,\eps$-flow tube} for $\x_0$ is the set $E_0\subset E$, defined by $E_0 = \flow{[0,T]}{B_\eps(\x_0)}$, i.e.\ the 
set of all trajectories up to time $T$ starting in a ball of radius $\eps$ centred in $\x_0$. Now, consider a $T,\eps$-flow tube 
$E_0$ for $\x_0$. For any $\x\in E_0$, let $T^+_{\x} = T^+_{\x}(E_0)  = \sup\{t~|~\flow{[0,t]}{\x}\in E_0\}$ be the time at which 
the trajectory starting in $\x$ leaves $E_0$. Furthermore, let $T^-_{\x} = T^-_{\x}(E_0) = \inf\{t~|~\flow{[t,0]}{\x}\in E_0\}$  be 
the time at which the trajectory starting in $\x$ enters $E_0$. 

A subset $D\subseteq \calS\times E_0$ is a \emph{d-set} for $E_0$ if and only if, 
(i) $D$ is closed (in $\calS\times E_0$), (ii) $D$ is the union of a finite number of  smooth manifolds\footnote{An smooth manifold is the zero set of a sufficiently smooth function, in this paper at least having continuous first-order derivatives.} of dimension $n-1$ or less, and (iii) for each $\x\in E_0$, it holds  that $\{s\} \times \flow{[T^-_{\x}(E_0),T^+_{\x}(E_0)]}{\x}\cap D$ contains at most $k$ points in each state $s$. In other words, 
a d-set is a union of piecewise analytic manifolds that intersects each trajectory in at most $k$ points. It can be easily checked that each d-set has (Lebesgue) 
measure zero.\footnote{
Any set of topological dimension $n-1$ or less has Lebesgue measure zero in $\bbR^n$.}
%For instance, we can reason as follows: each flow tube is locally diffeomorphic to $A\times [t_1,t_2]$ and, 
%for each such local representation, the set $D$ has measure zero by an application of Fubini's theorem \cite{STOC:Billingsley:1979:ProbabilityTheory}.}

We also introduce a notion of \emph{robust subset} of $\calS\times E_0$, for a $T,\eps$-flow tube $E_0$ in $\x_0$. 
Consider a subset $V\subset \calS\times E_0$. We say that $V$ is robust in $\calS\times E_0$ if and only if, (i) its boundary $\partial V$ is a d-set in $\calS\times E_0$, and (ii) for each  $(s,\x)\in\calS\times  E_0$, the time-varying set $V_{\x}[s](t) = \vr{1}\{(s,\flow{t}{\x})\in V\}$, $T^-_{\x}<t<T^+_{\x}$, is robust in 
the sense of Definition \ref{def:robustSet} (notice that it contains at most $k<\infty$ discontinuity points, where $k$ does not depend on $\x$, as  $\partial V$ is a d-set).
%Consider a subset $V_f\subset E$, defined by a piecewise analytic function $f:\calS\times E\rightarrow [0,1]$, solving the inequality $f(\x)\bowtie p$ in $E_0$. We say that $V_f$ is robust in $E_0$ if and only if, for each $x\in E_0$ and $0<T_1<T^+_{\x}$, the time-varying set $V_{f,\vr{x},T_1}(t) = \vr{1}\{\flow{t}{\x}\in V_f\}$, $t\in[0,T_1]$, is robust in the sense of Definition \ref{def:robustSet}, and contains at most $k<\infty$ discontinuity points.\\ 
%Furthermore, let $Disc(V) = \bigcup_{(s,\x)\in \calS\times E_0} \{(s,\flow{\bar{t}}{\x}) \mid V_\x[s](\bar{t})\ \text{is discontinuous}\}$ 
%be the discontinuity set of $V$, i.e.\ the union of the points  in which $V_\x$ is discontinuous along each trajectory in 
%$\calS\times E_0$. We require that $Disc(V)$, for a robust set $V$, is closed. It follows that $Disc(V)$  is a d-set in $E_0$. 
We sometimes denote $\partial V$ by $Disc(V)$. We also say that two robust subsets $V_1$ and $V_2$ of  $\calS\times E_0$ are \emph{compatible} if $\partial V_1\cap \partial V_2 = \emptyset$.

Similarly to Section \ref{sec:timeVaryingReach}, we say that a sequence of sets $V\N\subset \calS\times E_0$ \emph{converges 
robustly} to a robust set $V\subseteq  \calS\times E_0$, with $E_0$ a $T,\eps$-flow tube in $\x_0$, if and only if, for each open 
neighbourhood $U$ of $Disc(V)$, there is  $N_0>0$ such that, $\forall N\geq N_0$ and all $(s,\x)\in (\calS\times E_0)\setminus U$, $(s,\x)\in V\N$ 
if and only if $(s,\x)\in V$.

We are now ready to state the following lemmas, which are space-versions of Lemmas \ref{lemma:timeDepNextProb} and 
\ref{lemma:tameTimeDepReachab} on time-varying sets, and are the key to the induction step of Lemma \ref{lemma:convergenceSpaceMCictmc}. 

\begin{lemma}
\label{lemma:nextProbRobustSets}
Let $E_0\subset E$ be a $T,\eps_0$-flow tube for $\x_0$. Let $G$ be a robust subset of $\calS\times E_0$, and  $G\N$ be a 
sequence of subsets of $\calS\times E_0$ that converge robustly to $G$.

Let $\bP(s,\x)=P_{next}(\y,s,\x,T_a,T_b,G)$ be the probability that the first jump of $\y(t)$ is into a state in $G$ and happens 
at a time $t\in[T_a,T_b]$, given that $\y$ started at time $t=0$ in state $(s,\x)\in \calS\times E_0$, and let $\bP\N(s,\x)=
P\N_{next}(\Y\N,s,\prjN{\x},T_a,T_b,G\N)$ be defined similarly, with $G$ and $\x$ replaced by $G\N$ and $\prjN{\x}$, 
respectively. 
Furthermore, define $\phantom{aaa}$ $V = \{(s,\x)~|~\bP(s,\x)\bowtie p\}$ and   $V\N = \{(s,\x)~|~\bP\N(s,\x)\bowtie p\}$. Then there exists 
$\eps_1>0$ such that, in $E_1$, the $(T-T_b),\eps_1$-flow tube for $\x_0$:
\begin{enumerate}
\item $\bP\N(s,\x) \rightarrow \bP(s,\x)$ for all $\x\in E_1$, uniformly in $(s,\x)$.  
\item If $V_{\x_0}(t)$, $t\in[T^-_{\x_0}(E_1),T^+_{\x_0}(E_1)]$, is a robust time-varying set, then $V$ is robust in $E_1$ and 
$V\N$ converges robustly to $V$. 
\end{enumerate}
\end{lemma}

\begin{lemma}
\label{lemma:reachabilityRobustSets}
Let $E_0\subset E$ be a $T,\eps_0$-flow tube for $\x_0$. Let $U$ and $G$ two robust and compatible subsets of $\calS\times E_0$, and 
$U\N$, $G\N$ be sequences of subsets of $\calS\times E_0$ that converge robustly to $U$ and $G$, respectively.

Let $P(s,\x)=P_{reach}(\y,s,\x,T_1,T_2,U,G)$ be the probability that $\y(t)$ reaches a state in $G$ within time $[T_a,T_b]$, 
avoiding any unsafe state in $U$, given that $\y$ started at time $t=0$ in state $(s,\x)\in \calS\times E_0$, and let $P\N(s,\x)=
P\N_{reach}(\Y\N,s,\prjN{\x},T_a,T_b,U\N,G\N)$ be defined similarly, with $G$, $U$, $\x$ replaced by $G\N$, $U\N$, and 
$\prjN{\x}$, respectively. 
Furthermore, define $V = \{(s,\x)~|~P(s,\x)\bowtie p\}$ and   $V\N = \{(s,\x)~|~P\N(s,\x)\bowtie p\}$. Then there exists $\eps_1>0$ 
such that, in $E_1$, the $(T-T_b),\eps_1$-flow tube for $\x_0$:
\begin{enumerate}
\item $P\N(s,\x) \rightarrow P(s,\x)$ for all $\x\in E_1 \setminus D$, where $D$ is a d-set, uniformly in $(s,\x)$.  
\item If $V_{\x_0}(t)$, $t\in[T^-_{\x_0}(E_1),T^+_{\x_0}(E_1)]$, is a robust time-varying set, then $V$ is robust in $E_1$ and 
$V\N$ converges robustly to $V$. 
\end{enumerate}
\end{lemma}

The previous lemmas are the key arguments used in the structural induction to prove the following result.

\begin{lemma}
\label{lemma:convergenceSpaceMC}
Let $\calXN$ be a sequence of CTMC models, as defined in Section \ref{sec:modelingLanguage}, and 
let $Z\N_k$ and $z_k$ be defined from $\calXN$ as in Section \ref{sec:fastSimulation}.\\
Assume that there is a flow tube $E_0$ of $\x_0$ such that all trajectories in $E_0$ are piecewise analytic.\\
Let $\phi = \phi(\vr{p})$ be a robust CSL formula for the trajectory $\flow{t}{\x_0}$. Then, there is an $N_0$ such that, for all $N\geq N_0$, $$s,\x_0 \models_{\y} \phi\;\;  \Leftrightarrow\;\; s, \prjN{\x_0} \models_{\Y\N}  \phi.$$ 
\end{lemma}

We now turn to consider the relationship between the model checking problem of a CSL formula $\phi$ for $z_k(t)$ and the 
model checking problem for the same formula with respect to $\y(t)$.   In this case, it is easy to see that a formula is true for 
$z_k(t)$ if and only if it is true for $\y(t)$.  In fact, in this process the truth value of a formula in state $(s,\x_0)$ depends only 
on the trajectory $\flow{t}{\x_0}$ starting in $\x_0$. 
Furthermore, if we fix a time $\bar{t}$ and consider the point $\x_{\bar{t}} = \flow{\bar{t}}{\x_0}$, then the process $\bar{z}_k(t)$, 
defined with respect to the trajectory $\flow{t}{\x_{\bar{t}}}$ starting in point $\x_{\bar{t}}$ at time zero, equals the process 
$z_k(t+\bar{t})$, starting in $\x_0$ at time zero, due to the semi-group property of the flow $\flow{\cdot}{\cdot}$. 
Hence, any reachability probability for $\z_k$ with respect to the initial time $\bar{t}$ equals the reachability probability for 
$\bar{z}_k$ at time 0: We can always turn a time-dependent reachability problem into a more classical space-dependent one.  
From the previous discussion, the following lemma follows:

\begin{lemma}
\label{lemma:convergenceSpaceMCictmc}
Let $\calXN$ be a sequence of CTMC models, as defined in Section \ref{sec:modelingLanguage}, and 
let $Z\N_k$ and $z_k$ be defined from $\calXN$ as in Section \ref{sec:fastSimulation}.\\
Let $\phi = \phi(\vr{p})$ be a robust CSL formula for the piecewise analytic trajectory $\flow{t}{\x_0}$, and let $z_k$ be the 
ICTMC defined on $\calS$ with respect to trajectory $\flow{t}{\x_0}$. Then,  $$s,\x_0 \models_{\y} \phi \;\;\Leftrightarrow\;\; s
 \models_{z_k}  \phi.$$ \qed
\end{lemma}

Using the previous lemmas, we can easily show the following theorem.

\begin{theorem}
\label{th:convergenceCSLclassicTosingleAgentCSL}
Let $\calXN$ be a sequence of CTMC models, as defined in Section \ref{sec:modelingLanguage}.
%, and let $Z\N_k$ and $z_k$ be defined from $\calXN$ as in Section \ref{sec:fastSimulation}.\\
Assume that there is a flow tube $E_0$ of $\x_0$ such that all trajectories in $E_0$ are piecewise analytic.\\
Let $\phi = \phi(\vr{p})$ be a robust CSL formula for the trajectory $\flow{t}{\x_0}$, let $Z\N_k(t)$ and $z_k(t)$ be the stochastic 
processes on $\calS$ defined as in Section \ref{sec:fastSimulation}, and let $\y(t)$ and $\Y\N(t)$ be defined as in this section.  
Then, there is an $N_0$ such that, for all $N\geq N_0$,
$$s \models_{Z\N_k} \phi\;\; \Leftrightarrow\;\; s,\prjN{\x_0} \models_{\Y\N} \phi.$$
\end{theorem}

\proof There exists an $N_0$, such that, for all $N\geq N_0$,
$$s \models_{Z\N_k} \phi\;\; \Leftrightarrow\;\; s \models_{z_k}\;\; \Leftrightarrow\;\; s,\x_0 \models_{\y} \phi\;\; \Leftrightarrow\;\; 
s,\prjN{\x_0} \models_{\Y\N} \phi,$$
where the first equivalence follows from Theorem \ref{th:CSLconvergence}, the second equivalence from Lemma 
\ref{lemma:convergenceSpaceMCictmc}, and the third equivalence from Lemma~\ref{lemma:convergenceSpaceMC}, 
while $N_0$ can be chosen as the largest one between that of Theorem~\ref{th:CSLconvergence} and that of Lemma 
\ref{lemma:convergenceSpaceMCictmc}. \qed

Inspecting the proof of the previous theorem, the following corollary is straightforward.

\begin{corollary}
\label{cor:convergenceCSLclassicToFluidICTMC}
Let $\phi = \phi(\vr{p})$ be a robust CSL formula for the trajectory $\flow{t}{\x_0}$. Then, there is an $N_0$ such that, for all 
$N\geq N_0$,
$$s,\prjN{\x_0} \models_{\Y\N} \phi \;\;\Leftrightarrow\;\; s \models_{z_k} \phi .$$ \qed
\end{corollary}

%*****************************************************************************

\section{Conclusions}
\label{sec:Conc}

In this paper we exploited a corollary of fluid limit theorems to approximate properties of the behaviour of single agents in l
arge population models. In particular, we focussed on reachability and stochastic model checking of CSL formulae.
The method proposed requires us to model check a time-inhomogeneous CTMC of size equal to the number of internal states 
of the agent (which is usually rather small). Hence, it gives a large improvement in terms of computational efficiency.

We then focussed on the reachability problem for ICTMC, both in the case of time-constant and time-varying sets. We provided 
algorithms to tackle both cases, and we also proved convergence of the reachability probabilities computed for the single agent 
in a finite population of size $N$ to those of the limit fluid CTMC\@. 
Finally, we focussed on model checking CSL formulae for ICTMC proposing an algorithm that works for a subset of CSL including 
the time bounded next operator and the time bounded until operator. We also showed decidability and convergence results for 
robust formulae, proving that the set of non-robust formulae has measure zero. 

%for almost any set of constants $\vr{p}$ used in the until formulae, a CSL formula will have the same truth value on the limit CTMC and on the CTMC for sufficiently large populations.

There are many issues that we wish to tackle in the future. First, we would like to better understand the quality of convergence. 
This can be accomplished by trying to derive theoretical error bounds (which may be too loose to be of practical interest) and 
by  running many experiments to identify situations in which the approximation performs well (in terms of both classes of formulae 
and model structure).
In addition, we would like to provide a working implementation of the model checking algorithm for ICTMC, studying its 
computational cost empirically (and exploring how easy it is in practice to find a non computable instance). 
Furthermore, we want to investigate the connections between single agent properties and system level properties. We believe 
this approach can become a powerful tool to investigate the relationship between microscopic and  macroscopic characterisations 
of systems, and to understand their emergent behaviour.

As far as CSL model checking for ICTMC is concerned, we aim to extend it to include time unbounded and steady state operators, 
at least for those subsets of rate functions in which the algorithm can be shown to be decidable. We also need to consider 
rewards, at least for a finite time horizon (here we expect their inclusion to be relatively straightforward).
Then, we would like to show convergence results also for this larger subset of CSL, under the hypothesis required for steady 
state convergence of the fluid approximation.

Another line of investigation would be to consider different temporal logics, such as MTL\@. For this logic, asymptotic correctness 
is relatively easy to prove, along the lines of Proposition \ref{prop:reachability}. What is more difficult is to find an effective 
algorithm to model check MTL properties for ICTMC\@. One possibility may be to 
combine the approaches of \cite{MC:Mereacre:2011:MTLmc, MC:Mereacre:2009:LTLmcICTMC, MC:Mereacre:2011:DTAmc}, and 
exploit algorithms and techniques to compute reachability of PDMP \cite{STOC:Davis:1993:PDMP}.

%Finally, we also would like to investigate in more detail the idea of abstracting CTMC using a time dependent kernel. The method shown here, based on fluid approximation, works in general (under the hypothesis of the fluid approximation theorems), but 
%may require large populations to produce good results. 
%However, it may be the case that for certain classes of formulae and a fixed population level, we can find better time-dependent approximation kernels searching for optimal kernels with respect to information theoretic criteria, as in \cite{SB:Sanguinetti:2007:variationalInference}.

%*****************************************************************************
%
\bibliographystyle{elsarticle-num}
\bibliography{../../../../Bibliografia/biblio,extra}

\clearpage

\appendix

\section{Proofs}
\label{app:proofs}

In this appendix, we present the proofs of propositions, lemmas, and theorems of the paper. 

\subsection{Next-State Probability}

\begin{appproposition}[\ref{prop:measureSimpleZeroSetsPiecewiseAnalytic}]
Let $f:I\rightarrow \bbR$ be a piecewise analytic function, with $I\subseteq \bbR$ a compact interval.  
Let $E_f = \{x\in \bbR~|~\mu_\ell(f^{-1}(\{x\})=0\}$ be the set of all values $x$ such that $f$ is not locally constantly equal to $x$, 
where $\mu_{\ell}$ is the Lebesgue measure.
Furthermore, let $Z_x = f^{-1}(\{x\})$ be the set of solutions of $f(t) = x$ 
and let $DZ_f =\{x\in \bbR~|~\forall t\in Z_x, f'(t)\neq 0\}$. Then
\begin{enumerate}
\item $\forall x\in E_f$, $Z_x$ is \emph{finite}.
\item $\mu_\ell(E_f\cap DZ_f) = 1$
\end{enumerate}
\end{appproposition}

\proof Point 1 follows from basic properties of the piecewise analytic function $(f - x)$: in any analytic piece, either the function is constantly equal to zero, or it has only a finite number of zeros. Point 2, instead, follows from the fact that the derivative $f'(t)$ of $t$ is piecewise analytic, hence has only a finite number of zeros (in the analytic pieces in which $f$ is not constant).
\qed

\begin{appproposition}[\ref{prop:convergenceToRobustSet}]
Let $V\N(t)$ be a sequence of time varying sets converging robustly to a robust set $V(t)$, $t\in I$. Let $D\N_V = \{t~|~V\N(t)\neq V(t)\}$. Then $\mu_\ell(D\N_V)\rightarrow 0$, where $\mu_{\ell}$ is the Lebesgue measure on $\bbR$.
\end{appproposition}

\proof A straightforward consequence of the definition of robust convergence is that, for each open neighbourhood $U$ of $Disc(V)$, there exists an $N_0$ such that, for all $N\geq N_0$, $V\N(t) = V(t)$ for $t\in I\setminus U$. Now, as $V$ is robust, then $|Disc(V)|=m<\infty$. 
Fix $\eps>0$ and define $U_\eps = \bigcup_{\bar{t}\in Disc(V)}B(\bar{t},\eps)$, where $B(\bar{t},\eps)$ is the open ball centred in $\bar{t}$ of radius $\eps$. Then $\mu_\ell(U_\eps)\leq 2m\eps$. Now, fix $\eps_k\rightarrow 0$. 
For each $k$, there is an $N_k$ such that, for all $N\geq N_k$, $V\N(t) = V(t)$ for $t\in I\setminus U_{\eps_k}$, and therefore $D\N_V\subseteq U_{\eps_k}$.  \qed

\begin{applemma}[\ref{lemma:timeDepNextProb}]
Let $\calXN$ be a sequence of CTMC models, as defined in Section \ref{sec:modelingLanguage}, and 
let $Z\N_k$ and $z_k$ be defined from $\calXN$ as in Section \ref{sec:fastSimulation}, with piecewise real analytic rates, 
in a compact interval $[0,T']$, for $T'>t_1+T_b$. \\
Let $G(t)$, $t\in[t_0,t_1+T_b]$ be a \emph{robust} time-varying set, and let $G\N(t)$ be a sequence of time-varying sets 
converging robustly to $G$. \\ 
Furthermore, let $\bP(t) = P_{next}(z_k,t,T_a,T_b,G)$ and $\bP\N(t) = \linebreak P_{next}(Z_k\N,t,T_a,T_b,G)$, $t\in[t_0,t_1]$.\\
Finally, fix $p\in[0,1]$, $\bowtie\in\{\leq,<,>,\geq\}$, and let $V_p(t) = I\{\bP(t)\bowtie p\}$, $V\N_p(t) = I\{\bP\N(t)\bowtie p\}$. Then
\begin{enumerate}
\item $\bP\N(t)\rightarrow \bP(t)$, uniformly in  $t\in [t_0,t_1]$. 
\item For almost every $p\in[0,1]$, $V_p$ is robust and the sequence $V\N_p$ converges robustly to $V_p$.
\end{enumerate}
\end{applemma}

\proof
By a standard coupling argument, assume that $z_k$ and $Z\N_k$ are defined on the same probability space $\Omega$. 
Then, letting $Y$ be either $z_k$ or $Z\N_k$, for $\omega\in\Omega$, let $\chi(t,Y(\omega))$ be equal to one if trajectory 
$Y(\omega)$'s first jump, starting at time $t$, is into a state of $G$ at time $t'\in[t+T_a,t+T_b]$, and zero otherwise. Similarly, 
let $\chi\N(t,Y(\omega))$ be 1 if $Y(\omega)$'s first jump, starting at time $t$, is into $G\N$ at time $t'\in[t+T_a,t+T_b]$, and 
zero otherwise. 
Then $\bP(t) = \bbE[\chi(t,z_k)]$, and $\bP\N(t) = \bbE[\chi\N(t,Z\N_k)]$.  It follows that
\begin{eqnarray*}
|\bbE[\chi(t,z_k)] - \bbE[\chi\N(t,Z\N_k)]| & \leq & \underbrace{\bbE[|\chi(t,z_k) - \chi\N(t,z_k) |]}_{(1)}\\  
&+& \underbrace{\bbE[| 
\chi\N(t,z_k) -\chi\N(t,Z\N_k) |]}_{(2)}\\
\end{eqnarray*}

Consider term (2) above. We can partition trajectories into two measurable subsets: $\Omega_1 = \{\omega\in\Omega \mid 
z_k(t,\omega) = Z\N_k(t,\omega),t\leq t_1 + T_b\}$ and $\Omega_0 = \Omega \setminus \Omega_1$.
Let $\mu_{\Omega}$ be the probability measure in $\Omega$.  Applying
Theorem \ref{th:fastSimulation} up to time $t_1 +T_b$, we have that $\chi\N(t,Z\N_k(\omega)) = \chi\N(t,z_k(\omega))$ for  
$\omega\in\Omega_1$ and $\bbP(\Omega_0)\leq \eps_N$. Hence, for any $t\in [t_0,t_1]$, 
{\small
\begin{eqnarray*}
\bbE[|\chi\N(t,Z\N_k) - \chi\N(t,z_k)|] 
& = & \int_{\Omega_1} |\chi\N(t,Z\N_k) - \chi\N(t,z_k)| d\mu_{\Omega}\\
& +  &\int_{\Omega_0} |\chi\N(t,Z\N_k) - \chi\N(t,z_k)| d\mu_{\Omega}\\ 
& \leq & \eps_N\rightarrow 0.
\end{eqnarray*} 
}
Notice that $\eps_N$ does not depend on $t$. 
 
Let us focus now on term (1) in the inequality above. 
Let  $T_1<T_2<\ldots<T_h$ be all the points in $Disc(G)$
(which are finite in number as $G$ is robust). Fix $t\in[t_0,t_1]$.
As $G\N$ converges robustly to $G$, for $N\geq N_0$ they differ only in disjoint balls $B(T_i,\eps)$, for $\eps$ small enough. 
Furthermore, if $G$  has a discontinuity for state $s$ in $T_i$, then the value of $G$  on the left of  $B(T_i,\eps)$ is different 
from the value of $G$ on the right of $B(T_i,\eps)$. 

It follows that the only trajectories of $z_k$ for which $\chi(t,z_k)\neq\chi\N(t,z_k)$ are those jumping within the set 
$D\N_G$ (intersected with $[t,t+T_b]$).\footnote{Notice that robustness of $G$ is not necessary for this proof, but we enforce 
it for uniformity with the convergence of  reachability probabilities in Section \ref{sec:reachability}.}
As the rate functions of $z_k$ are piecewise analytic, they are bounded by a constant $\Lambda$, thus the probability of a 
trajectory jumping in $D\N_G$ is bounded by $\int_{D\N_G} \Lambda e^{-\Lambda t} dt \leq \int_{D_G\N} \Lambda dt = \Lambda \mu_\ell(D\N_G)\rightarrow 0$ 
(independently of $t$).
It follows that
$$|\bbE[\chi(t,z_k)] - \bbE[\chi\N(t,Z\N_k)]| \leq \delta_N,$$
with $\delta_N = \eps_N + \Lambda \mu_{\ell}(D\N_G) \rightarrow 0$ independently of $t$, which proves uniform convergence of 
$P\N(t)$ to $P(t)$ . 

Let us turn now to point 2 of the lemma.\\
Consider the set $H_{\bP}$ of values $p\in [0,1]$ for which either (i) $\bP(t)$ is constantly equal to $p$ in one analytic piece of $\bP$, or (ii) $\bP(t)=p$ and $\bP'(t) = 0$ for some $t$, or (iii) $\bP(t) = p$ and $\bP$ is not analytic in $t$. By Prop. \ref{prop:measureSimpleZeroSetsPiecewiseAnalytic} and the definition of piecewise analytic functions, the set $H_{\bP}$ is finite.
Fix a $p\not\in H_{\bP}$. For such a $p$, the function $\bP(t) - p$ defines a robust time-varying set, as it has a finite number of simple zeros, all in analytic points of $\bP$. 
\\
%
%Fix a $p\in E_{P}\cap DZ_{P}$, which is a set of Lebesgue measure 1 due to Prop. \ref{prop:measureSimpleZeroSetsPiecewiseAnalytic}.
%Then, as $\bP$ is piecewise analytic (it is piecewise the solution of ODEs with an analytic vector field),  the number of zeros 
%of the functions $\bP(t) - p$ is finite and those zeros are simple (i.e.\ the derivative of $\bP$ in those points is non null). 
%Hence, the set $A$ of points in which $V_p$ has a discontinuity is finite. In addition, if $V_p$ has a discontinuity for state $s$ 
%at time $T_i$, then it has to be either right or left-continuous. In fact, if in this point $\bP_s(T_i) - p = 0$, then $\bP_s$ crosses 
%zero, as its derivative is non-null ($p\in DZ_P$). It follows that $V_p$ is a robust set.\\
Call  $A$ the set of points in which $V_p$ has a discontinuity, which is finite.
Fix $\eps$ and define $A_\eps$ to be $\bigcup_{t\in A} B(t,\eps)$, where $B_\eps(t) = (t-\eps,t+\eps)$. Now, if $W$ is a 
neighbourhood of $A$, then for a small $\eps>0$, $A_\eps \subset W$. 
Let $f_p(t) = |\bP(t) - p|$  and consider the set $I_\eps = I\setminus A_\eps$. Now, $I_\eps$ is compact and $f_p(t)$ is different 
from zero in $I_\eps$, so that $\min\{f_p(t)~|~t\in I_\eps\} = m_\eps >0$ (by the Weierstrass theorem \cite{THMAT:Rudin:1976:analysis}). 
As $\bP\N$ converges uniformly to $\bP$, there is $N_0$ such that, for all $N\geq N_0$ and all $t\in I_\eps$, $|\bP\N(t) - 
\bP(t)|\leq \frac{m_\eps}{2}$, hence for all $N\geq N_0$ and all $t\in I_\eps$, $V_p(t) = V\N_p(t)$. 
It follows that $V\N_p(t)$ converges robustly to $V_p(t)$. \qed

\subsection{Reachability}

\begin{appproposition}[\ref{prop:reachability}]
Let $\calXN$ be a sequence of CTMC models, as defined in Section \ref{sec:modelingLanguage}, and 
let $Z\N_k$ and $z_k$ be defined from $\calXN$ as in Section \ref{sec:fastSimulation}. Assume that the infinitesimal 
generator matrix $Q(t)$ of $z_k$ is bounded and integrable in every compact interval $[0,T]$. Then 
\[ P_{reach}(Z\N_k,t,T,G,U) \rightarrow P_{reach}(z_k,t,T,G,U), \text{ uniformly in } [t_0,t_1], \text{ as } N\rightarrow\infty \] 
i.e.\ $\sup_{t\in [t_0,t_1]}\|P_{reach}(Z\N_k,t,T,G,U) - P_{reach}(z_k,t,T,G,U) \|\rightarrow 0$. 
%Let $\calXN$ be a sequence of CTMC models, as defined in Section \ref{sec:modelingLanguage}, and 
%let $Z\N_k$ and $z_k$ be defined from $\calXN$ as in Section \ref{sec:fastSimulation}. Assume that the infinitesimal generator matrix $Q(t)$ of $z_k$ is bounded and integrable in every compact interval $[0,T]$. Then 
%\begin{itemize}
%\item $P_{reach}(Z\N_k,t,T,G,U) \rightarrow P_{reach}(z_k,t,T,G,U)$, uniformly in $[t_0,t_1]$, as $N\rightarrow\infty$ (i.e. $\sup_{t\in [t_0,t_1]}\|P_{reach}(Z\N_k,t,T,G,U) - P_{reach}(z_k,t,T,G,U) \|\rightarrow 0$). 
%\end{itemize}
\end{appproposition}

\proof The proposition follows from a similar argument to that used in the first part of the proof of Lemma \ref{lemma:timeDepNextProb}. 
By a standard coupling argument, we can assume that the processes $Z\N_k$ and $z_k$ are defined on the same probability 
space $\Omega$. 
Therefore, there exists  a sequence $\eps_N\in\bbR_+$, $\eps_N \rightarrow 0$, such that \linebreak$\bbP\{\omega\in\Omega 
\mid \forall t\leq T', Z\N_k(\omega,t) = z_k(\omega,t)\}\geq 1-\eps_N$. This means that with probability $1-\eps_N$, the 
trajectories of the two processes are the same up to time $T'$. 

Now, we can define a (measurable) function $\chi = \chi_{t,T,G,U}$ on the trajectories of the CTMCs which is equal to 1 if 
they satisfy the reachability property, and 0 otherwise. Therefore, it holds that\linebreak
$P_{reach}(Z\N_k,t,T,G,U) = \bbE[\chi(Z\N_k)]$, and similarly for $z_k$. With a similar notation as in Lemma \ref{lemma:timeDepNextProb}, let $\Omega_1 = \{\omega \mid Z\N_k(t,\omega) = z_k(t,\omega),\forall t \leq t_0 + T\}$, 
$\Omega_0 = \{\omega \mid Z\N_k(t,\omega) \neq z_k(t,\omega)\}$, and $\mu_{\Omega}$ be the probability measure in 
$\Omega$ (i.e.\ in the trajectory space). Observe that $\chi(Z\N_k) = \chi(z_k)$ on $\Omega_1$ and $\bbP(\Omega_0)\leq 
\eps_N$, hence
\begin{eqnarray*}
|\bbE[\chi(Z\N_k)] - \bbE[\chi(z_k)]| & \leq & \bbE[|\chi(Z\N_k) - \chi(z_k)|]\\ 
& = & \int_{\Omega_1} |\chi(Z\N_k) - \chi(z_k)| d\mu_{\Omega}\\
& + & \int_{\Omega_0} |\chi(Z\N_k) - \chi(z_k)| d\mu_{\Omega}\\ 
& \leq & \eps_N\rightarrow 0.
\end{eqnarray*}
Uniform convergence follows from the fact that the sequence $\eps_N$ does not depend on the initial or the final time of the reachability property, if they are both less than $T+t_1$. \qed

\begin{applemma}[\ref{lemma:tameTimeDepReachab}]
Let $\calXN$ be a sequence of CTMC models, as defined in Section \ref{sec:modelingLanguage}, and 
let $Z\N_k$ and $z_k$ be defined from $\calXN$ as in Section \ref{sec:fastSimulation}, with piecewise analytic rates, in a 
compact interval $[0,T']$, for $T'$ sufficiently large. \\
Let $G(t)$, $U(t)$, $t\in[t_0,t_1+T]$ be  \emph{compatible} and \emph{robust} time-varying sets, and let $G\N(t)$, $U\N(t)$ be sequences of 
time-varying sets converging robustly to $G$ and $U$, respectively. \\ 
Furthermore, let $P(t) = P_{reach}(z_k,t,T,G,U)$ and\\ $P\N(t) = P_{reach}(Z\N_k,t,T,G\N,U\N)$, $t\in[t_0,t_1]$.\\
Finally, fix $p\in[0,1]$, $\bowtie\in\{\leq,<,>,\geq\}$, and let $V_p(t) = I\{P(t)\bowtie p\}$, $V\N_p(t) = I\{P\N(t)\bowtie p\}$. Then
\begin{enumerate}
\item For all but finitely many $t\in [t_0,t_1]$, $P\N(t)\rightarrow P(t)$, with uniform speed (i.e.\ independently of $t$). 
\item For almost every $p\in[0,1]$, $V_p$ is robust and the sequence $V\N_p$ converges robustly to $V_p$.
\end{enumerate}
\end{applemma}

\proof As in the proof of Lemma \ref{lemma:timeDepNextProb}, by
a standard coupling argument assume that $z_k$ and $Z\N_k$ are defined on the same probability space $\Omega$. Then, 
letting $Y$ be either $z_k$ or $Z\N_k$, for $\omega\in\Omega$, let $\chi(t,Y(\omega))$ be equal to 1 if trajectory $Y(\omega)$ 
satisfies the reachability problem with respect to $G$ and $U$ and zero otherwise, starting at time $t$ and $\chi\N(t,Y(\omega))$ 
be 1 if $Y(\omega)$ satisfies the reachability problem for $G\N$, $U\N$, and zero otherwise, starting at time $t$. 
Then $P(t) = \bbE[\chi(t,z_k)]$, and $P\N(t) = \bbE[\chi\N(t,Z\N_k)]$, and
\begin{eqnarray*}
|\bbE[\chi(t,z_k)] - \bbE[\chi\N(t,Z\N_k)]| & \leq & \underbrace{\bbE[|\chi(t,z_k) - \chi\N(t,z_k) |]}_{(1)}\\
& +& \underbrace{\bbE[| \chi\N(t,z_k) -\chi\N(t,Z\N_k) |]}_{(2)}\\
\end{eqnarray*}

Term (2) above is bounded by $\eps_N\rightarrow 0$, as in Lemma \ref{lemma:timeDepNextProb}, by a straighforward 
application of Theorem \ref{th:fastSimulation}. 
Term (1) is also treated similarly to Lemma \ref{lemma:timeDepNextProb}, with an extra argument to deal with pointwise 
discontinuities in the reachability probability.
Let  $T_1<T_2<\ldots<T_h$ be all the points in $Disc(G)\cup Disc(U)$ (which is finite as $G$ and $U$ are robust). If we 
suppose neither $t$ nor $t+T$ coincide with one of the previous points (i.e.\ all discontinuities are internal in the time domain), 
then by robust convergence of $G\N$ (resp.\ $U\N$) to $G$ (resp.\ $U$), for $N\geq N_0$ they differ only in small disjoint balls 
$B(T_i,\eps)$  internal to $[t,t+T]$. 
Reasoning as in Lemma \ref{lemma:timeDepNextProb}, it  follows that the only trajectories of $z_k$ for which 
$\chi(t,z_k)\neq\chi\N(t,z_k)$ are those jumping within the set $D\N = D\N_G \cup D\N_U$.\footnote{If $G$ or $U$ are not robust, 
then even if they have a finite number of discontinuity points, the previous argument may not hold. In fact, they may have a 
discontinuity point $T_i$ such that $G_s(T_i) = 1$ but $G_s(t) = 0$ in a neighbourhood $W\setminus \{T_i\}$ of $T_i$. In this case,  
it is possible that $G\N_s(t)=0$ on all $W$, which implies that $\chi(t,z_k))\neq\chi\N(t,z_k)$ for all those trajectories that are in 
state $s$ at time $T_i$.}
As the rate functions of $z_k$ are piecewise analytic, they are bounded by a constant $\Lambda$, thus the probability of a 
trajectory jumping in $D\N$ is bounded by $\int_{D\N} \Lambda e^{-\Lambda t} dt \leq \int_{D\N} \Lambda dt = \Lambda \mu_\ell(D\N)\rightarrow 0$ 
(independently of $t$).
It follows that, if $t\not\in T_d$, with $T_d = \{T_1,\ldots,T_h,T_1-T,\ldots,T_h-T\}$, then
$$|\bbE[\chi(t,z_k)] - \bbE[\chi\N(t,Z\N_k)]| \leq \delta_N,$$
with $\delta_N = \eps_N + \Lambda \mu_{\ell}(D\N) \rightarrow 0$. 

On the contrary, if $t\in T_d$, then a discontinuity of $G$ or $U$ happens exactly at the boundary of the time domain $[t,t+T]$ 
in which we have to verify the formula. In this case, the value of sets $G\N$ and $G$ (or $U\N$ and $U$) may never be the 
same at this extreme point $t^*$, whatever small neighbourhood of $t^*$ in $[t,t+T]$ one takes into account (e.g.\ if $t^* = T_i$ 
is the left extreme of the time domain, it may happen that  all changes of $G\N$ occur before this point). Therefore, there can 
be a set of trajectories of measure $>0$ that are accepted by $\chi\N$ and refused by $\chi$ (or vice versa). In particular, this 
can happen if and only if $P(t)$ has a discontinuity in one of those points (otherwise, convergence follows by continuity).  Hence, in these time points, convergence may not 
hold. However, the set $T_d$ is finite, hence point 1 of the Lemma is proved. 

Let us turn now to point 2 of the lemma, which is similar to Lemma \ref{lemma:timeDepNextProb}, with extra care for the 
discontinuities of $P$.\\
As in Lemma \ref{lemma:timeDepNextProb}, construct the set $H_{\bP}$ of values $p\in [0,1]$ for which either (i) $\bP(t)$ is constantly equal to $p$ in one analytic piece of $\bP$, or (ii) $\bP(t)=p$ and $\bP'(t) = 0$ for some $t$, or (iii) $\bP(t) = p$ and $\bP$ is not analytic in $t$. This set is finite, and for $p\not\in H_{\bP}$, the function $\bP(t) - p$ is easily seen to  define a robust time-varying set, as it has a finite number of simple zeros, all in analytic points of $\bP$. Consider now the set $A$ of discontinuity points of $V_p$. 
Fix $\eps$ and define $A_\eps$ to be $\bigcup_{t\in A} B(t,\eps)$, where $B_\eps(t) = (t-\eps,t+\eps)$. By reasoning as in the last 
part of the proof of Lemma \ref{lemma:timeDepNextProb}, letting $I_\eps = I\setminus A_\eps$ and $\min\{|P(t) - p| \mid 
t\in I_\eps\} = m_\eps >0$, as $P\N(t)$ converges in $I_\eps$ to $P(t)$ with uniform speed, \ifx\arxiv\undefined\footnote{
%
% footnote on discontinuity points.
%
The set $I_\eps$ may contain time instants $\tilde{t}$ in which the convergence of $P\N$ to $P$ does not hold, but that do not generate a discontinuity in $V_p$, because $P(\tilde{t}^+)$ and  $P(\tilde{t}^-)$ are both greater or both less than $p$. These points do not create problems, essentially because the function $P\N$, for $N$ large, remains close to $P$. In fact, convergence in $\tilde{t}$ fails because the jumps in $G\N$ and $U\N$ happen at time instants converging to the ones of jumps in $G$ and $U$, but not necessarily in $\tilde{t}$. This slightly puts out of synchronization the time at which the discontinuity happens, but the values of $P\N$ and $P$ around such discontinuity are close. This implies that, for $N$ large, $P\N$ will remain below $p$ if both $P(\tilde{t}^+)$ and  $P(\tilde{t}^-)$  are below it, and similarly for the symmetric case.

} \fi
there is $N_0$ such that, for all $N\geq N_0$ and all $t\in I_\eps$, $|P\N(t) - P(t)|\leq \frac{m_\eps}{2}$, hence for all $N\geq 
N_0$ and all $t\in I_\eps$, $V_p(t) = V\N_p(t)$. 
It follows that $V\N_p(t)$ converges robustly to $V_p(t)$. \ifx\arxiv\undefined \qed\fi
\ifdefined\arxiv 

However, here we need extra care as the set $I_\eps$ may contain time instants $\tilde{t}$ in which the convergence of $P\N$ to $P$ does not hold, but that do not generate a discontinuity in $V_p$, because $P(\tilde{t}^+)$ and  $P(\tilde{t}^-)$ are both greater or both less than $p$. These points do not create problems, essentially because the function $P\N$, for $N$ large, remains close to $P$. In fact, convergence at $\tilde{t}$ fails because the jumps in $G\N$ and $U\N$ happen at time instants converging to the ones of jumps in $G$ and $U$, but not necessarily at $\tilde{t}$. This slightly puts out of synchronization the time at which the discontinuity happens, but the values of $P\N$ and $P$ around such a discontinuity are close. This implies that, for $N$ large, $P\N$ will remain below $p$ if both $P(\tilde{t}^+)$ and  $P(\tilde{t}^-)$  are below it, and similarly for the symmetric case.

A formalisation of this argument requires a more careful inspection of the behaviour of $G\N$ (respectively $U\N$) near a discontinuity of $G$ (respectively $U$), and a clarification of the connection between discontinuities in $G$ and $U$ and discontinuities in $P$. 
For the former point, note that by the robust convergence property of $G\N$ to $G$, if $G$ has a discontinuity for state $s$ at time $t$, say from $0$ to $1$, then $G\N$ also has a discontinuity of the same kind near $t$. In fact, it can do more than one jump around $t$,  but for sure, for any small $\eps>0$  and $N$ large, it will equal 0 before $t-\eps$ and 1 after $t+\eps$. The point is that these additional jumps do not matter, as they happen so close to each other that almost no probability mass moves in between, hence they have a vanishing effect on $P\N$ (as $N$ grows).  
As for the connection between discontinuities in $G$ and $U$ and the function $P$, observe that we can have a discontinuity in $P$ at time $t$ only if either $G$ or $U$ has a discontinuity at time $t$ or at time $t+T$ (they cannot  both have such a discontinuity, due to the compatibility condition). 
There are many cases to take into account (a change from goal to non-goal, or from non-goal to goal, and so on), but only a few of them induce a discontinuity, specifically a change from non-goal to goal of a safe state $s$ at time $t+T$ (inducing a discontinuity in any safe and non-goal state at time $t$), a change from goal to non-goal of a safe state $s$ or form unsafe to safe of a non-goal state  $s$ at time $t$ (inducing a discontinuity in $s$), and  a change in the goal status of an unsafe state at time $t$. In the first case, we can have a discontinuous increase in $P$. 
In the second case, the value of $P$ in $s$ can drop from 1 to a value $p'<1$. In the third case, the value of $P$ can increase from 0 to a value $p' > 0$. In the fourth case, which is a rather strange case, the value of $P$ changes from 0 to 1, or vice versa. 
\\
To understand the connection between $P$, $P\N$, $G$ and $G\N$, consider a situation of the first kind, in which one or more safe states $s$ change from non-goal to goal at time $t+T$. This creates a discontinuity in $P(t)$ for any safe state $s'$, such that there is a non null-probability of going from  $s'$ to $s$ along a safe and non-goal path from $t$ to $t+T$. This probability, in fact, is added to $P_{s'}(t)$,  according to the discussion in Section \ref{sec:reachability}. Suppose for simplicity that only a single state $s$ changes status in $t+T$ from non-goal to goal. Then this happens close to $t+T$ also in $G\N$. In fact, $s$ can change status many times in $G\N$, near $t+T$, but only the first one really matters for the discontinuity of $P\N$. This happens because the probability added to $P_{s'}\N$ for subsequent jumps of $s$ from non-goal to goal state close to $t+T$ is only the probability of jumping into $s$ from another safe state in the short time interval in which $s$ is non-goal. 
To be more concrete, if $s$ jumps from non-goal to goal at time $t_1\N+T$, then from goal to non-goal at time $t_2\N+T$ and back to goal at time $t_3\N+T$, then the probability added to $P_{s'}\N(t_3\N)$ is bounded by the amount of probability mass that can flow into $s$ in between times $t_2\N+T$ and $t_3\N+T$, which is of the order of $t_3\N - t_2\N$, hence vanishes as $N$ grows (as $t_2\N$ and $t_3\N$ collapse to $t$). More precisely, if $\Lambda$ is an upper bound for the exit rate of the single agent (uniform in $N$, which can be found as the exit rate of $Z_k\N$ converges to the exit rate of $z_k$, which is itself uniformly bounded), then the jump size at time $t_3\N$ is bounded by $2\Lambda(t_3\N - t_2\N)$. Furthermore, the speed at which $P_{s'}\N$ can increase or decrease, excluding jumps, is also bounded by $2\Lambda$, so that the value of $P_{s'}\N$ cannot vary too much after the first jump in a small time interval of size $\Delta_t$ around $t$. 
In fact, combining these two arguments, the total variation (after the first jump) is bounded by $2\Lambda\Delta_t$.\footnote{The factor 2 comes from the fact that we are working with a combination of the backward and forward equation, both giving an upper bound of $\Lambda$ on the rate of change of probability mass.}
Note that, if more than one safe state changes goal status at time $t+T$ in $G$, then in $G\N$ these events can happen asynchronously, hence to see the full increase in $P_{s'}\N$ we need to wait that all those states do their first jump in $G\N$. Yet the bound in terms of $\Lambda$ and $\Delta_t$ remains valid.
The other discontinuous jump types are treated analogously (with the exception of the jump from 0 to 1 or from 1 to 0, which however contains any threshold $p$ in its interior).
We can now give a formal argument that discontinuities in $I_\eps$ are not a problem. Assume that $P$ has a discontinuity in $\tilde{t}$ such that $\mu=\max\{P_{s'}(\tilde{t}^+),P_{s'}(\tilde{t}^-)\} < p$, and call $\eps = p-\mu$. Now, choose $\delta$ such that $4\delta\Lambda < \eps/4$, $\|P_{s'}(\tilde{t}-\delta) - P_{s'}(\tilde{t}^-)\| < \eps/4$, and $\|P_{s'}(\tilde{t}+\delta) - P_{s'}(\tilde{t}^+)\| < \eps/4$. Then,  choose an $N_0$ such that, for $N\geq N_0$, all the jumps of $G\N$ are closer than $\delta$ to the jumps of $G$, and such that  $\|P_{s'}\N(\tilde{t}\pm\delta) - P_{s'}(\tilde{t}\pm\delta)\| < \eps/4$. Then, using the previous reasoning, we can see that $\sup_{t\in ]\tilde{t}-\delta,\tilde{t}+\delta[} P_{s'}\N(t) <  \max\{P_{s'}\N(\tilde{t}+\delta),P_{s'}\N(\tilde{t}-\delta) \} + 4\delta\Lambda \leq  \mu + 3/4\eps < p$. The first inequality holds because $\max\{P_{s'}\N(\tilde{t}+\delta),P_{s'}\N(\tilde{t}-\delta) \}$ is a value close to the value of $P\N$ after the first jump, and is combined with the bound $4\delta\Lambda$ on the variation.  Hence $P_{s'}\N$ ultimately does not cross the line $p$ around $\tilde{t}$. The case in which $\min\{P_{s'}(\tilde{t}^+),P_{s'}(\tilde{t}^-)\} > p$ is dealt with similarly. \qed
\fi

\subsection{CSL model checking}

\begin{apptheorem}[\ref{th:CSLdecidabilityICTMC}]
The CSL model checking for ICTMC, for piecewise analytic interval computable rate functions, is decidable for a robust CSL formula $\phi(p_1,\ldots,p_k)$. 
\end{apptheorem}

\proof  First of all, we prove that we can approximate the function $P(t)$ for any top next or until formula $\phi$ with arbitrary 
small precision. To start, notice that procedures for integrating ODEs and doing matrix multiplication (which are at the basis of 
the methods in Sections \ref{sec:next} and \ref{sec:reachability}) can be computed with arbitrary precision, due to the assumptions 
of interval computability. Hence, let us focus on the set of zeros of $P(t) - p$, for a dependent next or until formula $\phi_1$. 
We want to prove that we can find those zeros with arbitrary precision, and that in doing this we will be able to compute the 
probability of any next or until formula which contains $\phi_1$ as a sub-formula with arbitrary precision. 
If $\phi$ is robust, then the time-varying truth of formula $\phi_1$ is robust. This means that $P(t) - p$ has a finite number of 
simple zeros (i.e.\ their derivatives are not zero). Hence, it is possible to effectively encapsulate them in disjoint intervals of 
size as small as desired \cite{COMP:Taylor:20120:RealLambdaCalculus}\footnote{As the number of zeros is finite  and their first-order derivative is non-zero, the function $f_p(t) = P(t)-p$ crosses zero in those points. Furthermore, notice that the absolute minimum value of the derivative in those zero points is $>0$. Hence, there is an $\eps$ such that each interval of size $\eps$ containing a zero point has different signs at the extremes and the derivative is provably different from zero. 
By iterated bisection, we can always find such intervals after a finite number of steps. Furthermore, all intervals $J$ not containing a zero can be eventually discarded by bisection, computing an upper bound $L$ on the absolute value of the derivative in such intervals and  bisecting them until we can prove that they are disjoint from zero using the Lipschitz condition with Lipschitz constant $L$ (compute $f_p$ on a single point $x$ in $J$ of length $\delta$, and discard $J$ if $|f_p(x)| - L\delta > 0$).}. 
Therefore, we can compute the time-varying truth value of the set of states satisfying the formula $\phi_1$ with arbitrary precision,  
in the sense that for each $\eps>0$ small enough, we can provide intervals of size at most $\eps$, each containing a single 
discontinuity point of the set in which one or more states change truth status. 
The condition on compatibility ensures that we can combine such approximation of time-varying sets and still obtain 
robust sets.  (This may fail if we take the minimum (conjunction) of two truth sets which have a discontinuity for $s$ in the same 
time point $T$: we can obtain a function which is neither left nor right continuous, a situation that cannot originate from a simple zero.) Furthermore we have the further property that 
we can always assume that there is a zero in every approximation interval\footnote{If we take the minimum (conjunction) of two truth sets which have a discontinuity for $s$ in the same time point $T$, then even if the conjunction is robust, when we have an approximation of the time-varying truth function, we can never know if both discontinuities happen in the same time point or in different ones.}. 
Consider now the problem of computing the probability of an until formula, having two approximations of time-varying truth as 
described above. Reasoning as in the proof of Lemma \ref{lemma:tameTimeDepReachab}, we can see that if we choose an 
arbitrary point in each interval wrapping a discontinuity point in spite of the correct one, we commit an error in computing the 
probability of the until which is uniformly bounded by the total size of the approximation intervals. Hence, we can make such error 
as small as desired. 
A similar conclusion can be drawn for a next formula, invoking the line of reasoning of Lemma \ref{lemma:timeDepNextProb}.
Reasoning inductively, we can therefore compute with any arbitrary precision the probability $P(0)$ of any top until formula.

Given this value, we then have to solve the inequality $P_s(0)<p_i$ (or $P_s(0)>p_i$) for any $s$ and any top next or until formula 
$\phi_i$. By the robustness of the CSL formula $\phi$, it cannot be that $P_s(0)=p_i$, hence we can effectively solve that problem 
by computing $P_s(0)$ with precision $\eps_i< |P_s(0)-p_i|$. As we are doing interval arithmetic computations, we can increase 
the precision until each $p_i$ will be outside the approximation interval for $P_s(0)$. This proves that the algorithm presented is 
effective for robust formulae and eventually computes the exact answer.\qed

%%%%%%%%%%%%%%%%%%%%%%%%%%%%%%%%%%%%%%%

\begin{apptheorem}[\ref{th:satisfiabilityAE}]
Given a CSL formula $\phi(\vr{p})$, with $\vr{p}\in[0,1]^k$, then the set $\{\vr{p}~|~\phi(\vr{p})\ \mbox{is robust}\}$ is relatively open\footnote{A set $U\subset V$ is relatively open in $V\subset W$, where $W$ is a topological space, if it is open in the subspace topology, i.e. if there exists an open subset $U_1 \subseteq W$ such that $U = V \cap U_1$.} in $[0,1]^k$ and has Lebesgue measure 1. 
\end{apptheorem}
\proof  
%In order to fix the notation, let $I\subseteq\{1,\ldots,k\}$; by $U_I$ we denote a subset of $[0,1]^k$ which can be written as the product $U\times [0,1]^{k-|I|}$, where $U$ is a subset of $[0,1]^{|I|}$. 
%Given a set $U\subset [0,1]^{|I|}$, $U_I$ also denotes its embedding in $[0,1]^k$ by taking its product with $[0,1]$ for all coordinates not in $I$.\\
We will prove the theorem by structural induction on the formula $\phi$. We first need some preliminary definitions. Consider an until formula $\phi = \calP_{ \bowtie p} (\phi_1 \until{T_1}{T_2} \phi_2)$ or a next formula $\phi = \calP_{ \bowtie p} (\next{T_a}{T_b} \phi_1)$ and call $\vr{q}$ a generic tuple of values for the thresholds on which $\phi_1$ and (in the until case) $\phi_2$ depend on. Fix a $\vr{q}$ such that the time-varying sets for $\phi_1$ and $\phi_2$ are robust. An open neighbourhood $U_{\vr{q}}$ of $\vr{q}$ is \emph{robust} if the time-varying sets  for $\phi_1$ and $\phi_2$ are robust for each $\vr{q'}\in U_{\vr{q}}$. Observe that in $U_{\vr{q}}$ the number of discontinuities of time-varying truth sets of $\phi_1$ and $\phi_2$  does not change ($\phi_j$ is robust for each point in $U$, and a change in the number of discontinuities can happen only at a non-robust point) and the time-instants at which such discontinuities happen depend 
continuously on $\vr{q}$.
Now, we define the set-valued function $b:U_{\vr{q}}\rightarrow 2^{[0,1]}$ in the following way: Given $\vr{q'}$,  $b(\vr{q'})$ is the set of values 
$p$ which causes the time-varying truth set of $\phi$ to be non-robust. Therefore, $b(\vr{q'})$ contains 
the values of $P_s(t)$ for which $P_s'(t)$ is zero,  the values $P_s(t^-)$ and $P_s(t^+)$ for each non-analytic point $t$ of $P$, 
and the values of constant pieces of $P_s$, plus the value $P_s(0)$.  
Hence it is finite. 
\\
By possibly restricting $U_{\vr{q}}$, we can also assume that the number of points in 
$b(\vr{q}')$ is bounded by $|b(\vr{q})|$  in $U_{\vr{q}}$\footnote{
We have to restrict $U$ to avoid the appearance of further zeros of the 
derivatives away from current zeros. Note also that if a value $p\in b(\vr{q})$ corresponds to a non-simple zero at a time $t_0$ in which the derivative has a maximum or a minimum, a small perturbation of $\vr{q}$ can split it in two, or make it disappear. Just think about raising or lowering a curve having a local maximum or minimum with value zero. However, split zeros will be at points $t_1$ and $t_2$ arbitrarily close to $t_0$, and therefore, by continuity of $P_s(t,\vr{q})$ with respect to $\vr{q}$, the values of $P_s(t_i,\vr{q'})$, for $\vr{q'}$ close to $\vr{q}$ will be close to  $P_s(t_0,\vr{q})$. Zeros that disappear are not a problem for semicontinuity, as the empty set is contained in any set. Hence, we need to count those points twice in $|b(\vr{q})|$. } 
and the value of such points depends continuously on $\vr{q'}$. Therefore, the set-valued map $b:U_{\vr{q}}\rightarrow 
2^{[0,1]}$ is \emph{upper-semicontinuous} in $U_{\vr{q}}$, i.e.\ for each neighbourhood $U_{b(\vr{x})}$ of $b(\vr{x})$ in $[0,1]$, there is a neighbourhood $U_\vr{x}$ 
of $\vr{x}$ in $U_{\vr{q}}$ such that $b(U_\vr{x})\subseteq U_{b(\vr{x})}$. 
\\
Given a formula $\phi = \phi(\vr{p})$, $\vr{p}\in[0,1]^k$ , we define the set $R_{\phi} \subset [0,1]^k$ of all robust thresholds, i.e.\   $\vr{p_0}\in R$ if and only if $\phi(\vr{p_0})$ is robust for $\phi$. Hence, our goal is to show that $R_{\phi}$ is open and has measure 1 for any formula $\phi$.
\\
We are now ready for the inductive argument.

\begin{description}
\item[Base case:] The base case corresponds to (boolean combinations of) atomic formulae, which are robust for each 
$\vr{p}\in[0,1]^k$. 

\item[Boolean combinations:] The only non-trivial cases are the conjunction or disjunctions of until or next formulae. In these cases, we have 
to enforce the compatibility condition by guaranteeing that the discontinuity times of truth-valued functions are disjoint for each pair of until or next formulae. Consider two until or next formulae $\phi_1$ and $\phi_2$, and let $\vr{p}=(\vr{q_1},p_1,\vr{q_2},p_2)$, where $p_j$ is the threshold for formula 
$\phi_j$ and $\vr{q_j}$ is the set of constants which $\phi_j$ depends on. By inductive hypothesis, the robust sets $R_j=R_{\phi_j}$ for 
$\phi_j$ are open and have measure 1. 
Now, let $P_j = P_j(t,\vr{q_j})$ be the probability of the until or next path formula in $\phi_j$, and fix a robust point $\vr{q} = 
(\vr{q_1},p_1,\vr{q_2})$. Let $g(\vr{q})$ be the set valued function $g(\vr{q}) = P_2(\{t~|~P_1(t,\vr{q_1})=p_1\},\vr{q_2})\cup 
b_2(\vr{q_2})$, where $b_2$ is the set of non-robust points for $\phi_2$, as defined above. 
The set $g(\vr{q})$ contains all thresholds for $\phi_2$ for which $\phi_2$ is non-robust and all thresholds that would make 
 the boolean combination non-robust. By properties of the analytic functions, it follows that $g(\vr{q})$ is finite. Hence by 
arguments similar to the ones  above, the function $g$ is upper-semicontinuous\footnote{The number of solutions of $P_1(t,\vr{q_1})=p_1$ in a sufficiently small neighbourhood $U_1$ of $(p_1,\vr{q_1})$ is constant and hence
the set-valued function $g_1(p_1,\vr{q_1})=\{t~|~P_1(t,\vr{q_1})=p_1\}$ is upper semicontinuous. 
Furthermore, in a sufficiently small neighbourhood $U_2$ of $\vr{q_2}$, the function $P_2(t,\vr{q_2})$ is continuous in $\vr{q_2}$ 
for each continuity point $t$ of $P_2(t,\vr{q_2})$. 
Points for which $P_2(t,\vr{q_2})$ is not continuous are covered by $b$, hence both $P_2(t^+,\vr{q_2})$ and $P_2(t^-,\vr{q_2})$ 
are in $g(\vr{q})$.
Now, for each neighbourhood $V$ of $g(\vr{q})$, by piecewise analyticity and right/left continuity of $P_2$, we can find a 
neighbourhood $U_2$ of $\vr{q_2}$ and a neighbourhood $V_1$ of $g_1(p_1,\vr{q_1})$  such that $b_2(U_2)\subseteq V$ and 
both $\{p \mid p=P_2(t^+,\vr{q_2}),t\in V_1\}\subseteq V$ and $\{p \mid p=P_2(t^-,\vr{q_2}),t\in V_1\}\subseteq V$. 
Now, by upper-semicontinuity of  $g_1$, there is a neighbourhood $U_1$  of $(\vr{q_1},p_1)$ such that $g_1(U_1)\subseteq 
V_1$.  
It follows that $g(U_1\times U_2)\subseteq V$, hence $g$ is upper-semicontinuous in $\vr{q}$.} 
in a neighbourhood $U$ of $\vr{q}$. 
Therefore, letting $p_2\not\in g(\vr{q})$ and $V\cap g(\vr{q}) = \emptyset$ a neighbourhood of $p_2$ in $[0,1]$, we can find a 
neighbourhood $U$ of $\vr{q}$ such that $g(U)\cap V = \emptyset$, so that $W=U\times V$ is an open neighbourhood of 
$\vr{p} = (\vr{q_1},p_1,\vr{q_2},p_2)$ which contains only robust points, which proves that $R=R_{\phi}$ is open. 
\\
Furthermore, $R$ is \emph{a fortiori}  measurable.
Now, let $h_R:[0,1]^{k_1+k_2}\rightarrow \{0,1\}$ be the indicator function of the set $R$ in which the boolean combination $\phi$ of $\phi_1$ and $\phi_2$  is robust. Note that $R\subseteq R_1\times R_2$, and call $R_2'$ the set of thresholds $\vr{q_2}$ for which  sub-formulae of $\phi_2$ are robust, which is open and has measure 1 in $[0,1]^{k_2-1}$ by inductive hypothesis.  By Fubini's theorem:
\[\begin{split}
\mu_\ell(R) & = \int_{[0,1]^{k_1+k_2}} h_R(\vr{q_1},p_1,\vr{q_2},p_2)\mu_\ell(d\vr{q_1},dp_1,d\vr{q_2},dp_2)\\
& = \int_{[0,1]^{k_1+k_2-1}} \int_{[0,1]} h_R(\vr{q_1},p_1,\vr{q_2},p_2)\mu_\ell(dp_2) \mu_\ell(d\vr{q_1},dp_1,d\vr{q_2})\\ 
& = \int_{R_1\times R_2'} \mu_\ell(d\vr{q_1},dp_1,d\vr{q_2}) = \int_{R_1} \mu_\ell(d\vr{q_1},dp_1) \int_{R_2'} \mu_\ell(d\vr{q_2}) = 1,
\end{split}\]
which proves that $R$ has measure 1.
\\
If we have a boolean combination of $j>2$ until formulae, we simply reason pairwise and then take the intersection of the so-obtained robust sets, thus getting an open set of measure 1. 

\item[Until formulae:] Let $\phi = \calP_{ \bowtie p_k} (\phi_1 \until{T_1}{T_2} \phi_2)$.
%, and let $I = \{1,\ldots,k-1\}$.  
By inductive hypothesis, the set $R_{\phi_1}\times R_{\phi_2} \subset [0,1]^{k-1}$ for which $\phi_1$ and $\phi_2$ are robust is open and has measure 
1. By reasoning as in the boolean combination case (and considering all until and next conjunct/disjuncts of $\phi_1$ and $\phi_2$), we can immediately conclude that the set $R'\subseteq R_{\phi_1}\times R_{\phi_2} $ in which the time varying sets of $\phi_1$ and $\phi_2$ are robust and compatible is open and has measure 1.

Now, fix a point $\vr{q}\in R'$, let $U\subseteq R'$ be a robust neighbourhood of $\vr{q}$, and consider the set valued function $b:U\rightarrow 2^{[0,1]}$ as defined above. Now fix $p\not\in b(\vr{q})$, and choose a neighbourhood $V$ of $p$ such that $V \cap  b(\vr{q}) = \emptyset$. As $b$ is  upper-semicontinuous, there exists $W\subset U$ such that $b(W)\cap V = \emptyset$, hence $\phi$ is robust in  $W\times V$. By the arbitrary choice of $\vr{p} = (\vr{q},p)$, it follows that $R=R_{\phi}$ is open, and hence measurable.
Now, let $h_R:[0,1]^k\rightarrow \{0,1\}$ be the indicator function of the set $R$ in which $\phi$ is robust. By Fubini's theorem,
%\[\begin{split}
%\mu_\ell(R) & = \int_{[0,1]^k} h_R(\vr{q},p)\mu_\ell(d\vr{q},dp)= \int_{[0,1]^{k-1}} \int_{[0,1]} h_R(\vr{q},p)\mu_\ell(dp)
%\mu_\ell(d\vr{q})\\ & = \int_{R'} \mu_\ell(d\vr{q}) = 1,
%\end{split}\]
it follows that $R$ has measure 1.

\item[Next formulae:] The argument for a next formula $\phi = \calP_{ \bowtie p_k} (\next{T_a}{T_b} \phi_1)$ is essentially the 
same as for until formulae, with the only difference that the inductive hypothesis is applied only to $\phi_1$ and there is no need to ensure compatibility. \qed
\end{description}

%%%%%%%%%%%%%%%%%%%%%%%%%%%%%%%%%%%%%%%%%%%%%%%%5
\begin{apptheorem}[\ref{th:CSLconvergence}]
Let $\calXN$ be a sequence of CTMC models, as defined in Section \ref{sec:modelingLanguage}, and 
let $Z\N_k$ and $z_k$ be defined from $\calXN$ as in Section \ref{sec:fastSimulation}.\\
Assume that $Z\N_k$, $z_k$ have piecewise analytic infinitesimal generator matrices.\\ 
Let $\phi(p_1,\ldots,p_k)$ be a robust CSL formula. 
Then, there exists an $N_0$ such that, for $N\geq N_0$ and each $s\in\calS$
$$s,0 \vDash_{Z\N_k} \phi \Leftrightarrow  s,0 \vDash_{z_k} \phi.$$
\end{apptheorem}

\proof

We use structural induction to prove that, for each formula $\phi$, the time-varying truth sets $V\N_\phi$ of $\phi$ in $Z\N_k$ 
converge robustly to the robust time-varying truth set $V_\phi$ of $\phi$ in $z_k$.

\begin{description}
\item[Base case:] The case for atomic propositions is trivial, as $V\N_\phi$ and $V_\phi$ are constant and equal.
\item[Negation: ] Let $\phi = \neg \phi_1$. The result follows because $V_\phi(t) = 1-V_{\phi_1}$ and $V\N_\phi(t) = 1-V\N_{\phi_1}$.
\item[Conjunction/Disjunction: ] Let $\phi = \phi_1 \circ \phi_2$, $\circ\in\{\wedge,\vee\}$. Due to the  compatibility condition 
of robustness of $\phi$ with respect to $z_k$, the set $V_\phi(t) = mm\{V_{\phi_1}(t),V_{\phi_2}(t)\}$, $mm\in\{\min,\max\}$ is 
robust, with $Disc(V_\phi) \subseteq Disc(V_{\phi_1})\cup Disc(V_{\phi_2})$. Using the inductive hypothesis, it easily follows that 
$V\N_\phi(t)= mm\{V\N_{\phi_1}(t),V\N_{\phi_2}(t)\}$ converges robustly to $V_\phi(t)$.
\item[Next:] Let $\phi = \calP_{\bowtie p} (\next{T_a}{T_b} \phi_1)$. By inductive hypothesis, we can apply 
Lemma \ref{lemma:timeDepNextProb} and deduce that $V_\phi$ is robust and $V\N_{\phi}$ converges robustly to $V_\phi$. 
\item[Until:] Let $\phi = \calP_{\bowtie p} (\phi_1 \until{T_a}{T_b} \phi_2)$. By inductive hypothesis  (and the compatibility condition in the definition of robustness of $\phi$), we can apply 
Lemma \ref{lemma:tameTimeDepReachab}\footnote{We need to apply it twice for the two reachability problems involved in 
computing the probability of an until formula, noticing that the probability of the path formula within $\phi$ is an analytic 
combination of the two so-computed probabilities. Robustness of $V_\phi$ and robust convergence of $V\N_\phi$ to $V_\phi$ 
follows from the same arguments of Lemma \ref{lemma:tameTimeDepReachab}. Alternatively, one can modify 
Lemma \ref{lemma:tameTimeDepReachab} and tailor it to the reachability involved in the until case (which reduces the time 
window in which one can reach  the goal set), by a straightforward modification of the definition of $\chi\N$ and $\chi$ and adaptation of the arguments for robust convergence.} and 
deduce that $V_\phi$ is robust and $V\N_{\phi}$ converges robustly to $V_\phi$. 
\end{description}
The fact that $V\N_\phi$ converges robustly to the robust set $V_\phi$, combined with property 1 of robustness of $\phi$, let us 
conclude that the truth value of $\phi$ at level $N$ converges to the truth value of the limit ICTMC at time zero (if $0$ was a point 
in which convergence of probability fails, then a small perturbation in $\vr{p}$ could change the truth value of $\phi$ in the limit 
ICTMC, contradicting the robustness of $\phi$; furthermore, robustness of $\phi$ forbids that $P_s(0) = p$).  \qed
%%%%%%%%%%%%%%%%%%%%%%%%%%%%%%%%%%%%%%%%%%%%%%%%5

\subsubsection{Comparison of CSL model checking for $Z\N_k$ and $(Z\N_k,\nX\N_k)$}

We will prove now the lemmas in Section \ref{sec:spaceVsTime} of the paper. We will start by an auxiliary result, which is needed to adapt the proof style of Lemmas \ref{lemma:timeDepNextProb} and \ref{lemma:tameTimeDepReachab} to the processes $(Z\N_k,\nX\N_k)$ and $(z_k,\nx_k)$ discussed in this section. In particular, in Lemmas \ref{lemma:timeDepNextProb} and \ref{lemma:tameTimeDepReachab}  we used the fact that processes jump with a small probability in a small temporal neighbourhood of the discontinuity points of time-varying sets. In the space-based setting, however, we need to consider neighbourhoods of the boundaries of goals and unsafe sets. Therefore, to use the same proof style, we need to bound the time trajectories spend in such a neighbourhood, in a uniform way in space. 
The key point is that, as the convergence results we are interested in depend only on a neighbourhood of the trajectory $\flow{[0,T]}{\x_0}$, we can always choose a small flow tube such that the velocity with which a trajectory crosses the boundary of a goal or an unsafe set is close to that of $\flow{[0,T]}{\x_0}$, and so will be the time spent in a neighbourhood around such boundary. In the following, we will make this intuition formal. 

First of all, observe that the notion of robust set $V$ implies that when a trajectory crosses the boundary $\partial V$ in a point $\x$, the function $h$ defining the smooth manifold of the d-set $\partial V$ around $\x$ changes sign.
\\
We will now prove an upper bound for the time spent by a trajectory in a neighbourhood of the d-set $D=\partial G$ of a robust set $G$ in $\calS\times E_0$, a $T,\eps_0$-flow tube of  $\flow{[0,T]}{\x_0}$. 

For each trajectory $\flow{[T^-_{\x}(E_0),T^+_{\x}(E_0)]}{\x}$, $\x\in E_0$, consider the   points $Disc(s,\x) = \{(s,\x_1) \in D\mid\x_1 \in \flow{[T^-_{\x}(E_0),T^+_{\x}(E_0)]}{\x}\}$ in which it intersects $D$.
\\
We define the $\eps$-neighbourhood of $D$ in $\calS\times E_0$ as $D_\eps = \bigcup_{(s,\x)\in D\cap(\calS\times E_0)} B_\eps(s,\x)$, which is an open set. Note that  $D_\eps = \bigcup_{(s,\x)\in \calS\times E_0}\bigcup_{(s,\x_1)\in Disc(s,\x)} B_\eps(s,\x_1)$, as $\calS\times E_0$ is the union of a set of trajectories.

Now, by the robustness property of $G$ in $\calS\times E_0$, we have that $|Disc(s,\x)|\leq k$. 
Furthermore, by the robustness of $G$, the trajectory $\flow{[0,T]}{\x_0}$ will cross $D$ moving from the interior of $G$ to the interior of its complement, or vice versa, for any point $(s,\x^s_i)\in Disc(s,\x_0)$ and any $s\in\calS$. 
Consider a neighbourhood $W$ of $(s,\x^s_i)$  in which  $D=\partial G$ is a smooth manifold.
Therefore, there is a sufficiently smooth function $h$ in $W$ such that $D\cap W$ is the zero set of $h$. 
By the robustness property of $G$, the function $h(\flow{t}{\x_0})$ equals 0 in $t_i$ (the time such that $\flow{t}{\x_0} = \x^s_i$), and changes sign around $t_i$ (i.e. it is positive in $t_i-\delta$ and negative in $t_i+\delta$, for a $\delta>0$). 
It follows that the derivative of $h(\flow{t}{\x_0})$ in $t_i$ is non-null. As it equals  $\nabla h(\x^s_i)\cdot F(\x^s_i)$, we have that $\nabla h(\x^s_i)\neq \vr{0}$, and hence $| \frac{\nabla h(\x^s_i)}{\| \nabla h(\x^s_i)\|}\cdot F(\x^s_i)| > 0$.
\\
Now,  by choosing a  suitably small neighbourhood $W_1\subset W$ of $(s,\x^s_i)$ (we need to ensure that the manifold containing $(s,\x^s_i)$  is the closest one in $W_1$, among those constituting $D$), we obtain that the function $\rho(t,\x) = dist((s,\flow{t}{\x}),\partial G) = \inf_{(s,\y)\in \partial G} \|\flow{t}{\x} - \y \|$ is differentiable (in $t$ and $\x$), and its derivative in $(0,\x^s_i)$ is $\rho'(0,\x^s_i) = \frac{\partial \rho(0,\x^s_i)}{\partial t} = \frac{\nabla h(\x^s_i)}{\| \nabla h(\x^s_i)\|}\cdot F(\x^s_i)$, i.e. it is the projection of the vector field along the normal to the surface $\{h=0\}$.
Now, by continuity of $\rho'$, we find a neighbourhood $W_2\subset W_1$ of $(s,\x^s_i)$ such that $|\rho'(0,\x)| \geq\rho^s_i/2$, where $\rho^s_i = |\rho'(0,\x^s_i)|$. 

Now, choose $\eps_1<\eps_0$ and $\bar{\eps}$ that satisfies:
(i) the number of intersections between $D$ and a trajectory in $\calS\times E_1$ is constant and equal to the number $k$ of intersections of $\calS\times \flow{[0,T]}{\x_0}$ with $D$ (this is possible because the flow tube $\calS\times E_1$ is a small neighbourhood of $\calS\times \flow{[0,T]}{\x_0}$), and (ii)  $D_{\bar{\eps}}$, the $\bar{\eps}$-neighbourhood of $D$ in the $T,\eps_1$-flow tube $\calS\times E_1$, is contained in the neighbourhood $W_2$ of $(s,\x^s_i)$ identified above, for any $s\in\calS$ and $i\leq k$. 
\\
By the choice of $W_2$, it follows that the speed at which each trajectory of $\calS\times E_1$ travels in $D_{\bar{\eps}}$ (with respect to the distance from $D$) is bounded below by $\rho_0 = \min_{s,i}\rho^s_i/2$, and hence the total time $\tau_{\bar{\eps}}$ a trajectory of $\calS\times E_1$ spends in $D_{\bar{\eps}}$ is  bounded above by $\frac{2\bar{\eps} k}{\rho_0}$, as it has to travel a total distance of $2\bar{\eps} k$ with speed no less than 
$\rho_0$. Notice that this bound is independent of the specific trajectory considered.

With the previous discussion, we have proved the following 
\begin{lemma}
\label{app:lemma:speedOfCrossing}
Let $E_0\subset  E$ be a $T,\eps_0$-flow tube for $\x_0$. Let $G$ be a robust subset of $\calS\times E_0$. Then, there are positive constants $\eps_1$, $\bar{\eps}$, and $\rho_0$ such that, for any $\eps'<\bar{\eps}$, the total time $\tau_{\eps'}$ a trajectory in $\calS\times E_1$ ($E_1$ the $T,\eps_1$-flow tube for $\x_0$) spends in $D_{\eps'}$, the $\eps'$ neighbourhood of $D=\partial G$, satisfies $\tau_{\eps'} \leq \frac{2\eps'k}{\rho_0}$, where $k$ is the number of intersections of any trajectory with $D$.
\end{lemma}

%%%%%%%%%%%%%%%%%%%%%%%%%%%%%%%%

Equipped with this lemma, we can now prove the following one.

\begin{applemma}[\ref{lemma:nextProbRobustSets}]
Let $E_0\subset E$ be a $T,\eps_0$-flow tube for $\x_0$. Let $G$ be a robust subset of $\calS\times E_0$, and  $G\N$ be a sequence of subsets of $\calS\times E_0$ that converge robustly to $G$.

Let $\bP(s,\x)=P_{next}(\y,s,\x,T_a,T_b,G)$ be the probability that the first jump of $\y(t)$ is into a state in $G$ and happens at 
a time $t\in[T_1,T_2]$, given that $\y$ started at time $t=0$ in state $(s,\x)\in \calS\times E_0$, and let 
$\bP\N(s,\x)=P\N_{next}(\Y\N,s,\prjN{\x},T_a,T_b,G\N)$ be defined similarly, with $G$ and $\x$ replaced by $G\N$ and 
$\prjN{\x}$, respectively. \\
Furthermore, define $V = \{(s,\x)~|~\bP(s,\x)\bowtie p\}$ and \\  $V\N = \{(s,\x)~|~\bP\N(s,\x)\bowtie p\}$. Then there exists 
$\eps_1>0$ such that, in $E_1$, the $(T-T_b),\eps_1$-flow tube for $\x_0$:
\begin{enumerate}
\item $\bP\N(s,\x) \rightarrow \bP(s,\x)$ for all $\x\in E_1$, uniformly in $(s,\x)$.  
\item If $V_{\x_0}(t)$, $t\in[T^-_{\x_0}(E_1),T^+_{\x_0}(E_1)]$, is a robust time-varying set, then $V$ is robust in $E_1$ and 
$V\N$ converges robustly to $V$. 
\end{enumerate}
\end{applemma}

\proof First, notice that we can always restrict to an arbitrary small neighbourhood of $\flow{t}{\x_0}$, i.e.\ to a $T,\eps_1$-flow 
tube $E_1$ for $\x_0$, with $\eps_1$ as small as desired. This follows from the convergence in probability implied by Kurtz 
Theorem \ref{th:Kurtz}.  Given $\delta>0$, this guarantees that we can find an index $N_0$ such that, for any $N> N_0$,  with probability at least $1-\delta$, the trajectories of $\nXN(t)$ are contained in $E_1$. Furthermore, we can choose such an index $N_0$  independently of $\x$. Hence, given $E_0$, if we consider a flow tube $E_1$ for $\x_0$ with radius $\eps_1<\eps_0/2$, each flow 
tube of radius $\eps_1$ wrapping a trajectory in $E_1$ will be contained in $E_0$. This guarantees that the next-step probability 
for $\Y\N$ for any point in $\calS\times E_1$ will ultimately depend only on the goal sets $G\N$ within  $\calS\times E_0$. 

We first prove point 1 of the lemma in a $(T-T_b),\eps_1$-flow tube $E_1$ for $x_0$, for an $\eps_1<\eps_0/2$ to be fixed in the following. We will prove convergence of $\bP\N$ to $\bP$ for each $(s,\x)\in \calS\times E_1$.

We will now use an argument similar to the one of Lemma \ref{lemma:timeDepNextProb}. Couple $\y$ and $\Y\N$ on the same probability space $\Omega$, and let $\chi$  (resp.\ $\chi\N$) be random 
variables defined on sample trajectories and equal to one if the trajectory's first jump is in $G$ (resp.\ $G\N$). 
Then, as $P(s,\x) = \bbE[\chi(s,\x,\y(\omega))]$, where $\y(0) = (s,\x)$, and similarly for $P\N(s,\x)$, to show convergence we just 
need to prove that $|\bbE[\chi(s,\x,\y)] - \bbE[\chi\N(s,\x,\Y\N)] |\rightarrow 0$. It holds that: 
{\small  
\begin{eqnarray*}
|\bbE[\chi(s,\x,\y)] - \bbE[\chi\N(s,\x,\Y\N)]| & \leq & \underbrace{\bbE[|\chi(s,\x,\y) - \chi\N(s,\x,\y) |]}_{(1)}\\
& + & \underbrace{\bbE[| \chi\N(s,\x,\y) -\chi\N(s,\x,\Y\N) |],}_{(2)}\\
\end{eqnarray*} }

To treat term (1), invoke Lemma \ref{app:lemma:speedOfCrossing}, assume $\eps_1$ is smaller than the one required by the lemma, and let $\bar{\eps}$ and $\rho_0$ the other two constants obtained from it. 
Now, as in Lemma \ref{lemma:timeDepNextProb}, observe that for each $\eps'<\bar{\eps}$, the only trajectories of  $\y$ for which $\chi$ and $\chi\N$ can have a different value are those jumping  at a time at which $\y(t)$ is in $D_{\eps'}$. 
Now, the total amount of time $\y$ spends in  $D_{\eps'}$ is uniformly bounded by $\tau_{\eps'}\leq \frac{2\eps' k}{\rho_0}$, where $k$ is the number of intersections of a trajectory in $\calS\times E_1$  with $D$. It follows that term (1) can be bounded by $\frac{2\eps' \Lambda  k}{\rho_0}$, where $\Lambda$ is an upper bound for the jump rate of $z_k$ in $\calS\times E_0$.

%
%Term (1) can be treated as in Lemma \ref{lemma:timeDepNextProb}, observing that for each $\eps'<\bar{\eps}$, the only trajectories of 
%$\y$ for which $\chi$ and $\chi\N$ can have a different value are those jumping 
%in a time in which $\y(t)$ is in $D_{\eps'}$. Now, the length of this set  is bounded by $\tau_{\eps'}\leq \frac{2\eps' k}{\rho_0}$, and so is term (1).

%in the $\eps'$-neighbourhood $Disc(s,\x,\eps')$ 
%of $Disc(s,\x)$, i.e.\ in $Disc(s,\x,\eps') = \bigcup_{t\in Disc(s,\x)} B_\eps'(t)$, and this event has a probability that is bounded by 
%$2\eps'|\bigcup_s Disc(s,\x)|\leq 2 n k \eps'$, where $k$ is the uniform bound for the robustness of $G$.  
The bound on term (2), instead, follows from the convergence of $\Y\N$ to $\y$, but it requires a slightly different treatment than 
in Lemma \ref{lemma:timeDepNextProb}, as now the time varying sets for $\Y\N$ depend on the sample trajectories of $\nXN$,  hence they are random quantities. 
Call $G\N_{\x,\nXN}(t)$ the time-varying sets relative to $G\N$, but defined  with respect to trajectories of $\Y\N(t)$. The time varying sets for $\y$, with respect to $G\N$, are denoted by $G\N_{\x}(t)$, while that relative to $G$ is $G_{\x}(t)$.
We will need now to control two things: first, we will construct a neighbourhood of $D$ in such a way that all the time varying sets are the same outside it, for $N$ large enough. Then, we will bound the time taken by $\nX\N(t)$ to cross such a neighbourhood (again for $N$ large enough). 

Assume $\eps'<\bar{\eps}/2$, and consider the $\eps'$-neighbourhood $D_{\eps'}$ of $D$. Invoking robust convergence, choose $N_0$ such that for $N\geq N_0$,  $G\N$ coincides with $G$ outside $D_{\eps'}$. We now want to find a neighbourhood $[\bar{t}-\tau',\bar{t}+\tau']$ of the time $\bar{t}$ in which $\nx(\bar{t})\in D$, such that we are guaranteed that if $t$ falls outside this neighbourhood, both $\nx(t)$ and $\nX\N(t)$ are outside $D_{\eps'}$. For any such time $t$, it clearly holds that $G\N_{\x,\nXN}(t)$ and $G\N_{\x}(t)$ coincide.
To find such neighbourhood of $\bar{t}$, let $N_1$ be such that, for $N\geq N_1$, $\|\nX\N(t) - \nx(t)\|$ is less than $\eps'$ with probability $1-\delta$ ($\delta$ to be fixed later). Call $\Omega_{\eps'}$ this event. Condition on it and consider $D_{2\eps'}$. If $\nx(t)\not\in D_{2\eps'}$, then it follows that $\nX\N(t)$ will not belong to $D_{\eps'}$. Hence, we just need to bound the time $\tau_{2\eps'}$ that $\nx(t)$ spends in $D_{2\eps'}$. By Lemma \ref{app:lemma:speedOfCrossing}, this time is  no more than $\frac{4\eps' k}{\rho_0}$.

Now, using Theorem \ref{th:fastSimulation}, choose an $N_2$ such that, for $N\geq N_2$, $Z_k\N(t)$ coincides in $[0,T]$ with $z_k(t)$ with probability at least $1-\delta$. To bound term 2, observe that if $Z\N_k$ and $z_k$ are the same, and conditional on event $\Omega_{\eps'}$, if both $Z\N_k$ and $z_k$ jump at time instants in which $\nx(t)$ is outside $D_{2\eps'}$, then $\chi\N(s,\x,\y)$ and $\chi\N(s,\x,\Y\N)$ will have the same value. 

Therefore, we can bound term (2) by the probability of  $\chi\N(s,\x,\y) \neq \chi\N(s,\x,\Y\N)$, which is itself bounded by 
\[\bbP\{ z_k\ jumps\ in\ D_{2\eps'} \} + \bbP\{\Omega_{\eps'}^c\} + \bbP\{Z\N_k\neq z_k \} \leq \frac{4\eps' \Lambda k}{\rho_0} + 2\delta.\]
 Now, fix $\eps>0$ and choose $\eps' < \min\{ \bar{\eps}/4, \frac{\eps \rho_0}{12 \Lambda k}\}$, and $\delta < \eps/4$. By combining the bounds on term (1) and term (2), we obtain that \[\limsup_{N\rightarrow\infty} |\bbE[\chi(s,\x,\y)] - \bbE[\chi\N(s,\x,\Y\N)]| \leq \frac{6\eps' \Lambda k}{\rho_0} + 2\delta \leq \frac{\eps}{2} + \frac{\eps}{2} = \eps,\] which by the arbitrariness of $\eps$ implies that  \[\lim_{N\rightarrow\infty} |\bbE[\chi(s,\x,\y)] - \bbE[\chi\N(s,\x,\Y\N)]| =0.\] 
Therefore, we obtain that $\bP\N(s,\x) \rightarrow \bP(s,\x)$, and this convergence is uniform with respect to $(s,\x)\in \calS\times E_1$, as the bound derived above is independent of it.

%However, we can reason in the following way. 
%Consider $D_{2\eps'}$, and let $N$ be large enough so that trajectories of $\nXN(t)$, starting from $\prjN{\x}$, are $\eps'$-close 
%to $\flow{t}{\x}$ with probability 1. 
%It follows that, with probability one, the time-varying sets $G\N_{\x,\nXN}(t)$, defined with respect to trajectories of $\Y\N(t)$, 
%coincide with $G_{\x}(t)$ outside $D_{2\eps'}$, and thus with $G\N_{\x}(t)$ (again, for $N$ large enough). 
%Now, with probability converging to one (independently of $(s,\x)$), the trajectories of $z_k$ and $Z\N_k$ are the same, hence 
%for $N$ large enough this probability is larger than $1-\eps'$. 
%For these trajectories, we can see that the only ones for which $\chi\N(s,\x,\y) \neq \chi\N(s,\x,\Y\N)$ are those jumping in 
%$Disc(s,\x,2\eps')$\footnote{This is a consequence of the robustness of $G_{\x}(t)$: in any point  $\bar{t}\in \bigcup_s 
%Disc(s,\x)$, one state $s'$ changes status (either for $G$ or $U$), hence the value of $G_{\x}(t)$ for that state is different before 
%$\bar{t}-2\eps'$ and after $\bar{t}+2\eps'$, for $\eps'$ small enough, and so are the values of $G\N_{\x}(t)$, for $N$ large 
%enough.}, and the probability of this event can be bounded by  $4 n k \eps'$. 
%Concluding, for $N$ large enough, we have that $|\bbE[\chi(s,\x,\y)] - \bbE[\chi\N(s,\x,\Y\N)]| \leq (6nk+1)\eps'$, which proves that 
%$P\N(s,\x) \rightarrow P(s,\x)$, and this convergence is uniform with respect to $(s,\x)$, as the bound derived above is 
%independent of it. 

As for point 2 of the lemma, observe that by the fact that the time-varying set  $V_{\x_0}(t)$  associated with the fluid trajectory $\flow{t}{\x_0}$ is robust,  and by piecewise analyticity of $\bP$, we can choose an $\eps_1$ sufficiently small not only to satisfy the constraints to derive convergence discussed above, but also such that all trajectories in the flow tube $E_1$ are robust, i.e.\  their time varying set with respect to $\bP$ are robust (just observe that the function $\bP(s,\flow{t}{\x})$ is piecewise analytic in $t$ 
and $\x$ for each $s$). Furthermore, we have chosen $\eps_1$ so that the number of intersections of $\flow{t}{\x}$ with $V$ in each state $s$, i.e.\ the number of times $\bP(s,\flow{t}{\x}) - p$ changes sign, is the same as that of $\flow{t}{\x_0}$. 
Now, consider the boundary $\partial V$ in $\calS\times E_1$, which is the zero set of the function $h(s,\x) = \bP(s,\x) - p$. By continuity of $\bP$ and by the robustness property of $V_{\x_0}(t)$, we have that the trajectory $(s,\flow{[0,T]}{\x_0})$ intersects 
$\partial V$ in points $\x^s_i$ in which the function $\bP$ is analytic. Hence, $\bP$ will be analytic in a neighbourhood of $\x^s_i$, and, by a suitable choice of $\eps_1$, $\bP$ will be analytic in the whole component of $\partial V$ containing $\x^s_i$. It follows that $\partial V$ is the union of smooth manifolds (analytic in this case). 
\\
It follows that, by choosing $\eps_1$ suitably small,  $\partial V$ is a d-set and  $V$ is robust.
\\
As for the robust convergence of $V\N$ to $V$, by the uniform convergence of $\bP\N$ to $\bP$ outside an open neighbourhood of $\partial V$, we obtain the robust convergence of $V\N$ to $V$. \qed

%
%
%, notice that $\partial V$ is a d-set. In fact, it is closed being the zero set of a  continuous function, and it is the union of a finite number of smooth manifolds, each one identified by a point in which $\flow{t}{\x_0}$ intersects $\partial V$.\footnote{This is due to the local nature of $\calS\times E_1$, that guarantees that we can restrict to a small neighbourhood of $\partial V$ around each such an intersection point.} Hence, by the uniform convergence of $\bP\N$ to $\bP$ outside an open neighbourhood of $\partial V$, we obtain the robust convergence of $V\N$ to $V$. \qed

%%%%%%%%%%%%%%%%%%%%%%%%%%%%%%%%%%%%%%%%%%%%%%%%5

\begin{applemma}[\ref{lemma:reachabilityRobustSets}]
Let $E_0\subset E$ be a $T,\eps_0$-flow tube for $\x_0$. Let $U$ and $G$ two robust and compatible subsets of $\calS\times E_0$, and $U\N$, 
$G\N$ be sequences of subsets of $\calS\times E_0$ that converge robustly to $U$ and $G$, respectively.

Let $P(s,\x)=P_{reach}(\y,s,\x,T_a,T_b,U,G)$ be the probability that $\y(t)$ reaches a state in $G$ within time $[T_a,T_b]$, 
avoiding any unsafe state in $U$, given that $\y$ started at time $t=0$ in state $(s,\x)\in \calS\times E_0$, and let 
$P\N(s,\x)=P\N_{reach}(\Y\N,s,\prjN{\x},T_a,T_b,U\N,G\N)$ be defined similarly, with $G$, $U$, $\x$ replaced by $G\N$, $U\N$, 
and $\prjN{\x}$, respectively. 
Furthermore, define $V = \{(s,\x) \mid P(s,\x)\bowtie p\}$ and   $V\N = \{(s,\x) \mid P\N(s,\x)\bowtie p\}$. Then there exists 
$\eps_1>0$ such that, in $E_1$, the $(T-T_b),\eps_1$-flow tube for $\x_0$:
\begin{enumerate}
\item $P\N(s,\x) \rightarrow P(s,\x)$ for all $\x\in E_1 \setminus D$, where $D$ is a d-set, uniformly in $(s,\x)$.  
\item If $V_{\x_0}(t)$, $t\in[T^-_{\x_0}(E_1),T^+_{\x_0}(E_1)]$, is a robust time-varying set, then $V$ is robust in $E_1$ and 
$V\N$ converges robustly to $V$. 
\end{enumerate}
\end{applemma}

\proof First, notice that, as in Lemma \ref{lemma:nextProbRobustSets}, we can always restrict on an arbitrary small neighbourhood 
of $\flow{t}{\x_0}$, i.e.\ on a $T,\eps_1$-flow tube $E_1$ for $\x_0$, with $\eps_1$ as small as desired,
implying that the reachability problem for $\Y\N$ for any point in $\calS\times E_1$ will eventually depend only on the goal sets $G\N$ and $U\N$ within $\calS\times E_0$. 

We first prove point 1 of the lemma in a $(T-T_b),\eps_1$-flow tube $E_1$ for $\x_0$, for an $\eps_1<\eps_0/2$ ($\eps_1$ will 
be fixed in the following). 
Consider the set $D$ in $E_0$, $D = Disc(G)\cup Disc(U)\cup \flowi{T_a}{Disc(G)} \cup \flowi{T_a}
{Disc(U)} \cup \flowi{T_b}{Disc(G)} \cup \flowi{T_b}{Disc(U)}$, containing the discontinuity points of $G$ and $U$ and all points 
that are mapped by the flow to $Disc(G)\cup Disc(U)$ after $T_a$ or $T_b$ units of time. $D$ is easily seen to be a d-set. In fact,  it is 
closed and it intersects each trajectory a finite number of times, as $Disc(G)$ and $Disc(U)$ are d-sets. Furthermore, each smooth manifold of $G$ or $U$, defined as the zero set of the  function $h(\x)$, will be mapped by $\flowi{T_j}{\cdot}$, $j=a,b$, into the smooth manifold defined by the function $h(\flow{t_j}{\x})$. This function is smooth as $\flow{t}{\x}$ is piecewise analytic and it is at least of class $\calC^1$.

Differently from Lemma \ref{lemma:nextProbRobustSets}, we will prove convergence of $P\N$ to $P$ for each 
$(s,\x)\in (\calS\times E_1)\setminus D$.\\
Consider now the set $D_0  = Disc(G)\cup Disc(U)$, and define the $\eps$-neighbourhood $D_\eps$ of it as done in Lemma \ref{lemma:nextProbRobustSets}. It clearly holds that $D_{\eps}\rightarrow D_0$, as $\eps\rightarrow 0$. 

%
%Due to robustness of $G$ and $U$, the time-varying sets $G_{\x}(t)\subset\calS$ and $U_{\x}(t)\subset\calS$ (defined with respect 
%to $\y$) are robust (in the sense of Definition \ref{def:robustSet}). 
%Furthermore, as $G\N$ and $U\N$ converge robustly to $G$ and $U$, if follows that the time-varying sets 
%$G\N_{\x}(t)\subset\calS$ and $U\N_{\x}(t)\subset\calS$ converge robustly to $G_{\x}(t)$ and  $U_{\x}(t)$, respectively.

We will now use an argument similar to the one of Lemma \ref{lemma:tameTimeDepReachab}. Let $\chi$  (resp.\ $\chi\N$) be 
random variables defined on sample trajectories and equal to one if the trajectory satisfies the reachability problem of the Lemma with respect to $G,U$ (resp.\ $G\N,U\N$). 
Then, as $P(s,\x) = \bbE[\chi(s,\x,\y(\omega))]$, where $\y(0) = (s,\x)$, and similarly for $P\N(s,\x)$, to show convergence we just 
need to prove that $|\bbE[\chi(s,\x,\y)] - \bbE[\chi\N(s,\x,\Y\N)] |\rightarrow 0$. It holds that: 
{\small  
\begin{eqnarray*}
|\bbE[\chi(s,\x,\y)] - \bbE[\chi\N(s,\x,\Y\N)]| & \leq & \underbrace{\bbE[|\chi(s,\x,\y) - \chi\N(s,\x,\y) |]}_{(1)}\\
& + & \underbrace{\bbE[| \chi\N(s,\x,\y) -\chi\N(s,\x,\Y\N) |],}_{(2)}\\
\end{eqnarray*} }

From Lemma \ref{app:lemma:speedOfCrossing},  we obtain constants $\bar{\eps}_1$ and $\bar{\eps}$ that bound the size of the flow tube $E_1$, and of the $D_{\eps'}$ neighbourhood of $D_0$. Under these constraints, we can reason exactly as in the proof of Lemma \ref{lemma:nextProbRobustSets} to bound terms (1) and (2) by $\frac{6\eps' \Lambda k}{\rho_0} + 2\delta$, where $\delta$ and $\eps'<\bar{\eps}/2$ can be chosen arbitrary small for $N$ large enough, concluding that $|P\N(s,\x) - P(s,\x)|$ converges to zero, uniformly in $(\calS\times E_1)\setminus D$.

%
%
%we let $\bar{\eps}$ and $\rho_0$ the constants that it provides, further assuming that the value of $\eps_1$ we consider is smaller than the one required by Lemma \ref{app:lemma:speedOfCrossing}. Choosing a neighbourhood $D_{\eps'}$ for $\eps'<\bar{\eps}$, term (1) can be treated as in Lemma \ref{lemma:nextProbRobustSets}, bounding it by $\frac{2\eps' \Lambda k}{\rho_0}$, where $\lambda$ is an upper bound for the jump rate of $z_k$ in $\calS\times E_0$ and $k$ is the 
% number of intersections of a trajectory in $\calS\times E_1$ with $G$ and $U$.
%
%As for the bound on term (2), we can reason as in Lemma \ref{lemma:nextProbRobustSets}, obtaining for $\eps'<\bar{\eps}/2$ a bound of $\frac{6\eps' \Lambda k}{\rho_0} + 2\delta$, for and $N$ sufficiently large so that $\nX\N(t)$ is $\eps'$-close to $\nx(t)$ with probability $1-\delta$, and $Z\N_k(t) = z_k(t)$ with probability $1-\delta$. From this, we conclude that 
%$|P\N(s,\x) - P(s,\x)|$ converges to zero, uniformly in $(\calS\times E_1)\setminus D$.

As for point 2 of the lemma, robustness of $V$ follows by the same argument as Lemma \ref{lemma:nextProbRobustSets}. 
Notice that $Disc(V)$ is closed, as it is the union of the zero sets of continuous functions (the analytic pieces of $P$), plus the 
subset $D_p$ of discontinuity points of $P$ such that $\liminf P(s,\x) \leq p$ and $\limsup P(s,\x) \geq p$, which is also closed. Furthermore, by robustness of $V_{\x_0}(t)$, we can choose $\eps_1$ such that all points in $D_p$ satisfy $\liminf P(s,\x) < p$ and $\limsup P(s,\x) > p$ (we need this because $V_{\x}(t)$ has to be robust for all $\x$ in $\calS\times E_1$). 
%
%
%observe that by the fact that the fluid trajectory $\flow{t}{\x_0}$ is robust for $V$ and by piecewise analyticity of $P$, we can choose an $\eps_1$ sufficiently small such that all trajectories in the flow tube $E_1$ are robust, i.e. their time varying set with respect to $P$ is robust (just observe that the function $P(s,\flow{t}{\x})$ is piecewise analytic in $t$ and $\x$ for each $s$). Furthermore, we can choose $\eps_1$ so that the number of intersections of $\flow{t}{\x}$ with $V$ in each state $s$, i.e. the number of time s $P(s,\flow{t}{\x}) - p$ changes sign, is the same as that of $\flow{t}{\x_0}$. It follows that $V$ is robust for this choice of $\eps_1$.
For the robust convergence of $V\N$ to $V$, $Disc(V)$ is a d-set, hence we can use uniform convergence of $P\N$ to $P$ 
outside an open neighbourhood of $Disc(V)$. Additionally, notice that, as in the proof of Lemma \ref{lemma:tameTimeDepReachab}, the points in which we do not have convergence of $P\N$ to $P$ and that are not in $Disc(V)$, 
do not create problems, as in a small neighbourhood of those points, $P$ is always strictly above or below $p$, and the lim sup 
or the lim inf of $P\N$  in those points will  uniformly satisfy the inequality defining $V\N$ (thanks to the compatibility condition of $G$ and $U$).\qed

%%%%%%%%%%%%%%%%%%%%%%%%%%%%%%%%%%%%%%%%%%%%%%%%5

\begin{applemma}[\ref{lemma:convergenceSpaceMC}]
Let $\calXN$ be a sequence of CTMC models, as defined in Section \ref{sec:modelingLanguage}, and 
let $Z\N_k$ and $z_k$ be defined from $\calXN$ as in Section \ref{sec:fastSimulation}.\\
Assume that there is a flow tube $E_0$ of $\x_0$ such that all trajectories in $E_0$ are piecewise analytic.\\
Let $\phi = \phi(\vr{p})$ be a robust CSL formula for the trajectory $\flow{t}{\x_0}$. Then, there is an $N_0$ such that, for all $N\geq N_0$, $$s,\x_0 \models_{\y} \phi\;\;  \Leftrightarrow\;\; s, \prjN{\x_0} \models_{\Y\N}  \phi.$$ 
\end{applemma}

\proof We will prove by structural induction on the formula $\phi$, that there is a $T,\eps$-flow tube $E_\phi$ of $\x_0$ such that $V\N_{\phi}$  converges to $V_{\phi}$ robustly, where $V_{\phi}$ is the set of points $(s,\x)\in \calS\times E_\phi$ such that $s,\x \models_{\y} \phi$ and $V\N_{\phi}$ is the set of points $(s,\x)\in \calS\times E_\phi$ such that $s, \prjN{\x_0} \models_{\Y\N}  \phi$.  

\begin{description}
\item[Base case:] the result for atomic formulae is trivial as their truth value depends only on $s$, hence we can choose $E_{\phi} = E_0$ and  $V_{\phi} = V\N_{\phi} = \calS_{\phi}\times E_0$, where $\calS_\phi = \{s~|~s\models \phi\}$. 
\item[Negation:] If $\phi = \neg \phi_1$, we can choose $E_\phi = E_{\phi_1}$, and simply observe that robust convergence of $V\N_{\phi_1}$ to $V_{\phi_1}$ implies robust convergence of $V\N_{\phi} = E_{\phi_1}\setminus V\N_{\phi_1}$ to $V_{\phi} = E_{\phi_1}\setminus V_{\phi_1}$.
\item[Conjunction and Disjunction:]
If $\phi = \phi_1 \circ \phi_2$, $\circ \in \{\wedge,\vee\}$, consider $E_{\phi_i}$, a $T,\eps_i$-flow tube, and sets $V\N_{\phi_i}\rightarrow V_{\phi_i}$. 
As $\phi$ is robust for the trajectory starting in $\x_0$ and $\x_0\in E_{\phi_i}$, by the  compatibility condition of $\phi$ there exists an $\eps$ such that the d-sets of $V_{\phi_1}$ and $V_{\phi_2}$ are disjoint in $\calS\times E_\phi$, for $E_\phi$ the $T,\eps$-flow tube in $\x_0$. 
It easily follows that $V_\phi =  V_{\phi_1} \bullet V_{\phi_2}$ is robust in $\calS\times E_\phi$, $\bullet\in\{\cap,\cup\}$,  and $V\N_{\phi_1}\bullet V\N_{\phi_2} \rightarrow V_{\phi_1} \bullet V_{\phi_2}$ robustly. 
\item[Next:] If $\phi = \calP_{\bowtie p}( \next{T_1}{T_2} \phi_1)$, let $E_{\phi_1}$ be a $T,\eps$-flow tube for $\phi_1$ and let $V_{\phi_1}$ be a robust set, such that $V\N_{\phi_1}\rightarrow V_{\phi_1}$ robustly.
By considering the $\eps,T$-flow tube $E_j$ for $\x_0$, and using the robustness of $\phi$, we satisfy the hypothesis of Lemma \ref{lemma:nextProbRobustSets}, hence there is an $(T-T_2),\eps$-flow tube $E_\phi$ for $\x_0$ such that $V_\phi$ is robust in $\calS\times E_\phi$ and $V\N_{\phi}\rightarrow V_{\phi}$ robustly.

\item[Until:] If $\phi = \calP_{\bowtie p}(\phi_1 \until{T_1}{T_2} \phi_2)$, let $E_{\phi_i}$ be  $T,\eps_i$-flow tubes for $\phi_i$, $i=1,2$, and robust sets $V_{\phi_i}$, such that $V\N_{\phi_i}\rightarrow V_{\phi_i}$ robustly. By letting $\eps_0 < \min\{\eps_1,\eps_2\}$, such that $\partial V_{\phi_i} \cap \partial V_{\phi_2}=\emptyset$ (which can be found by the compatibility condition enforced by  robustness of $\phi$), considering the $\eps_0,T$-flow tube $E_j$ for $\x_0$, and using the robustness of $\phi$, we satisfy the hypothesis of Lemma \ref{lemma:reachabilityRobustSets}, hence there is an $(T-T_2),\eps$-flow tube $E_\phi$ for $\x_0$ such that $V_\phi$ is robust in $\calS\times E_\phi$ and $V\N_{\phi}\rightarrow V_{\phi}$ robustly.
\end{description}
Given a formula $\phi$, and the flow-tube $E_\phi$ for $\x_0$, such that $V_\phi$ is robust in $\calS\times E_\phi$ and $V\N_{\phi}\rightarrow V_{\phi}$ robustly, then the lemma follows by observing that, due to robustness of $\phi$, $\x_0$ does not belong to the d-set  $\partial V_\phi$, hence there is an $N_0$ such that, for all $N\geq N_0$, $(s,\x_0)\in V\N_\phi \Leftrightarrow (s,\x_0)\in V_\phi$. \qed

%%%%%%%%%%%%%%%%%%%%%%%%%%%%%%%%%%%%%%%%%%%%%%%%5

\end{document}